\documentclass[aps,prd,preprintnumbers,superscriptaddress,showpacs,showkeys,floatfix,amsmath,amssymb,amsfonts,twocolumn]{revtex4-2}

\usepackage{graphicx}
\usepackage{makecell}
\usepackage{multirow}
\usepackage{xparse} 
\usepackage{float}
\usepackage{xcolor}
\NewDocumentCommand\Nf{mgg}{N\textsubscript{f}=#1\IfNoValueTF{#2}{}{+#2}\IfNoValueTF{#3}{}{+#3}}
\NewDocumentCommand\vol{mg}{#1\textsuperscript{3}\IfNoValueTF{#2}{}{×#2}}

\newcommand{\tins}{t_\mathrm{ins}}
\newcommand{\xins}{x_\mathrm{ins}}
\newcommand{\vxins}{\vec{x}_\mathrm{ins}}
\usepackage[normalem]{ulem}

\usepackage{hyperref}
\usepackage{slashed}
\begin{document}

\title{Proton and neutron electromagnetic form factors in the continuum limit
using lattice QCD ensembles with physical pion masses}

\author{Constantia Alexandrou} \affiliation{Computation-based Science and Technology Research Center, The Cyprus Institute, 20 Kavafi Str., Nicosia 2121, Cyprus}\affiliation{Department of Physics, University of Cyprus, P.O. Box 20537, 1678 Nicosia, Cyprus}
\author{Simone Bacchio} \affiliation{Computation-based Science and Technology Research Center, The Cyprus Institute, 20 Kavafi Str., Nicosia 2121, Cyprus}
\author{Giannis Koutsou} \affiliation{Computation-based Science and Technology Research Center, The Cyprus Institute, 20 Kavafi Str., Nicosia 2121, Cyprus}
\author{Bhavna Prasad} \affiliation{Computation-based Science and Technology Research Center, The Cyprus Institute, 20 Kavafi Str., Nicosia 2121, Cyprus}
\author{Gregoris Spanoudes}\affiliation{Department of Physics, University of Cyprus, P.O. Box 20537, 1678 Nicosia, Cyprus}

\date{\today}

\begin{abstract} 
  We compute the electromagnetic form factors of the proton and
  neutron using lattice QCD. We employ \Nf{2}{1}{1} twisted mass
  clover-improved fermions with quark masses tuned to their physical
  values. Three ensembles with lattice spacings of $a$=0.080~fm,
  0.068~fm, and 0.057~fm, and approximately the same physical volume
  allow us to obtain the continuum limit directly at the physical pion
  mass. For each ensemble, we use several values of the sink-source
  time separation, ranging from 0.5~fm to 1.5~fm, to allow for a
  thorough analysis of excited state effects via multi-state fits. The
  disconnected contributions are also analyzed using high statistics
  combined with techniques to mitigate stochastic noise in the
  estimation of the fermion loop. These techniques include low-mode
  deflation, dilution in the color and spin components, and
  hierarchical probing. We study the momentum transfer dependence of
  the form factors using the $z$-expansion and dipole Ans\"atze,
  thereby enabling the extraction of the electric and magnetic radii
  and the magnetic moments, as well as the Zemach and Friar radii in
  the continuum limit.  Results for the proton and neutron electric
  and magnetic mean square radii are $\sqrt{\langle r_E^2\rangle^p} =
  0.860(38)(23)$~fm, $\langle r_E^2\rangle^n = -0.147(48)$~fm$^2$,
  $\sqrt{\langle r_M^2\rangle^p} = 0.870(53)(15)$~fm and
  $\sqrt{\langle r_M^2\rangle^n} = 0.913(67)(19)$~fm, and for the
  proton and neutron magnetic moments $\mu^p=2.849(92)(52)$ and
  $\mu^n=-1.819(76)(29)$, respectively. In all cases, the first error
  is statistical and the second systematic, where the latter includes
  an estimate of the error from the fits to the momentum dependence of
  the form factors and from the continuum extrapolation.

\end{abstract}

\maketitle

\section{Introduction}
The study of proton and neutron electromagnetic form factors offers
insights into the rich internal electromagnetic structure of the
hadronic world. The proton, being a stable particle, has been used as
the prototype system for experimental investigations of hadronic bound
states. Thus, the proton electromagnetic form factors are known to
high precision from electron-proton scattering~\cite{A1:2013fsc}. The
neutron form factors on the other hand are accessed indirectly
experimentally through electron-deuteron or electron-helium scattering
and, therefore, quantities, such as the neutron electric charge
radius, remain less well known compared to the proton. Despite the
many years of experimental and theoretical
studies~\cite{Punjabi:2015bba} the electromagnetic form factors of the
proton and the neutron still remain an active area of research. In
particular, new experiments are reaching high precision in the
extraction of charges, moments, and radii of the
proton~\cite{A1:2013fsc,Pohl:2010zza,Golak:2000nt} but also improving
the accuracy of the corresponding neutron
quantities~\cite{Xiong:2019umf}. The computation of these fundamental
quantities from lattice QCD, especially for the proton, thus, provides
a benchmark for the lattice formulation validating non-perturbative
calculations. This in turn and most importantly enables us to use
lattice QCD to evaluate hadron structure quantities that are less well
known experimentally.

In the past five years, lattice QCD simulations are being carried out with quark mass  parameters approximately equal to their physical values, thus enabling us to extract hadron structure quantities without the need of a chiral extrapolation. However,  to date computations of the electromagnetic form factors
either lack a continuum extrapolation or rely on the inclusion of
ensembles with heavier-than-physical quark masses to carry out the
continuum and chiral
extrapolation~\cite{Djukanovic:2023beb,Djukanovic:2023jag,Alexandrou:2018sjm,Tsuji:2023llh,Jang:2019jkn}.
In this work, we provide a calculation of the electromagnetic form
factors of the proton and neutron using lattice QCD on three ensembles
of clover-improved twisted mass fermions with two degenerate light,
strange, and charm quarks (\Nf{2}{1}{1}) with masses tuned to their
physical values (physical point). The lattice spacings span
$a$=0.080~fm, 0.068~fm, and 0.057~fm, allowing us to take the
continuum limit directly at the physical pion mass, thus, eliminating
the need for chiral extrapolations.

The remainder of this paper is organized as follows: In
Sec.~\ref{sec:theory} we define the nucleon electromagnetic form
factors, radii and magnetic moments; In Sec.~\ref{sec:lattice} we
provide details on the setup of our lattice QCD calculation, including
details of the ensembles used and statistics; In
Sec.~\ref{sec:extraction}, we explain the techniques used to extract
the form factors from the lattice QCD correlators, including the
methods used for disconnected contributions, our approaches for
identifying ground-state dominance, and the fits of the momentum
transfer-dependence of the form factors. The results on the proton and
neutron form factors for each gauge ensemble together with the fits to
their momentum transfer-dependence are carried out in
Sec.~\ref{sec:fits}, where we also include their continuum
extrapolation. Final results and comparisons with the literature are
carried out in Sec.~\ref{sec:final_results} and in
Sec.~\ref{sec:conclusions}, we summarize and provide our
conclusions. For completeness, we include in
Appendix~\ref{sec:appendix_equations} the decomposition of the nucleon
matrix elements in terms of the form factors and in
Appendix~\ref{sec:appendix_results}, we provide tables with the
numerical results for the electric and magnetic form factors as well
as the fit parameters that reproduce our final result.

\section{Nucleon electromagnetic form factors}
\label{sec:theory}
We consider the electromagnetic form factors of the nucleon in the
SU(2) flavor isospin symmetric limit, i.e. neglecting isospin-breaking
effects due to QED interactions and the u–d quark mass difference. The
proton and neutron are thus degenerate, and the form factors are given
in terms of the matrix element of the electromagnetic current with
nucleon states in Minkowski space:
\begin{eqnarray}
  && \langle N(p',s') \vert j_\mu \vert N(p,s) \rangle = \sqrt{\frac{m_N^2}{E_N(\vec{p}\,') E_N(\vec{p})}} \times   \label{eq:me} \\
  && \bar{u}_N(p',s') \left[ \gamma_\mu F_1(q^2) + \frac{i \sigma_{\mu\nu} q^\nu}{2 m_N} F_2(q^2) \right] u_N(p,s)\,\nonumber,
\end{eqnarray}
where $N(p,s)$ is the nucleon with initial (final) momentum $p$ ($p'$)
and spin $s$ ($s'$), with energy $E_N(\vec{p})$ ($E_N(\vec{p}\,') $)
and mass $m_N$, $u_N$ is the nucleon spinor, 
$j_\mu$ is the vector current, and $q^2{\equiv}q_\mu q^\mu$ is the momentum transfer
squared with $q_\mu{=}(p_\mu'-p_\mu)$. In Eq.~(\ref{eq:me}), the form factors of the nucleon are given
in terms of the Dirac ($F_1$) and Pauli ($F_2$) form factors. Alternatively, we can compute the electric and magnetic Sachs form
factors $G_E(q^2)$ and $G_M(q^2)$, which  are related to the Dirac and Pauli form factors via
\begin{eqnarray}
  G_E(q^2) &=& F_1(q^2) + \frac{q^2}{4m_N^2} F_2(q^2)\,,\nonumber\\
  G_M(q^2) &=& F_1(q^2) + F_2(q^2)\,.\label{eq:sachs}
\end{eqnarray}
The local vector current is given by,
\begin{equation}
  j_\mu = \sum_{\mathsf{f}=u,d,s,c} e_\mathsf{f} j^\mathsf{f}_\mu = \sum_{\mathsf{f}=u,d,s,c} e_\mathsf{f} \bar{\mathsf{\psi_f}} \gamma_\mu \mathsf{\psi_f}\,,\label{eq:current}
\end{equation}
where the sum over $\mathsf{f}$ runs over the up-, down-, strange- and
charm-quark flavors (u, d, s and c, respectively) and $e_\mathsf{f}$
is the electric charge of the quark with flavor $\mathsf{f}$. In the
twisted mass formulation, the lattice conserved vector current after
symmetrization is given by,
\begin{eqnarray}
  j_\mu^\mathsf{f}(x) = \frac{1}{4} &[&\bar{\mathsf{\psi_f}}(x{+}\hat{\mu}) U^\dagger_\mu (x) (1{+}\gamma_\mu) \mathsf{\psi_f}(x) \nonumber\\
    &+& \bar{\mathsf{\psi_f}}(x) U^\dagger_\mu(x{-}\hat{\mu}) (1{+}\gamma_\mu) \mathsf{\psi_f}(x{-}\hat{\mu}) \nonumber \\
     &-&\bar{\mathsf{\psi_f}}(x) U_\mu(x) (1{-}\gamma_\mu) \mathsf{\psi_f}(x{+}\hat{\mu})\nonumber\\
    &-& \bar{\mathsf{\psi_f}}(x{-}\hat{\mu}) U_\mu(x{-}\hat{\mu}) (1{-}\gamma_\mu) \mathsf{\psi_f}(x) \;]\,,\label{eq:conserved}
\end{eqnarray}
where $U_\mu(x)$ is the gauge link of lattice site $x$ in direction
$\hat{\mu}$. Within SU(2) considered in this work, we can extend the
notation of the flavor composition of the current and use $j_\mu^{u-d}
= j_\mu^u - j_\mu^d$ to denote the isovector current and $j_\mu^{u+d}
= j_\mu^u + j_\mu^d$ to denote the isoscalar current. The standard
electromagnetic current is then given as,
\begin{equation}
  j_\mu = \frac{2}{3}j^u_\mu - \frac{1}{3} j^d_\mu = \frac{1}{2}j^{u-d}_\mu + \frac{1}{6}j^{u+d}_\mu
\end{equation}
and the proton ($G^p_E$ and $G^p_M$) and neutron ($G^n_E$ and $G^n_M$)
electromagnetic form factors are given in terms of the isovector
($G^{u-d}_E$ and $G^{u-d}_M$) and isoscalar ($G^{u+d}_E$ and
$G^{u+d}_M$) form factors, via
\begin{eqnarray}
  G_X^p(q^2) &= &\frac{1}{2}G_X^{u-d}(q^2) + \frac{1}{6}G_X^{u+d}(q^2)\nonumber\\
  G_X^n(q^2) &= -&\frac{1}{2}G_X^{u-d}(q^2) + \frac{1}{6}G_X^{u+d}(q^2),\label{eq:vstopn}
\end{eqnarray}
where $X=E,\,M$.

At zero momentum transfer ($q^2=0$), the electric form factor yields
the charge of the hadron and the magnetic form factor the magnetic
moment, i.e.
\begin{eqnarray}
  G_E^p(0) =& 1\,,\,\, G_E^n(0) &= 0\,, \nonumber\\
  G_M^p(0) =& \mu_p\,,\,\, G_M^n(0) &= \mu_n.
\end{eqnarray}
When using the lattice conserved current, these relations hold by
symmetry, with no renormalization of the currents required.

The electric and magnetic mean-squared radii are defined by the slope
of the corresponding Sachs form factor as $q^2\rightarrow 0$, namely
\begin{equation}
  \langle r_X^2 \rangle^{\mathsf{f}} = \frac{-6}{G_X^\mathsf{f}(0)}
  \frac{\partial G_X^\mathsf{f}(q^2)}{\partial q^2} \Big
  \vert_{q^2=0}\,.
  \label{eq:radius}
\end{equation}

\section{Lattice setup}
\label{sec:lattice}
\subsection{Lattice correlation functions}
Nucleon matrix elements are computed within lattice QCD via
appropriate combinations of two- and three-point correlation
functions. These are defined in a Euclidean space-time, and we
will thus use notation for Euclidean quantities from here on,
including expressing the form factors in terms of $Q^2=-q^2$. We use
the standard nucleon interpolating field,
\begin{equation}
  \chi_N(\vec{x},t)=\epsilon^{abc}u^a(x)[u^{b\intercal}(x)\mathcal{C}\gamma_5d^c(x)]\,,
\end{equation}
where $\mathcal{C}{=}\gamma_0 \gamma_2$ is the charge conjugation
matrix.  The two-point function in momentum space is given by
\begin{align}
C(\Gamma_0,\vec{p};t_s,t_0) = \sum_{\vec{x}_s} &
e^{{-}i (\vec{x}_s{-}\vec{x}_0) \cdot \vec{p}}{\times} \label{eq:twop}\\
&\mathrm{Tr} \left[ \Gamma_0 {\langle}\chi_N(t_s,\vec{x}_s) \bar{\chi}_N(t_0,\vec{x}_0) {\rangle} \right] \,,\nonumber
\end{align}
and the three-point function is given by
\begin{align}
  C_\mu(\Gamma_\nu,\vec{q},\vec{p}\,';t_s,\tins,t_0) {=} &
  \sum_{\vxins,\vec{x}_s}  e^{i (\vxins {-} \vec{x}_0)  \cdot \vec{q}}  e^{-i(\vec{x}_s {-} \vec{x}_0)\cdot \vec{p}\,'} {\times} \nonumber \\
  \mathrm{Tr} [ \Gamma_\nu \langle \chi_N(t_s,\vec{x}_s) & j_\mu(\tins,\vxins) \bar{\chi}_N(t_0,\vec{x}_0) \rangle].
  \label{eq:thrp}
\end{align}
The initial lattice site where the nucleon is created is denoted by  $x_0$ and referred to as the
\textit{source}, the lattice site where the current couples to a quark  $\xins$ as the \textit{insertion}, and the site where the nucleon is annihilated  $x_s$ as
the \textit{sink}. $\Gamma_\nu$ is a projector acting on Dirac
indices, with $\Gamma_0 {=} \frac{1}{2}(1{+}\gamma_0)$ yielding the
unpolarized and $\Gamma_k{=}\Gamma_0 i \gamma_5 \gamma_k$ the
polarized matrix elements. Without loss of generality we will take
$t_s$ and $\tins$ relative to the source time $t_0$ in what follows.

Inserting a complete set of states in Eq.~(\ref{eq:twop}) yields,
\begin{align}
  C(\vec{p},t_s) =& \sum_{n}c_n(\vec{p}) e^{-E_n(\vec{p}) t_s},\,\mathrm{where}\label{eq:twop_spec}\\
  c_n(\vec{p}) =& \sum_{s}\mathrm{Tr}[ \Gamma_0 \langle \chi_N | n, \vec{p}, s \rangle \langle n, \vec{p}, s \vert \bar{{\chi}}_N  \rangle],\nonumber
\end{align}
with $| n, \vec{p}, s \rangle$ a QCD eigenstate with the quantum
numbers of the nucleon, spin $s$, momentum $\vec{p}$, and energy
$E_n(\vec{p})$. Similarly, inserting two complete sets of states in
Eq.~(\ref{eq:thrp}) yields,
\begin{align}
  C_{\mu}(\Gamma_\nu,\vec{q},\vec{p}\,';t_s,\tins) =& \label{eq:thrp_spec}\\
  \sum_{n,m}  \mathcal{A}^{n,m}_{\mu}(\Gamma_\nu,\vec{q},\vec{p}\,') &
    e^{-E_n(\vec{p}\,')(t_s-\tins)-E_m(\vec{q})\tins},\,\mathrm{where}\nonumber\\
  \mathcal{A}^{n,m}_{\mu}(\Gamma_\nu,\vec{q},\vec{p}\,') =  &\nonumber\\
  \sum_{s,s'}\mathrm{Tr}[\Gamma_\nu \langle \chi_N | n, \vec{p}\,',s'\rangle& \langle n, \vec{p}\,',s' | j_\mu | m, \vec{p}, s \rangle \langle m, \vec{p}, s | \bar{\chi}_N  \rangle].\nonumber
\end{align}
The desired nucleon matrix element is obtained for $m=n=0$ in
Eq.~(\ref{eq:thrp_spec}) and after canceling the overlaps $\langle
\chi_N | 0, \vec{p}\,',s'\rangle$ and $\langle 0, \vec{p}, s |
\bar{\chi}_N \rangle$. One way to achieve this is by using an
optimized ratio composed of two- and three-point functions, given by

\begin{align}
R_{\mu}(\Gamma_{\nu},\vec{p},\vec{p}\,';t_s,\tins) = \frac{C_{\mu}(\Gamma_{\nu},\vec{p},\vec{p}\,';t_s,\tins\
)}{C(\Gamma_0,\vec{p}\,';t_s)} \times \nonumber\\
\sqrt{\frac{C(\Gamma_0,\vec{p};t_s-\tins) C(\Gamma_0,\vec{p}\,';\tins) C(\Gamma_0,\vec{p}\,';t_s)}{C(\Gamma_0,\vec{p}\,';t_s-\tins) C(\Gamma_0,\vec{p};\tins) C(\Gamma_0,\vec{p};t_s)}}.
\label{eq:full_ratio}
\end{align}
In the limit of large time separations $(t_s-\tins) \gg$ and $\tins
\gg$, the ratio in Eq.~(\ref{eq:full_ratio}) converges to the nucleon
ground state matrix element, which we can express in terms of the
amplitudes in Eqs.~(\ref{eq:twop_spec}) and~(\ref{eq:thrp_spec}) as
\begin{equation}
  \Pi^\mu(\Gamma_\nu;\vec{p}\,', \vec{q}) =
  \frac{\mathcal{A}^{0,0}_{\mu}(\Gamma_\nu,\vec{q},\vec{p}\,')}{\sqrt{c_0(\vec{p})
      c_0(\vec{p}\,')}}.
      \label{eq:gsmatrix}
\end{equation}
In Sec.~\ref{sec:extraction}, we detail the analysis we follow to
extract the ground state ($n=m=0$) matrix element via a combination of
multi-state fits to the two- and three-point functions.  The  ratio in Eq.~(\ref{eq:full_ratio}) will be used only for visualization
purposes of the fits to the two- and three-point functions.

\subsection{Gauge ensembles and statistics}
\label{sec:ens}
We use the twisted-mass fermion discretization
scheme~\cite{Frezzotti:2003ni,Frezzotti:2000nk}, which provides
automatic ${\cal O}(a)$-improvement~\cite{ETM:2010iwh}.  The bare
light quark parameter $\mu_l$ is tuned to reproduce the isosymmetric
pion mass $m_\pi=135$~MeV~\cite{Alexandrou:2018egz,
  Finkenrath:2022eon}, while the the heavy quark parameters, $\mu_s$
and $\mu_c$, are tuned using the kaon mass and an appropriately defined
ratio between the D-meson mass and decay constant as well as the ratio
between charm and strange quark mass, following the procedure of
Refs.~\cite{Finkenrath:2022eon,Alexandrou:2018egz}. A clover term is
included in the action that reduces isospin-breaking effects.  The
parameters of the three ensembles analyzed in this work can be found
in Table~\ref{tab:ens}. The lattice spacings and pion masses are taken
from Ref.~\cite{ExtendedTwistedMass:2022jpw}. We quote the values of
the lattice spacing obtained from the meson sector, which are
compatible with the values determined from the nucleon mass in
Ref.~\cite{ExtendedTwistedMass:2021gbo}.

\begin{table}[h]
    \caption{Parameters of the three \Nf{2}{1}{1} ensembles analyzed
      in this work. From the leftmost to rightmost columns, we provide
      the name of the ensemble and its short acronym in parenthesis, the lattice volume, $\beta=6/g^2$ with
      $g$ the bare coupling constant, the lattice spacing, the pion
      mass, and the value of $m_\pi L$. Lattice spacings and pion
      masses are taken from
      Ref.~\cite{ExtendedTwistedMass:2022jpw}.}
    \label{tab:ens}
    \begin{tabular}{cccccc}
    \hline\hline
       Ensemble   & $(\frac{L}{a})^3{\times}(\frac{T}{a})$ & $\beta$ & \makecell[c]{$a$\\$[$fm$]$} & \makecell[c]{$m_\pi$\\ $[$MeV$]$}  & $m_\pi L$ \\
       \hline 
        \texttt{cB211.072.64} (\texttt{B})& $64^3 {\times} 128$ & 1.778 &  0.07957(13) & 140.2(2) & 3.62 \\
        \texttt{cC211.060.80} (\texttt{C})& $80^3 {\times} 160$ & 1.836 &  0.06821(13) & 136.7(2) & 3.78 \\
        \texttt{cD211.054.96} (\texttt{D})& $96^3 {\times} 192$ & 1.900 &  0.05692(12) & 140.8(2) & 3.90 \\
        \hline
    \end{tabular}
\end{table}

To increase the overlap of the interpolating field with the nucleon
ground state, so that excited states are suppressed at earlier time
separations, we use Gaussian smeared point
sources~\cite{Gusken:1989qx,Alexandrou:1992ti},
\begin{equation}
    \psi^\mathrm{sm}(\vec{x}, t) = \sum_{\vec{y}} [\mathbf{1} +
      a_G H(\vec{x}, \vec{y}; U(t))]^{N_G} \psi(\vec{y},t),
\end{equation}
where the hopping matrix is given by
\begin{eqnarray}
H(\vec{x},\vec{y};U(t)) = \sum_{i=1}^3 \left[ U_i(x) \delta_{x,y-\hat{i}} + U_i^\dag(x-\hat{i}) \delta_{x,y+\hat{i}}  \right].
\end{eqnarray}
The parameters $a_G$ and $N_G$ are
tuned~\cite{Alexandrou:2018sjm,Alexandrou:2019ali} in order to
approximately give a root mean square (r.m.s) radius for the nucleon of $ 0.5$~fm. For the links
entering the hopping matrix, we apply APE
smearing~\cite{APE:1987ehd} to reduce statistical errors due to
ultraviolet fluctuations. The smearing parameters are tuned separately
for each value of the lattice spacing~\cite{Alexandrou:2022dtc} and are given in
Table~\ref{tab:smearing}.

\begin{table}
  \caption{Number of Gaussian smearing iterations ($N_G$) and Gaussian
    smearing parameter $a_G$ used for each ensemble. We also provide
    the number of APE-smearing iterations $n_{\rm APE}$ and parameter
    $\alpha_{\rm APE}$ applied to the links that enter the Gaussian
    smearing hopping matrix. The resulting quark r.m.s. radius is
    given in the last column as $\langle r^2\rangle_{\psi^\mathrm{sm}}=\sum_{\vec{r}} r^2
    |\psi^\mathrm{sm}(\vec{r})|^2/\sum_{\vec{r}}
    |\psi^\mathrm{sm}(\vec{r})|^2$, where the error is due to
    the uncertainty in the lattice spacing.}\label{tab:smearing}
  \begin{tabular}{cccccc}
    \hline\hline
    Ensemble & $N_G$ & $a_G$ & $n_{\rm APE}$ & $\alpha_{\rm APE}$ & $\sqrt{\langle r^2\rangle_{\psi^\mathrm{sm}}}$ [fm]\\
    \hline
    \texttt{cB211.072.64} & 125 & 0.2 & 50 & 0.5 & 0.461(2) \\
    \texttt{cC211.060.80} & 140 & 1.0 & 60 & 0.5 & 0.516(2) \\
    \texttt{cD211.054.96} & 200 & 1.0 & 60 & 0.5 & 0.502(3) \\    
    \hline
  \end{tabular}
\end{table}

\begin{figure}
  \includegraphics[width=0.7\linewidth]{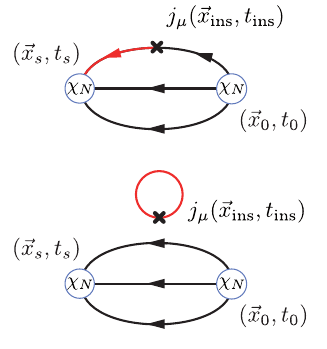}
  \caption{Connected (top) and disconnected (bottom) contributions to
    the nucleon three-point function. The red lines denote all-to-all
    propagators.}\label{fig:thrpdiagram}
\end{figure}

Carrying out the quark field contractions for the three-point function
in Eq.~(\ref{eq:thrp}) for a general current insertion gives rise to
both connected and disconnected contributions, as shown in
Fig.~\ref{fig:thrpdiagram}. For the case of the isovector combination,
only the connected diagram contributes since the disconnected
contribution cancels due to the degeneracy between the up- and
down-quark fermion loops at the continuum limit. For the connected
contribution, we restrict to $\vec{p}\,'=0$, meaning the source
momentum $\vec{p}$ is determined via momentum conservation by the
momentum transfer as $\vec{p} = -\vec{q}$. The all-to-all propagator
required due to the sums over $\vec{x}_\mathrm{ins}$ and $\vec{x}_s$
is computed using a sequential inversion through the sink, i.e. using
the so-called~\emph{fixed-sink} method. Thus, the same sequential
propagator yields all insertion momenta $\vec{q}$, however new
inversions are required for each sink time $t_s$ and projector
$\Gamma_\nu$. To reliably extract the nucleon matrix element of
interest, we compute the connected three-point function for multiple
values of $t_s$ and seek convergence to the asymptotic ground state
matrix element via multi-state fits. We increase statistics with
increasing $t_s$ to counter the expected exponential increase of
statistical uncertainty. The statistics used for each ensemble and
each value of $t_s$ are shown in Table~\ref{tab:statistics}, which
also conveys our strategy of keeping the number of configurations
fixed while increasing the number of randomly chosen source-positions
per configuration with increasing $t_s$.

\begin{table}[t!]
  \caption{Statistics used for the computation of the connected
    contributions for \texttt{cB211.072.64} (left),
    \texttt{cC211.060.80} (center), and \texttt{cD211.054.96}
    (right). For each ensemble, we give the number of configurations
    ($n_\mathrm{conf}$), the sink-source separations in lattice units
    ($t_s/a$, first column) and physical units ($t_s$, second column),
    and the number of source positions per configuration
    ($n_\mathrm{src}$, third column).}
  \label{tab:statistics}
   \begin{minipage}[t]{0.32\linewidth}
     \begin{tabular}{rrr}
       \hline\hline
       \multicolumn{3}{c}{\texttt{cB211.072.64}} \\
       \multicolumn{3}{c}{$n_\mathrm{conf}$=750} \\
       \hline
       $t_s/a$ & $t_s$[fm] & $n_{src}$ \\
       \hline
       8 & 0.64 &  1 \\
       10 & 0.80 &  2 \\
       12 & 0.96 &  5 \\
       14 & 1.12 &  10 \\
     16 & 1.28 & 32 \\
     18 & 1.44 & 112 \\
     20 & 1.60 & 128 \\
    \hline\hline
    \multicolumn{3}{c}{} \\
    \multicolumn{3}{c}{} \\
    \multicolumn{3}{c}{} \\
    \end{tabular}
   \end{minipage}\hfill
   \begin{minipage}[t]{0.32\linewidth}
    \begin{tabular}{rrr}
    \hline\hline
      \multicolumn{3}{c}{\texttt{cC211.060.80}} \\
      \multicolumn{3}{c}{$n_\mathrm{conf}$=400} \\
    \hline
      $t_s/a$ & $t_s$[fm] & $n_{src}$ \\
    \hline
      6 & 0.41 &   1 \\
      8 & 0.55 &   2 \\
     10 & 0.69 &   4 \\
     12 & 0.82 &  10 \\
     14 & 0.96 &  22 \\
     16 & 1.10 &  48 \\
     18 & 1.24 &  45 \\
     20 & 1.37 & 116 \\
     22 & 1.51 & 246 \\
     \hline\hline
    \multicolumn{3}{c}{} \\
    \end{tabular}
  \end{minipage}\hfill
  \begin{minipage}[t]{0.32\linewidth}
    \begin{tabular}{rrr}
    \hline
    \hline
      \multicolumn{3}{c}{\texttt{cD211.054.96}} \\
      \multicolumn{3}{c}{$n_\mathrm{conf}$=500} \\
    \hline
      $t_s/a$ & $t_s$[fm] & $n_{src}$ \\
    \hline
      8 & 0.46 &   1 \\
     10 & 0.57 &   2 \\
     12 & 0.68 &   4 \\
     14 & 0.80 &   8 \\
     16 & 0.91 &  16 \\
     18 & 1.03 &  32 \\
     20 & 1.14 &  64 \\
     22 & 1.25 &  16 \\
     24 & 1.37 &  32 \\
     26 & 1.48 &  64 \\
    \hline    \hline
    \end{tabular}
  \end{minipage}
\end{table}

The disconnected three-point functions involve   correlating the two-point
function of Eq.~(\ref{eq:twop}) with the quark loop with the electromagnetic current insertion, which results in
\begin{equation}
  L_\mathsf{f}(\tins,\vec{q}) = \sum_{\vec{x}_\mathrm{ins}} \mathrm{Tr}
  \left[ D_\mathsf{f}^{-1}(x_\mathrm{ins};x_\mathrm{ins}) \gamma_\mu \right].
  e^{i \vec{q} \cdot \vec{x}_\mathrm{ins}},\label{eq:looptrace}
\end{equation}
for each quark flavor $\mathsf{f}$. Therefore, the selection of
$\vec{q}$ and $t_s$ is independent of the computation of the insertion
operator, and we can thus obtain all insertion sites and sink times. In
practice however, this is limited by statistical uncertainties, which
grow with $t_s$. Note that the disconnected three-point function can
also be obtained for non-zero sink momentum $\vec{p}\,'$. We compute
the isoscalar combination $L_{u+d}(\tins,\vec{q})$ stochastically, via
the \emph{generalized one-end trick}~\cite{McNeile:2006bz}, where the
trace of the sum of the up- and down-quark propagators are written in
terms of the trace of a product, namely 
\begin{align}
L_{u+d}(t_\mathrm{ins},\vec{q})=&  \nonumber\\ 
\sum_{\vec{x}_\mathrm{ins}}e^{i\vec{q}\cdot\vec{x}_\mathrm{ins}}\big[ \mathrm{Tr}[&D^{-1}_u(x_\mathrm{ins};x_\mathrm{ins})\gamma_\mu]+\mathrm{Tr}[D^{-1}_d(x_\mathrm{ins};x_\mathrm{ins})\gamma_\mu]\big]=  \nonumber\\
\sum_{\vec{x}_\mathrm{ins}}e^{i\vec{q}\cdot\vec{x}_\mathrm{ins}}
\mathrm{Tr}[&D^{-1}_u(x_\mathrm{ins};x)D_W(x;y)D^{-1}_d(y;x_\mathrm{ins})\gamma_\mu],
\label{eq:gen-one-end}\end{align}
where $D_u$ ($D_d$) is the twisted mass operator for $+\mu_l$ ($-\mu_l$) and $D_W$ is the standard Wilson operator. The trace in the last line of Eq.~(\ref{eq:gen-one-end}) can be obtained using the one-end trick with volume stochastic sources, as also used in our previous
studies~\cite{Alexandrou:2013wca,Alexandrou:2017hac,Alexandrou:2017qyt,Alexandrou:2017oeh,Alexandrou:2018sjm}.

\begin{table}
  \caption{Parameters for the computation of the  disconnected three-point function. For the three ensembles (indicated in the first
    column), we give the number of configurations
    ($n_\mathrm{conf}$, second column) and the number of vectors that
    are inverted per configuration ($n_\mathrm{vec}$) resulting from
    color and spin dilution ($n_\mathrm{dil}$, third column),
    hierarchical probing ($n_\mathrm{had}$, fourth column), and the
    number of stochastic vectors used ($n_\mathrm{stoch}$, fifth
    column). We indicate the number of low-lying eigenvectors treated
    exactly ($n_{ev}$, sixth column) for the two ensembles where
    deflation is used and the number of source positions for the
    two-point functions ($n_\mathrm{src}$).  }\label{tab:stoch}
  \begin{tabular}{ccc@{$\times$}c@{$\times$}lcc}
    \hline\hline
    \multirow{2}{*}{Ensemble} & \multirow{2}{*}{$n_\mathrm{conf}$}  & \multicolumn{3}{c}{$n_\mathrm{vec}$=} & \multirow{2}{*}{$n_{ev}$} &\multirow{2}{*}{$n_\mathrm{src}$}                   \\
                              &                                    & $n_\mathrm{dil}$                      & $n_\mathrm{had}$ & $n_\mathrm{stoch}$ &     & \\
    \hline
    \texttt{cB211.072.64}     & 750                                & 12                                    & 512              & 1                  & 200 & 477 \\
    \texttt{cC211.060.80}     & 400                                & 12                                    & 512              & 1                  & 450 & 650 \\
    \texttt{cD211.054.96}     & 500                                & 12                                    & 512              & 8                  & {-} & 480 \\    
    \hline
  \end{tabular}
\end{table}

The parameters used for the computation of the disconnected three-point function 
are summarized in Table~\ref{tab:stoch}. For all three ensembles we
use the same number of configurations as in the connected three-point
functions and fully dilute in color and spin
($n_\mathrm{dil}$=12).

Furthermore, we use hierarchical probing~\cite{Stathopoulos:2013aci}
to distance eight in the 4-dimensional volume, which results in 512
Hadamard vectors ($n_\mathrm{had}$=512). Hierarchical probing
suppresses stochastic noise from the off-diagonal spatial indices in
the trace of Eq.~(\ref{eq:looptrace}) and therefore, as indicated in
Eq.~(\ref{eq:looptrace}), for the disconnected contribution we use the
non-conserved local vector current
$\mathsf{\bar{\psi_f}}\gamma_\mu\mathsf{\psi_f}$ which only couples
the diagonal components of the spatial indices of the current. For the
disconnected contributions we therefore need to renormalize the
current, as will be explained in Sec.~\ref{subsec:renorm}.

For the ensembles, \texttt{cB211.072.64} and \texttt{cC211.060.80}, we
use 1 stochastic source ($n_\mathrm{stoch}$=1) and
deflate~\cite{Gambhir:2016uwp} the lowest 200 and 450 eigenvectors
($n_{ev}$), respectively. For the \texttt{cD211.054.96} ensemble with
the largest lattice size, deflation was impractical due to memory
requirements at the time of its analysis, and we therefore use instead
$n_\mathrm{stoch}$=8. On each configuration, we compute two-point
functions on $\mathcal{O}(10^2)$ source positions ($n_\mathrm{src}$),
as indicated in the last column of Table~\ref{tab:stoch}. Beyond
increasing the statistics of the disconnected three-point function,
these two-point functions improve the analysis of multi-state fits to
the connected three-point functions.

\subsection{Renormalization}
\label{subsec:renorm}

The connected parts of the isovector and isoscalar matrix elements are
calculated using the conserved vector current, for which no
renormalization is required. However, the disconnected part of the
isoscalar combination is computed using the local operator, which
needs renormalization. To address this, we determine the
renormalization factors of both flavor nonsinglet and flavor singlet
vector operator, denoted as $Z_V^{ns}$ and $Z_V^s$, respectively,
using the RI$'$/MOM scheme~\cite{Martinelli:1994ty}. In the massless
limit, where SU($N_f=4$) flavor symmetry holds, the isoscalar
combination $j_\mu^{u+d}$ can potentially mix with $j_\mu^{s+c}$ (or equivalently
with the singlet combination $j_\mu^{u+d+s+c}$), where u,d,s,c are the
$N_f=4$ quark flavors. In particular, the renormalized isoscalar
vector current $(j_\mu^{u+d})^R$ is obtained through:
\begin{eqnarray}
&& \hspace{-0.8cm}(j_\mu^{u+d})^R = \nonumber \\
&& \frac{1}{2}(j_\mu^{u+d+s+c})^R + \frac{1}{6} (j_\mu^{u+d+s-3c})^R + \frac{1}{3} (j_\mu^{u+d-2s})^R \nonumber \\
    &=& \frac{1}{2} Z_V^{s} j_\mu^{u+d+s+c} + \frac{1}{6} Z_V^{ns} j_\mu^{u+d+s-3c} + \frac{1}{3} Z_V^{ns} j_\mu^{u+d-2s} \nonumber \\
    &=& Z_V^{ns} \ j_\mu^{u+d} + \frac{1}{2} (Z_V^s - Z_V^{ns}) \ j_\mu^{u+d+s+c}, \label{u+d_R}
\end{eqnarray}
where the nonsinglet combinations $j_\mu^{u+d+s-3c}$ and $j_\mu^{u+d-2s}$, as well as the singlet combination $j_\mu^{u+d+s+c}$ renormalize multiplicatively with $Z_V^{ns}$ and $Z_V^{s}$, respectively. Thus, the calculation of the matrix element of $j_\mu^{u+d+s+c}$ is required in addition to the matrix element of $j_\mu^{u+d}$, and the renormalization factors $Z_V^{ns}$, and $Z_V^s$. However, as shown in Ref.~\cite{ExtendedTwistedMass:2022jpw} using symmetry arguments and confirmed by our analysis, $Z_V^{s}$ and $Z_V^{ns}$ coincide, and thus the second term of Eq. \eqref{u+d_R} is eliminated, and the computation of the matrix element of $j_\mu^{u+d+s+c}$ is no longer necessary.   

The RI$'$/MOM condition used for extracting $Z_V^{ns}$ and $Z_V^s$ is~\cite{ETMC-ren:2025}:
\begin{equation}
{(Z_f^{{\rm RI}'})}^{-1} Z_V^{{\rm ns (s)}} \frac{1}{36} \sum_\mu {\rm Tr} \left[\Lambda^{\rm ns (s)}_{V_\mu} (p)(\gamma_\mu - \frac{\slashed{\tilde{p}}}{4 \tilde{p}_\mu})\right] \Bigg |_{p^2 = \mu_0^2} = 1,
\end{equation}
where $\tilde{p}_\mu \equiv \sin(a p_\mu)\neq 0$, $\mu_0$ represents
the RI$'$/MOM scale and $\Lambda^{\rm ns (s)}_{V_\mu} (p)$ is the
amputated vertex function of the nonsinglet (singlet) vector operator
with external quark fields.  Note that a disconnected diagram enters
the calculation of $\Lambda^{\rm s}_{V_\mu} (p)$. $Z_f^{{\rm RI}'}$ is
the renormalization factor of the quark field, calculated through the
quark propagator.
 
The vertex function $\Lambda^{\rm ns (s)}_{V_\mu} (p)$ is calculated
using the momentum source approach~\cite{Gockeler:1998ye}, which gives
high statistical accuracy using only $\mathcal{O} (10)$
configurations. For the disconnected part of $\Lambda^{\rm s}_{V_\mu}
(p)$, we employ the generalized one-end trick~\cite{McNeile:2006bz}
along with the hierarchical probing
algorithm~\cite{Stathopoulos:2013aci} to mitigate stochastic noise.

Our strategy can be summarized as follows: we first perform chiral
extrapolations using a set of ensembles specifically simulated for our
renormalization program (see Ref.~\cite{Alexandrou:2024ozj}), which
features four mass-degenerate quarks ($N_f = 4$) at the same $\beta$
values as the three physical-point ensembles used in our analysis of
the matrix elements. A linear fit with respect to the twisted-mass
parameter $\mu_{\rm sea}$ adequately removes the mild dependence on
quark mass, consistent with findings from previous studies, e.g.,
Ref.~\cite{Alexandrou:2015sea}.

To minimize lattice artifacts and rotational $O(4)$ breaking effects,
we employ spatially isotropic momenta of the form:
\begin{equation}
(a p) \equiv 2 \pi \left\{\frac{n_t + 0.5}{T/a}, \frac{n_x}{L/a}, \frac{n_x}{L/a}, \frac{n_x}{L/a}\right\}, 
\end{equation}
where $n_i \in \mathbb{Z}$ and $L \ (T)$ is the spatial (temporal)
extent of the lattice. We impose momentum cuts defined by $\sum_{\mu}
p_\mu^4/(\sum_\mu p_\mu^2)^2 < 0.3$. Additionally, we improve our
nonperturbative results by subtracting one-loop discretization errors
calculated through lattice perturbation theory, leading to a reduced
dependence of the renormalization factors on $(a^2 p^2)$. Further
details can be found in similar studies by our
group~\cite{Alexandrou:2015sea}.

The vector current is scheme- and scale-independent for both
nonsinglet and singlet operators. Consequently, unlike other bilinear
operators, we do not encounter systematic errors from perturbative
conversion and evolution functions.

\begin{figure}[ht!]
\centering
\includegraphics[width=\columnwidth]{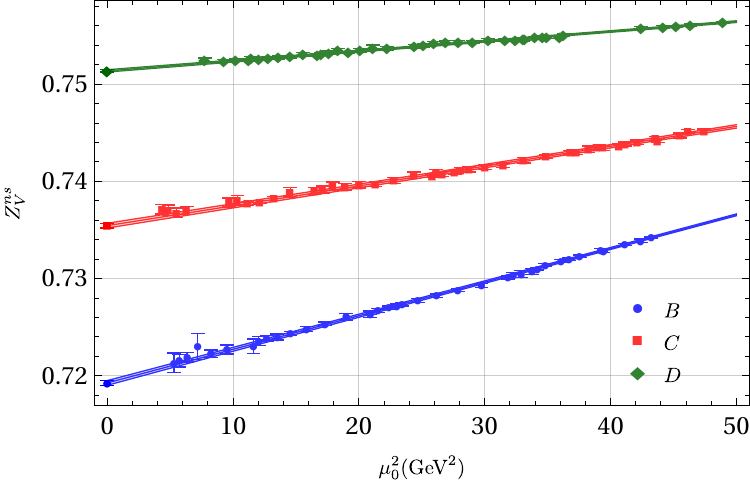} \\
\includegraphics[width=\columnwidth]{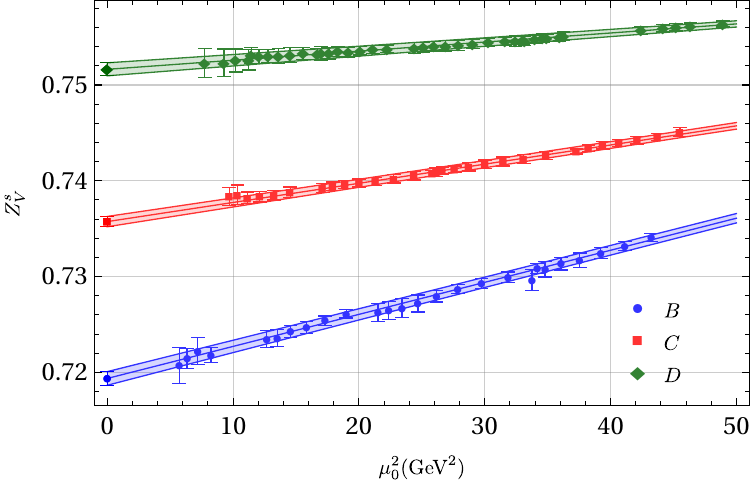} \\
    \caption{Renormalization factors of the flavor nonsinglet (top)
      and flavor singlet (bottom) vector operator for the three
      physical-point ensembles: \texttt{cB211.072.64} (``B''),
      \texttt{cC211.060.80} (``C''), \texttt{cD211.054.96} (``D'').}
    \label{fig:Zfactors}
\end{figure}

Next, we apply a linear fit in $\mu_0^2$ to eliminate any residual
dependence on the renormalization scale resulting from discretization
effects. Fig.~\ref{fig:Zfactors} displays the momentum fits of
$Z_V^{ns}$ and $Z_V^{s}$ for the three different $\beta$ values. The
fit range is set to 12 - 50 GeV$^2$, where the linear fit can
sufficiently describe the remaining momentum dependence.

The extrapolated values of $Z_V^{ns}$ and $Z_V^{s}$ at $\mu_0^2 = 0$
are given in Table \ref{tab:renorm}. The two values are in strong
agreement within the small statistical uncertainties, as expected. Our
analysis provides robust support for the theoretical proof established
in Ref.~\cite{ExtendedTwistedMass:2022jpw}. A more comprehensive
analysis of the nonsinglet quark bilinear operators using a total of
five lattice spacings will be provided in a forthcoming publication by
ETMC~\cite{ETMC-ren:2025}.

\begin{table}[ht!]
  \caption{Renormalization factors of the flavor nonsinglet and flavor
    singlet vector operator for the three physical-point ensembles.}
  \label{tab:renorm}  
    \centering
    \begin{tabular}{cccc}
        \hline
        \hline
        & \texttt{cB211.072.64} \ & \texttt{cC211.060.80} \ & \texttt{cD211.054.96} \\
        \hline
        $Z_V^{ns}$ & 0.7193(2) & 0.7354(2) & 0.7514(1) \\
        $Z_V^s$ & 0.7194(6) & 0.7357(5) & 0.7516(7) \\
        \hline
    \end{tabular}
\end{table}

Obtaining the disconnected contributions using the local current
suppresses stochastic errors in the quark loop that would arise from
the product of off-diagonal space-time points had the conserved
current been used. The use of the conserved current for the connected
matrix elements and the renormalized local current for the
disconnected may affect the approach to the continuum. Practically
though, any systematic arising from this is much smaller than our
statistical errors. Indeed, the difference between the isoscalar
matrix elements obtained using the renormalized local current and the
conserved current is at most 30\% of the statistical error for all
$Q^2$ and values of $a$ considered in this work.

\section{Extraction of nucleon matrix elements}
\label{sec:extraction}
The bare form factors at each value of the momentum transfer squared,
$Q^2$, are obtained by appropriate combinations of $\Gamma_\nu$ and
$\mu$ depending on the momenta $\vec{p}\,'$ and $\vec{q}$ in
$\Pi^\mu(\Gamma_\nu;\vec{p}\,', \vec{q})$ of Eq.~(\ref{eq:gsmatrix})
in order to isolate $G_E(Q^2)$ and $G_M(Q^2)$.  For the case of the
connected contributions, the fixed sink
approach~\cite{Alexandrou:2018sjm} requires us to fix the sink
momentum, which we set to $\vec{p}\,'=0$, as explained in the previous
section. For this case, the expressions yielding $G_E(Q^2)$ and
$G_M(Q^2)$ can be disentangled via,

\begin{equation}
  \Pi^0(\Gamma_0,\vec{q}) = C \frac{E_N + m_N}{2m_N} G_E(Q^2),
  \label{eq:GEpp0_1}
\end{equation}

\begin{equation}
  \Pi^i(\Gamma_0,\vec{q}) = C \frac{q_i}{2m_N} G_E(Q^2),
    \label{eq:GEpp0_2}
\end{equation}

\begin{equation}
  \Pi^i(\Gamma_k,\vec{q}) = C \frac{\epsilon_{ijk} q_j}{2m_N} G_M(Q^2)\,
    \label{eq:GMpp0}
\end{equation}
where,
\begin{equation}
  C= \sqrt{\frac{2m_N^2}{E_N(E_N+m_N)}},
\end{equation}
and $E_N = E_N(\vec{q})$. In practice, combining
Eqs.~(\ref{eq:GEpp0_1}) and~(\ref{eq:GEpp0_2}) does not improve the
statistical precision of our results for $G_E(Q^2)$ compared to using
Eq.~(\ref{eq:GEpp0_1}) alone, and therefore in what follows we extract
$G_E(Q^2)$ from Eq.~(\ref{eq:GEpp0_1}) and $G_M(Q^2)$ from
Eq.~(\ref{eq:GMpp0}).  In the disconnected contributions, the sink
momentum can be chosen in the two-point function independently of the
disconnected loop and at no additional computational cost. We, thus,
take sink momenta $\vec{p}\,'=\frac{2\pi}{L}\vec{k}$ with
$\vec{k}^2=1$ and $2$ in addition to the case $\vec{p}\,'=0$. With
these additional momenta, the expressions yielding $G_E$ and $G_M$
cannot be disentangled (see for example
Appendix~\ref{sec:appendix_equations}), i.e.  each combination of
$\Gamma_\nu$, $\mu$, and sink and insertion momenta in
$\Pi^\mu(\Gamma_\nu, \vec{p}\,', \vec{q})$ relates to a linear
combination of $G_E(Q^2)$ and $G_M(Q^2)$. The set of equations is
over-determined, and we, thus, resort to using a Singular Value
Decomposition (SVD) to solve for the electromagnetic form factors. Our
procedure, including how statistical errors are propagated, is the
same as the one used in Ref.~\cite{Alexandrou:2019ali} where we
extracted the Generalized Form Factors of the nucleon.

To benefit from the availability of kinematics with $\vec{p}\,'\neq 0$
in the disconnected case, as well as the additional separations, we
proceed by carrying out the excited state analysis by fitting the
connected and disconnected contributions separately and then adding
the obtained ground state contributions. As will be shown in
Sec.~\ref{sec:disc}, the additional kinematics are crucial for
obtaining a good signal in the disconnected case. The alternative
approach, whereby the connected and disconnected three-point functions
are added first and the excited state analysis follows, would require
us to restrict our analysis to using the disconnected contributions
with sink momentum zero ($\vec{p}\,'=0$) and only for the sink-source
separations that are also available for the connected case. For
completeness, in Appendix~\ref{sec:appendix_excited}, we compare the
two approaches, showing that they lead to ground state matrix elements
which are consistent within errors.

\subsection{Connected}
\begin{figure}
    \includegraphics[width=0.35\textwidth]{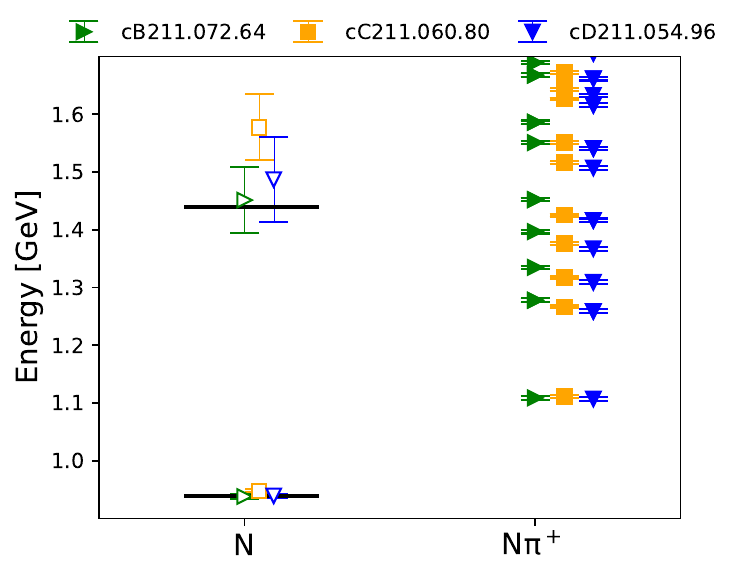}
    \caption{The ground- and first excited state energies obtained
      from a three-state fit to the nucleon two-point function (open
      symbols) and the energies of the non-interacting $\pi^{+}N$
      states (filled symbols, green for the B, orange for the C and
      blue for the D ensemble ) for all three ensembles and for total
      momentum $\vec{p}^2 = (\frac{2\pi}{L})^2$. The black lines
      correspond to the energies of the nucleon and Roper resonance,
      respectively.}
    \label{fig:E_comp}
\end{figure}

With multiple sink-source time separations and high statistics for
both connected three- and two-point functions, we can carry out a
thorough analysis of excited state effects. In particular, our data
allow for fits to the three-point functions including the first
excited state, namely with $(m,n)$=(0,0), (0,1), (1,0), and (1,1) in
Eq.~(\ref{eq:thrp_spec}), and fits to the two-point functions
including up to the second excited state, namely with $n=0$, 1, and 2
in Eq.~(\ref{eq:twop_spec}). An important component of our multi-state
fit analysis is to allow for different excited state energies in the
two- and three-point function. Indeed, a strong coupling of the
electromagnetic current to a multi-particle excited state may amplify
those terms in the spectrum of the three-point function while
otherwise being suppressed in the two-point function. This has also
been pointed out in Ref.~\cite{Bar:2021crj}, where a study in chiral
perturbation theory suggests that this may be the case for
pion-nucleon states coupling to the isovector electromagnetic
current. As an example, in Fig.~\ref{fig:E_comp}, we show the ground
and first excited state energies obtained via three-state fits to the
nucleon two-point function and compare with the non-interacting $\pi^+
N$ energies using the mass of $\pi^+$ taken from Table~\ref{tab:ens}
for each ensemble. The first excited state energies of the two-point
function for all ensembles considered coincide with the Roper
resonance and are thus higher than several possible pion-nucleon
states.

Our multi-state fitting procedure for the connected contributions
($\vec{p}\,'=0$) proceeds as follows; for each value of the momentum
transfer squared $Q^2 = 2m_N(E_N(\vec{q}) - m_N)$, we fit together the
two-point functions $C(\vec{0}, t_s)$ and $C(\vec{q}, t_s)$ and the
isovector and isoscalar three-point functions $C^{u-d}_\mu(\Gamma_\nu,
\vec{q}; t_s,\tins)$ and $C^{u+d}_\mu(\Gamma_\nu, \vec{q};
t_s,\tins)$, respectively, for both the electric and magnetic form
factors. We allow different fit parameters between two- and
three-point functions. The fit parameters for a single value of
$\vec{q}$ are described below:
\begin{itemize}
\item The nucleon ground state energy at rest $E_N(0) = m_N$ is a
  common fit parameter and the dispersion relation is used to set the
  nucleon ground state energy at momentum $\vec{q}$,
  i.e. $E_N(\vec{q}) = \sqrt{m_N^2 + \vec{q}^2}$.
\item For each two-point function, we do a three-state fit that has
  three coefficient parameters and two parameters corresponding to the
  first and second excited-state energies. This leads to 11 parameters
  so far.
\item The two-state fits to the isovector and isoscalar three-point
  functions are carried out in the same fit as the two-point functions
  above, thus sharing common ground state energies $E_N(0)$ and
  $E_N(\vec{q})$. However, we allow for the isovector and isoscalar
  three-point functions to have different excited state energies,
  i.e. two excited state energies for the isovector and two for the
  isoscalar, which are different from those of the two-point
  functions. We thus have 4 parameters for the energies of the excited
  states of the three-point functions. Furthermore, we have 16
  amplitudes as fit parameters, i.e. four for each $G^{u -
    d}_{E}(Q^2)$, $G^{u - d}_{M}(Q^2)$, $G^{u + d}_{E}(Q^2)$, and
  $G^{u + d}_{M}(Q^2)$.

\end{itemize}
Thus, each fit has 31 fit parameters. While we do not use any priors
in the final fit with the 31 fit parameters, a series of fits precedes
the final fit to determine the initial parameters. Namely, we first
fit to the individual two-point functions and then fit the three-point
functions omitting the $\mathcal{A}^{1,1}$ term of
Eq.~(\ref{eq:thrp_spec}), using the obtained parameters at each stage
as starting values for subsequent fits. An exception is a few cases
with $Q^2>0.6$~GeV$^2$ where the two-state fit to the three-point
function is unstable, likely due to weak excited state
contamination. In these cases we use a prior for $\mathcal{A}^{0,0}$
centered around the result from a plateau fit and with width 10 times
the error of the plateau fit. In all fits, we use the covariance
matrix taking into account correlations between three-point functions
of different separations ($t_s$) and correlations with the two
two-point functions.

We vary the following time separations in our fitting procedure:
\begin{itemize}
\item $t_s^{\rm low,3pt}$: the smallest sink-source separation of the
  connected three-point function included. The largest is the maximum
  separation listed in Table~\ref{tab:statistics}.
\item $t_s^{\rm low,2pt}$: the smallest sink-source separation of the
  two-point function included.
\item $\tins^{\rm source}$: the smallest insertion-source separation
  of the connected three-point function included.
\item $\tins^{\rm sink}$: the smallest sink-insertion separation of
  the connected three-point function included.  
\end{itemize}
The current insertion times included in a given fit are thus,
$\tins\in[\tins^{\rm source}, t_s-\tins^{\rm sink}]$.  Since the sink
is at rest and the source has momentum, the magnitude of the excited
state contamination at a given time distance from the source and sink
of the three-point function may be different which is why we allow
$\tins^{\rm source}\neq \tins^{\rm sink}$. The values used for
each ensemble are provided in Table~\ref{tab:fitmodels}.

\begin{table}[h]
  \caption{The values of $t_s^{\rm low,3pt}$, $t_s^{\rm low,2pt}$ and
    $\tins^{\rm sink}$ varied in our multi-state fits of the two- and
    connected three-point functions. For each value of $\tins^{\rm
      sink}$, we use three values for $\tins^{\rm source}$ namely:
    $\tins^{\rm source}$=$\tins^{\rm sink}$, $\tins^{\rm sink}+1$, and
    $\tins^{\rm sink}+2$.}
  \label{tab:fitmodels}  
    \centering\begin{tabular}{ccccccccc}
    \hline\hline
       Ensemble  & $t_s^{\rm low,3pt}/a$ & $t_s^{\rm low,2pt}/a$ &  $\tins^{\rm sink}/a$ \\
       \hline 
        \texttt{cB211.072.64} & 8, 10, 12, 14 & 1, 2, 3  & 2, 3, 4 \\ 
        \texttt{cC211.060.80} & 8, 10, 12, 14 & 1, 2, 3, 4  & 2, 3, 4 \\
        \texttt{cD211.054.96} & 8, 10, 12, 14 & 1, 2, 3, 4, 5 & 2, 3, 4 \\
        \hline
    \end{tabular}
\end{table}

\begin{figure*}
    \includegraphics[width=\linewidth]{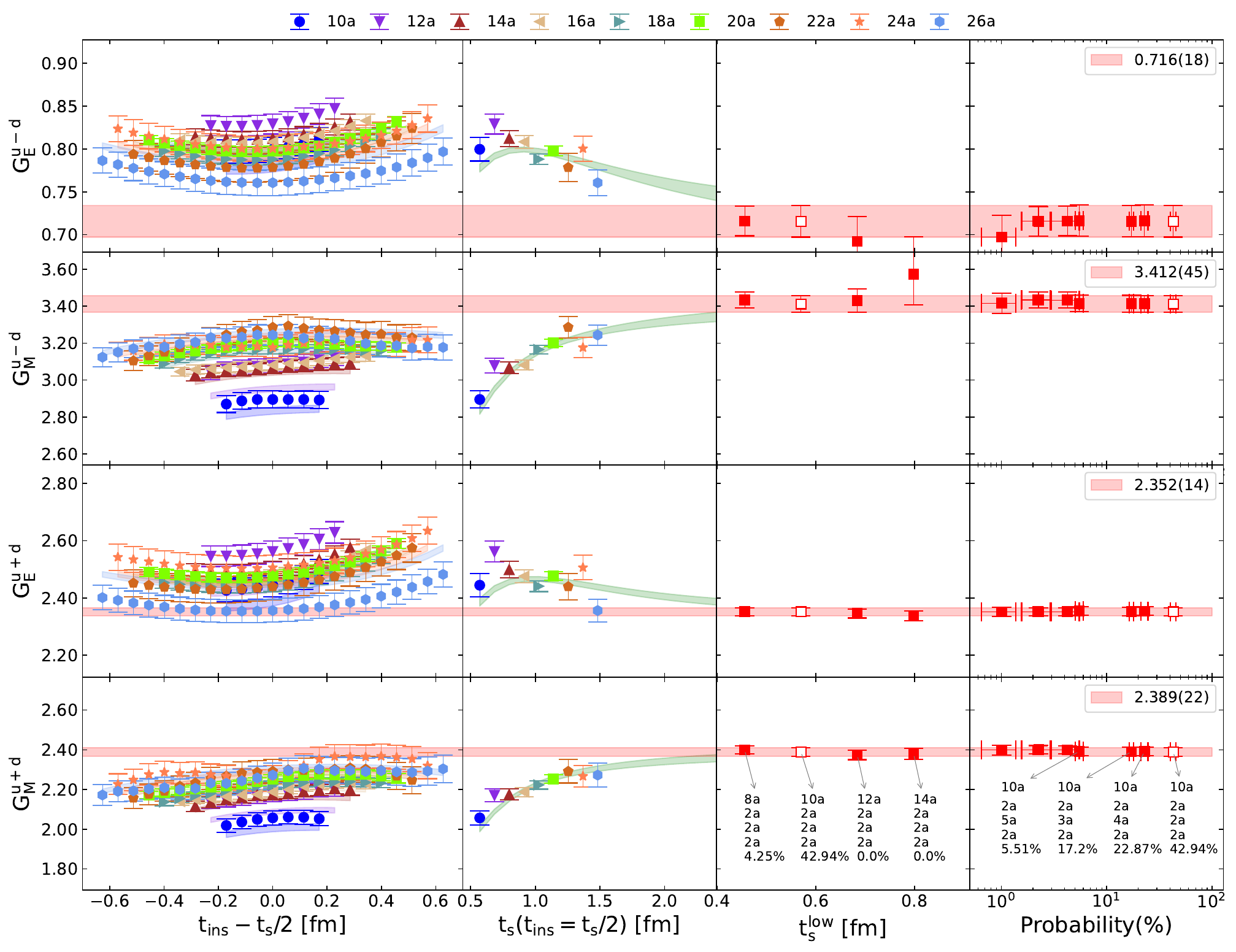}
    \caption{The left column shows the ratio of three- to two-point
      functions as defined in Eq.~(\ref{eq:full_ratio}) versus
      $\tins-t_s/2$ from which the isovector (isoscalar) electric and
      magnetic form factors are extracted in the first (third) and
      second (fourth) rows respectively. Different symbols denote
      different $t_s$ as shown in the header. Results are for $Q^2 =
      0.0996 \;\rm GeV^2$ and the \texttt{cD211.054.96}
      ensemble. The second column shows the value of the ratio for
      $\tins = t_s/2$. The third column shows the result of the most
      probable fit for a given $t_s^{\rm low,3pt}$, and the right
      column gives the result of each fit versus its fit probability,
      evaluated as explained in the text. The bands in the left column
      and the light green band in the second column are obtained by
      constructing the ratio using the fit parameters of the most
      probable fit.  The open symbol in the third and fourth columns
      and the horizontal red band is the value for the form factor
      obtained using the fit parameters of the most probable fit. The
      annotation in the third and fourth columns in the last row
      corresponds to the vector $(t_s^{\rm low,3pt}, \tins^{\rm
        source}, \tins^{\rm sink},t_s^{\rm low,2pt},\rm
      Probability(\%))$. The horizontal errors in last column are
      obtained via jackknife.}
    \label{fig:cD96_Q2_2_s}
\end{figure*}

An example of this fitting procedure is shown in
Fig.~\ref{fig:cD96_Q2_2_s}, for the case of the connected isoscalar
and isovector form factors for the \texttt{cD211.054.96}. In
particular, in the left columns we show the ratio of
Eq.~(\ref{eq:full_ratio}) after multiplying with the relevant
kinematic factor for isolating $G_E(Q^2)$ and $G_M(Q^2)$ according to
Eqs.~(\ref{eq:GEpp0_1}) and~(\ref{eq:GMpp0}). The data in the second
column shows the value of the ratio at the mid-point $\tins=t_s/2$, in
order to visualize the convergence of the data as $t_s$ is
increased. We note that the data fitted to in our multi-state fits are
the two- and three-point functions and not the ratios shown in
Fig.~\ref{fig:cD96_Q2_2_s}, i.e. the ratio is formed for visualization
purposes only. In the third column, we show results for the most
probable model in the subset of models obtained after fixing $t^{\rm
  low, 3pt}_s$ to demonstrate the variation in the results while
varying the $t^{\rm low, 3pt}_s$. In the last column, we show our
result for the form factor as a function of the fit probability, which
we determine via the Akaike Information Criterion (AIC) explained in
what follows. The fit with the highest probability is shown with the
open symbol and the horizontal band spanning the x-axis. Using the fit
parameters of the most probable fit, we construct the ratio as a
function of $\tins$ and $t_s$, yielding the bands in the first and
second columns. As can be seen, the fit range combinations having
probabilities greater than $1\%$ are compatible with each other and
with the most probable fit.

\subsubsection{Model Averaging}
For each combination of fit ranges, as given in
Table~\ref{tab:fitmodels}, we obtain the $\chi^2$ value of the fit and
evaluate its probability via the AIC~\cite{Jay:2020jkz,Neil:2022joj}
to obtain combined statistical and systematic errors of the
ground-state matrix elements. We follow the model-average procedure
that was used for the analysis of the axial form factors using the
same three ensembles~\cite{Alexandrou:2023qbg}. In brief, for each
combination of fit ranges ($j$) and jackknife sample ($i$) we obtain
$\mathcal{O}^i_j$, the desired form factor at the given value of
$Q^2$. We associate a weight, $w_{i,j}$, defined as,
\begin{equation}\label{eq:weight}
  \log(w_{i,j}) = -\frac{\chi_{i,j}^2}{2} + N^{\rm dof}_{i,j},
\end{equation}
where $N^{\rm dof}_{i,j} = N^{\rm data}_{i,j} - N^{\rm params}_{i,j}$
is the number of degrees of freedom, given as the difference between
the number of data, $N^{\rm data}_{i,j}$, and the number of
parameters, $N^{\rm params}_{i,j}$, used in the corresponding fit.
$\chi^2_{i,j}$ is defined as
\begin{equation}
  \chi^2_{i,j} = {(\vec{r}_{i,j})}^{\,\top} (C)_j^{-1}
  \vec{r}_{i,j}\quad\text{with}\quad
  \vec{r}_{i,j}=\vec{y}_{i,j}-f(\vec{x}_{i,j}),
\end{equation}
where, for each fit $j$, $C_j$ is the covariance matrix of the data
$\vec{y}\,_{i,j}$ determined via jackknife resampling, $\vec{y}_{i,j}$
is the jackknife sample $i$, of the data and $\vec{r}_{i,j}$ the
residual, computed using the selected fit function and ranges
$f(\vec{x}_{i,j})$. The probability associated with each combination
of fit ranges is
\begin{equation}
  p_{i,j} = \frac{w_{i,j}}{Z_i}\quad\text{with}\quad Z_i=\sum_j{w_{i,j}},
\end{equation}
and the model-averaged value of our observable, $\mathcal{O}^i$, in
jackknife bin $i$ is,
\begin{equation}\label{eq:model_average}
  \mathcal{O}^i = \sum_j p_{i,j} \mathcal{O}^i_j.
\end{equation}
The central value and error of the observable is then obtained via the
standard jackknife procedure. 

It should be noted that with this procedure, all three-point functions
for the isovector and isoscalar currents with the different projectors
in Eqs.~(\ref{eq:GEpp0_1}) and~(\ref{eq:GMpp0}) are fitted together,
meaning that a single probability is associated for a given fit
range. The advantage is that when combining the isoscalar and
isovector form factors to obtain the proton and neutron form factors,
the combination can be done per fit range and the same value of the
probability can be carried through and assigned to the proton and
neutron combinations.

\subsubsection{Energy spectrum and dispersion relation}
Since in our fits we allow for different excited state energies in the
two- and the isoscalar and isovector three-point functions, it is
worthwhile checking and comparing the spectrum with the Roper and
$\pi^+N$ free energies. In Fig.~\ref{fig:Energies_vs}, we collect the
excited state energies obtained from the fits to the isovector and
isoscalar connected three-point functions. For each value of the
momentum $\vec{q}^2=\vec{p^2}$, the fit yields two excited state
energies, one from the sink which is at rest ($E_{1,\vec{0}}$) and one
from the source which carries the opposite momentum of the insertion
($E_{1,\vec{p}}$, $\vec{p}=-\vec{q}$). From this plot, we observe the
following:
\begin{figure*}
    \centering
    \includegraphics[width=\linewidth]{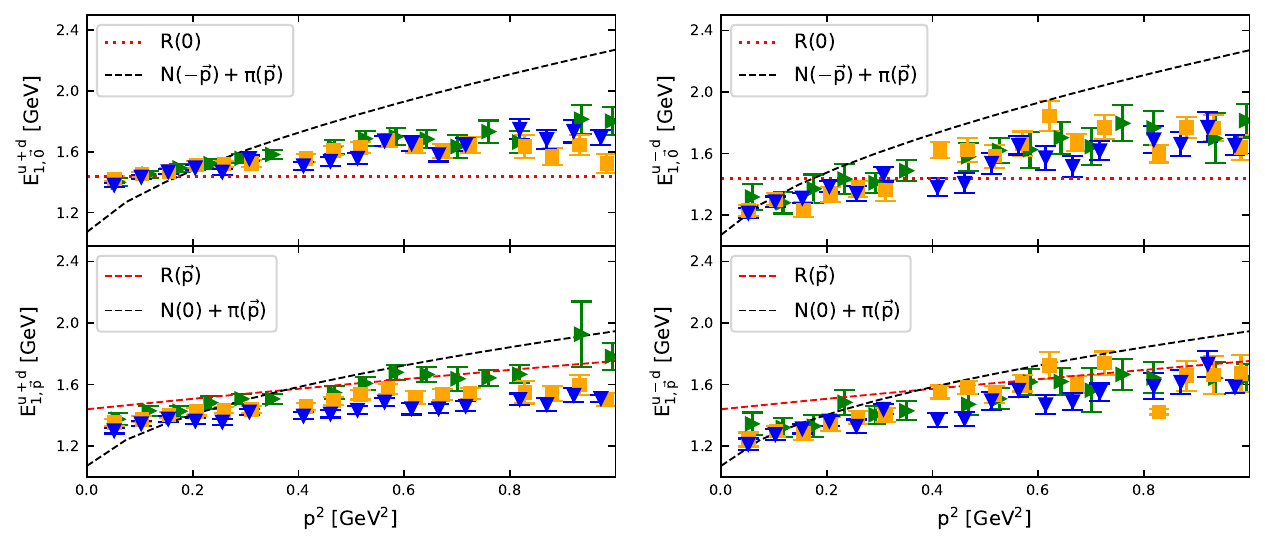}
    \caption{Excited state energies in the three-point function, as
      obtained from the combined fits to the two- and three-point
      functions explained in the text, for the B (green right-pointing
      triangles), C (yellow squares), and D (blue down-pointing
      triangles) ensemble. In the top (bottom), we show the first
      excited state at sink (source) for the isoscalar (left)
      $E^{u+d}_{1, \vec{0}}$ ($E^{u+d}_{1, \vec{p}}$) and isovector
      (right) $E^{u-d}_{1, \vec{0}}$ ($E^{u-d}_{1, \vec{p}}$)
      three-point functions. The red line shows the Roper mass (top)
      and the Roper energy with momentum $\vec{p}$ (bottom). Top: The
      dashed black lines show the energy of the non-interacting
      nucleon-pion state with total momentum zero and back-to-back
      relative momentum $\vec{p}$. Bottom: The dashed (dotted) black
      lines show the non-interacting nucleon-pion energy with total
      momentum $\vec{p}$ when $\vec{p}$ is carried by the pion
      (nucleon).}
    \label{fig:Energies_vs}
\end{figure*}

\begin{itemize}
    \item The first excited state energy at sink in the isoscalar
      case, $E^{u+d}_{1,\vec{0}}$, is compatible with the Roper for
      values of $\vec{p}^2\stackrel{\sim}{<}0.4$ GeV$^2$. Then it
      increases as $\vec{p}^2$ increases. This behavior may indicate
      residual admixture of two particle states. We note that the
      excited energy from the two-point function on the other hand
      follows the dispersion relation.

    \item The first excited state energy at the source for the
      isoscalar case $E^{u+d}_{1,\vec{p}}$ lies between the Roper
      moving with finite momentum and the nucleon-pion state with
      total momentum $\vec{p}$, when the momentum is carried by the
      nucleon.

    \item The first excited state energy at the sink in the isovector
      case $E^{u-d}_{1,\vec{0}}$ is compatible with the nucleon-pion
      state with relative back-to-back momentum $\vec{p}$ for values
      of $\vec{p}^2$ where this energy is smaller than the Roper
      mass. When $\vec{p}^2$ is such that the nucleon-pion energy
      crosses the Roper mass, $E^{u-d}_{1,\vec{0}}$ lies between the
      Roper mass and the nucleon-pion energy indicating a possible
      admixture between the Roper and the $\pi^+N$ state.

    \item The first excited state energy at the source for the
      isovector case $E^{u-d}_{1,\vec{p}}$, unlike the isoscalar case,
      $E^{u-d}_{1,\vec{p}}$ follows the nucleon-pion state at lower
      $\vec{p}^2$.

\end{itemize}

The deviation from the Roper energy, which is the excited state energy
obtained in the two-point function, shows that the choice of allowing
for different excited state energies between the two and three-point
functions is justified. The difference between the excited-state
energies of isoscalar and isovector cases, especially at low $Q^2$
values, also explains the need for different excited-state energy
parameterization in multi-state fits for the two channels.

\subsection{Disconnected contributions}
\label{sec:disc}
For the disconnected contributions we take sink momenta
$\vec{p}\,'=\frac{2\pi}{L}\vec{k}$ with $\vec{k}^2=1$ and $2$ in
addition to the case $\vec{p}\,'=0$. The set of equations yielding
$G_E$ and $G_M$ cannot be disentangled and are over-determined (see
for example Appendix~\ref{sec:appendix_equations}) and we thus use the
SVD of the system of equations to solve for the electromagnetic form
factors. To summarize, for a given $Q^2$, with $N$ equations, we have
\begin{equation}
    \Pi^\mu(\Gamma_{\nu};\vec{p}\,', \vec{p}) =
    \mathcal{G}^{\mu\nu}(\vec{p}\,', \vec{p}) \; F(Q^2)
    \label{eq:ratio_matrix_eqn}
\end{equation}
where the matrix $\mathcal{G}$ is an $N\times2$ array of kinematic
factors and $F(Q^2)^T = [G_E(Q^2), G_M(Q^2)]$. Taking a SVD of the
coefficient matrix $\mathcal{G}$,
\begin{equation}
    \mathcal{G} = U\Sigma V
    \label{eq:SVD}
\end{equation}
We can then extract the form factors by taking, 
\begin{equation}
    F(Q^2) = V^{\dagger}\Sigma^{-1}U^{\dagger}\Pi(\vec{p},\vec{p}')
    \label{eq:SVD_extraction}
\end{equation}

\begin{figure}
    \includegraphics[width=\linewidth]{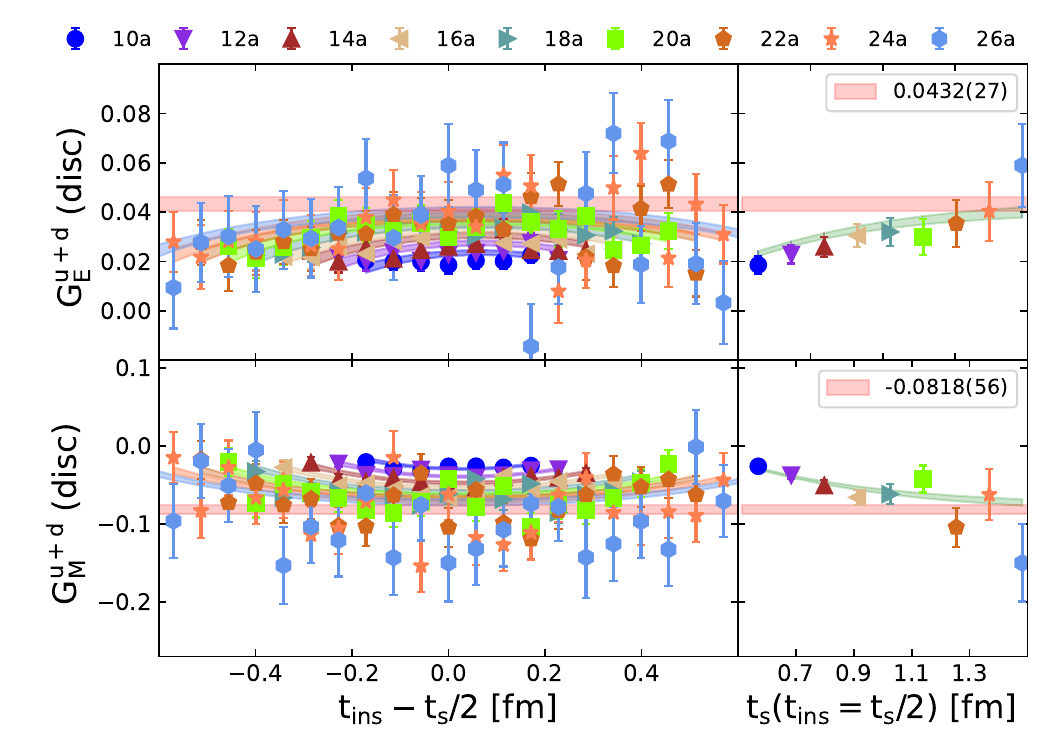}
    \caption{The renormalized isoscalar disconnected $G_E(Q^2)$ and
      $G_M(Q^2)$ for $Q^2 = 0.307$ GeV$^2$ for the
      \texttt{cD211.054.96} ensemble. The left column shows $G_E$
      (top) and $G_M$ (bottom) for source-sink separation as indicated
      in the header, for all insertion times $\tins$. The colored
      bands in the left and the light green band in the right column
      corresponds to the fit results from the two-state fit. The right
      column shows the value of the ratio for $\tins = t_s/2$.}
    \label{fig:D_disc}
\end{figure}

\begin{itemize}
    \item For each $Q^2$, we segregate the equations corresponding to
      unique $(\vec{p}^2, \vec{p}'^2)$.
    \item The nucleon mass and energy that appears in the coefficient
      matrix, $\mathcal{G}$, is set from the fits carried out for the
      connected contributions.
    \item After we obtain the SVD of $\mathcal{G}$ for a given
      $(\vec{p}^2, \vec{p}'^2)$, compute the rotated three-point
      function $\tilde{C}_{\mu\nu}(\vec{p},\vec{p}';t_s,\tins) \equiv
      U^{\dagger}C^{u+d,\mathrm{disc}}_\mu(\Gamma_\nu, \vec{p},
      \vec{p}'; t_s,\tins)$.
    \item Only the first two rows of
      $\tilde{C}_{\mu\nu}(\vec{p},\vec{p}';t_s,\tins)$ contribute to
      the form factors, therefore we fit these rows with the two-point
      functions $C(\vec{p}, t_s)$ and $C(\vec{p}', t_s)$.
    \item This results in a 14 parameter fit, with three-state fits to
      each of the two-point functions (3 coefficient parameters, 2
      excited-state energy parameters) and a two-state fit to the
      rotated three-point function (4 coefficient parameters for each
      row). The first excited-state energies between the two- and
      three-point functions are shared in this case, since the larger
      statistical errors for the disconnected do not allow following
      the procedure of the connected analysis. The nucleon ground
      state energy at rest is fixed and the dispersion relation is
      used to set the nucleon ground state energy at source or sink
      momentum. We obtain the ground-state matrix element using
      Eq.~(\ref{eq:gsmatrix}).
    \item Following the fit, we multiply the resulting two-element
      vector with $V^{\dagger}\Sigma^{-1}$.
    \item We average over the results obtained for multiple
      $(\vec{p},\vec{p}')$ pairs contributing to the same $Q^2$.
    
\end{itemize}
To demonstrate this fitting procedure, in Fig.~\ref{fig:D_disc}, we
show the ratio formed for the isoscalar disconnected form factor for
the \texttt{cD211.054.96} at $Q^2 = 0.307$ GeV$^2$ for $G_E(Q^2)$ and
$G_M(Q^2)$. In order to show convergence in $t_s$, in the second
column, the value of the ratio at the mid-point $\tins=t_s/2$ is
shown.

\begin{figure*}
    \includegraphics[width=\linewidth]{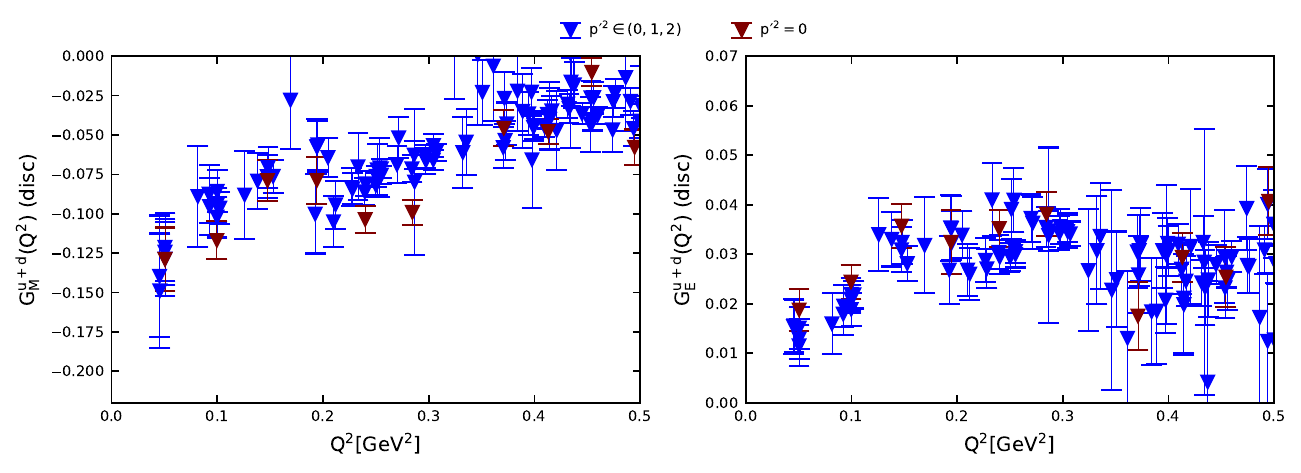}
    \caption{$G_M(Q^2)$ (left) and $G_E(Q^2)$ (right), disconnected
      isoscalar form factors as a function of $Q^2$ for the ensemble
      \texttt{cD211.54.96} obtained while allowing finite sink momenta
      $p'^2\in(0,1,2)$ (blue downward-pointing triangles) and while
      setting sink momenta to zero (maroon downward-pointing
      triangles).}
    \label{fig:disc_pp0_pp2}
\end{figure*}

The inclusion of momenta at sink leads to a total number of $Q^2$
values in the range of [0,1) to 224 for ensemble
  \texttt{cB211.072.64}, and, 262 each for ensembles
  \texttt{cC211.060.80} and \texttt{cD211.054.96} respectively. In
  Fig.~\ref{fig:disc_pp0_pp2}, we present a comparison of the
  disconnected data obtained for ensemble \texttt{cD211.054.96} when
  $\vec{p}'^2=0$ (maroon points) and when $p'^2\in(0,1,2)$ (blue
  points) for $Q^2\in[0,0.5] \rm\; GeV^2$. Furthermore, we find that
  the range of fit-ranges which yield stable fits are limited, and
  that systematic effects from varying within these fit-ranges are
  suppressed compared to the relatively larger statistical errors. We
  therefore take our result for the disconnected form factors from the
  single fit-range combinations listed in
  Table~\ref{tab:disc_range}. For all fits, the maximum source-sink
  separation used is kept the same as in the connected case.

\begin{table}
    \centering
    \caption{Values of lowest source-sink separation in three point
      function $t_s^{\rm low,3pt}$, minimum insertion time used
      $\tins^{\rm source}$ ($\tins^{\rm sink}$=$\tins^{\rm source}$)
      and lowest source-sink separation in two point function
      $t_s^{\rm low,2pt}$ used in the fits to the disconnected data.}
    \label{tab:disc_range}
    \begin{tabular}{cccc}
    \hline\hline Ensemble & $t_s^{\rm low,3pt}/a$ & $\tins^{\rm source}$/a  & $t_s^{\rm low,2pt}/a$ \\
    \hline cB211.72.64 & 8 & 2 & 2\\
     cC211.60.80 & 10 & 2 & 3\\
     cD211.54.96 & 10 & 2 & 4 \\
    \hline 
    \end{tabular}
\end{table}

\begin{figure*}
    \includegraphics[width=\linewidth]{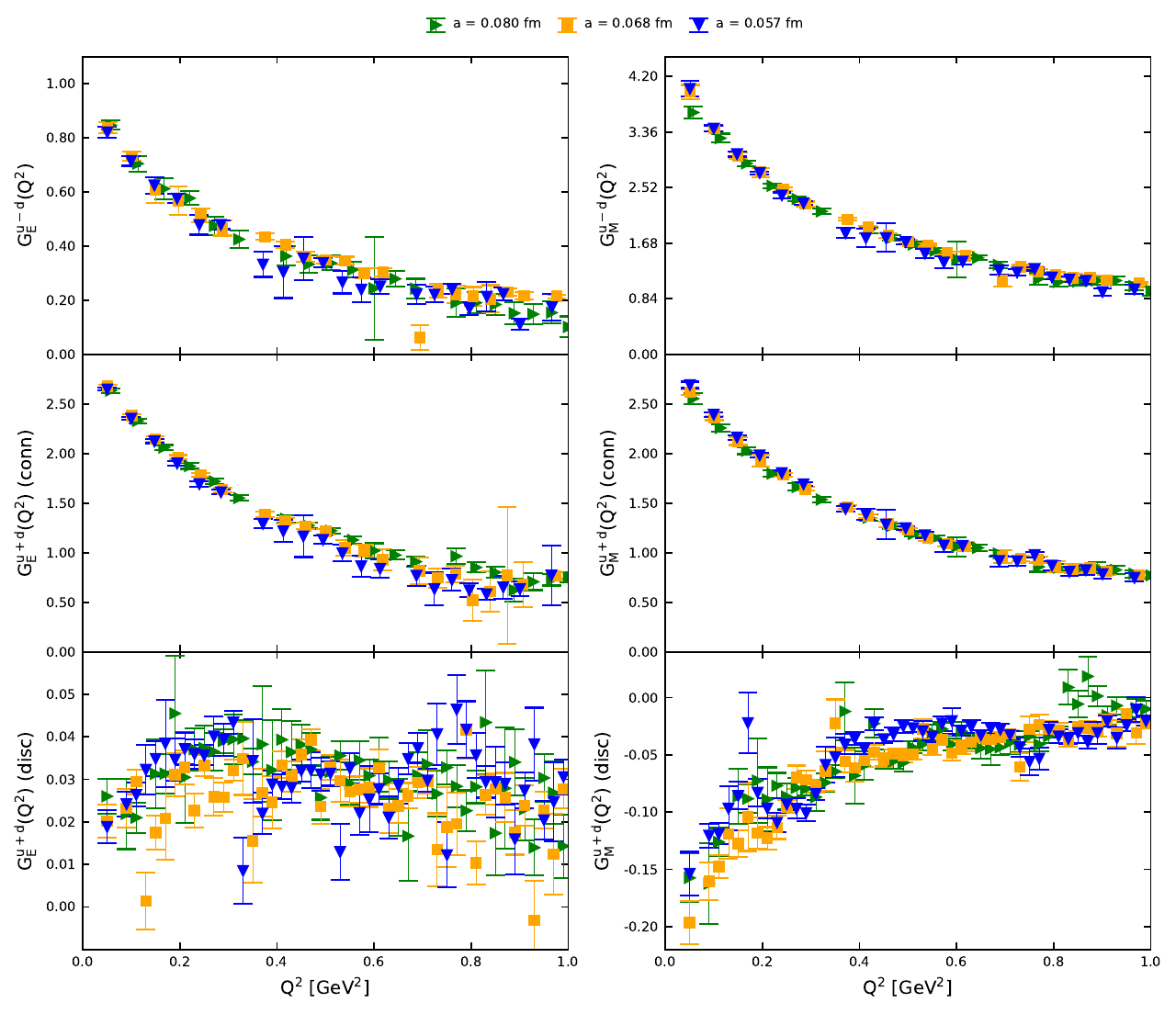}
    \caption{$G_E(Q^2)$ (left) and $G_M(Q^2)$ (right), connected isovector
      (top), connected isoscalar (center) and disconnected isoscalar
      (bottom) form factors as a function of $Q^2$ for the three
      ensembles analysed here, namely \texttt{cB211.72.64} (green
      right-pointing triangles), \texttt{cC211.60.80} (orange
      squares), and \texttt{cD211.54.96} (blue downward-pointing
      triangles).}
    \label{fig:GEMvs}
\end{figure*}

The isovector and connected and disconnected contributions to the
isoscalar form factors are shown in Fig.~\ref{fig:GEMvs} as a function
of $Q^2$ for the three ensembles analyzed in this work. The errors are
as described above, namely for the connected they are a result of the
model-average procedure, while for the disconnected they are
statistical from the fit-ranges indicated in
Table~\ref{tab:disc_range}. Furthermore, for the case of the
disconnected contributions, the large number of combinations of
$\vec{q}$ and $\vec{p}\,'$ yield values of $Q^2$ that would be too
close to each other to discern in Fig.~\ref{fig:GEMvs}. Therefore, for
the purpose of a clear visualization, the $Q^2$-axis has been
subdivided into 50 bins and the form factor values are averaged in
their respective bins, weighted by their errors (see
Appendix~\ref{sec:appendix_results}).

\section{Analysis of extracted form factors}
\label{sec:fits}
In this section we describe our approach for fitting the
$Q^2$-dependence of the form factors and taking the continuum
limit. This is carried out for the proton and neutron form factors
rather than the isovector and isoscalar. These are obtained via the
combinations in Eq.~(\ref{eq:vstopn}), which we construct per
jackknife bin. Within our errors, fitting the isoscalar and isovector
form factors and combining them to the proton and neutron form factor
at the continuum limit yields consistent results. 

For adding the isoscalar connected and disconnected contributions,
which are obtained at different $Q^2$ values due to the inclusion of
additional sink momenta in the disconnected, we first fit the
$Q^2$-dependence of the disconnected data and take the values that
match the connected from the fit. We use the Galster-like form for the
electric and the dipole form for the magnetic form factor, which are
explained in the remainder of this section.

In general, we carry out the analysis of the form factors in two
ways. In what we will refer to as the \textit{``two-step approach''},
we fit the $Q^2$-dependence of each ensemble individually, and then in
a second step we take the continuum limit as an extrapolation in $a^2$
of either the form factor at a given set of $Q^2$ values or the
observables of interest, i.e. the radii and magnetic moments. In what
we will refer to as a \textit{``one-step approach''} we fit the form
factors of the three ensembles together in a single fit, either
including $a^2$-dependent terms in the fit of the $Q^2$ dependence or
with no $a^2$-dependence, i.e. assuming no cut-off effects. In either
case, the jackknife mean and covariance matrix are used to take into
account correlations between data within the same ensemble. For
fitting the $Q^2$-dependence, we use both $z$-expansion and dipole
forms. For the case of the neutron electric form factor which is zero
for $Q^2=0$, we use the Galster-like parameterization.

\subsection{Dipole form}
The dipole form is given by
\begin{equation}
    G(Q^2) = \frac{g}{\left(1+\frac{Q^2}{M^2}\right)^2},
    \label{eq:dipole}
\end{equation}
with $g$ and $M^2$ the fit parameters. Employing
Eq.~(\ref{eq:radius}) for the radius, we have $\langle r^2\rangle =
\frac{12}{M^2}$.

For the one-step approach, we incorporate cut-off effects by including
the $a^2$-dependence of the observables, namely the charge and the
radius,
\begin{equation}
    g(a^2)=g_0+a^2g_2\,\,\text{and}\,\, \langle
    r^2(a^2)\rangle=\langle r\rangle_0^2+a^2\langle r^2\rangle_2
    \nonumber
    \label{eq:dipole_r_mu}
\end{equation}
which, when substituting in Eq.~(\ref{eq:dipole}) gives,
\begin{equation}
G(Q^2,a^2) = \frac{g(a^2)}{\left[1+\frac{Q^2}{12}\langle r^2(a^2)\rangle
    \right]^2}.
\label{eq:dipole_a}
\end{equation}
In the one-step approach, we fit all three ensembles to
Eq.~(\ref{eq:dipole_a}).  The radius and charge at the
continuum limit are given by the fit parameters $g_0$ and $\langle r^2\rangle_0$.

\subsection{Galster-like parameterization}
The neutron electric form factor and disconnected contribution to the
isoscalar electric form factor vanishes at $Q^2=0$. We use the
Galster-like parameterization~\cite{Galster:1971kv}, which is zero at
$Q^2=0$,
\begin{equation}
  G(Q^2) = \frac{Q^2 A}{4 m_N^2 + Q^2 B}
  \frac{1}{\left(1+\frac{Q^2}{0.71 {\rm GeV}^2}\right)^2},
  \label{eq:Galster-like}
\end{equation}
where $A$ and $B$ are the fit parameters. In this case, the radius
is given by
\begin{equation}
  \langle r^2 \rangle = - \frac{3A}{2 m_N^2}.
  \label{eq:Galster_radii}
\end{equation}
While the $a^2$-dependence can be incorporated in the same way as for
the dipole, as we will see, errors for the neutron electric form
factor are too large to identify any dependence on the lattice
spacing, and we thus fit all three ensembles simultaneously to
Eq.~(\ref{eq:Galster_radii}).

\subsection{$z$-expansion}
The $z$-expansion is given by,
\begin{equation}
G(Q^2) = \sum_{k=0}^{k_{\rm max}} c_k z^k(Q^2), 
\label{eq:zexp}
\end{equation}
where $c_k$ are the fit parameters and
\begin{equation}
z(Q^2) = \frac{\sqrt{t_{\rm cut} + Q^2} - \sqrt{t_{\rm cut}+t_0} }{
  \sqrt{t_{\rm cut} + Q^2} + \sqrt{t_{\rm cut}+t_0} }
\label{eq:zQ2}
\end{equation}
with $t_{\rm cut}$ being the particle production threshold and $t_0$
an arbitrary choice in $(-t_{\rm cut}, \infty)$. We set $t_0=0$, which
simplifies the relation of the charge and radii to the fit parameters,
namely
\begin{equation}
    g = c_0\quad\text{and}\quad \langle r^2\rangle = -\frac{3c_1}{2c_0t_{\rm
        cut}}\quad\text{with}\quad t_0=0.
\end{equation}
The dependence on the lattice spacing is introduced in the
coefficients of the $z$-expansion, similar to the procedure adopted in
Ref.~\cite{Alexandrou:2023qbg},
\begin{equation}
    c_k(a^2)=c_{k,0}+a^2c_{k,2}
\end{equation}
which modifies the $z$-expansion as
\begin{equation}
G(Q^2,a^2) = \sum_{k=0}^{k_{\rm max}} c_k(a^2) z^k(Q^2).
\label{eq:zexp_a2}
\end{equation}
We furthermore enforce smooth convergence of the form factor to zero
at $Q^2\rightarrow \infty$~\cite{Lee_2015}, which requires
\begin{equation}
    \sum_{k=0}^{k_{\rm max}} c_k \frac{d^nz^k}{dz^n}\Bigg|_{z=1} =
    0\quad\text{with}\quad n=0,1,2,3.
    \label{eq:inf_Q2_convergence}
\end{equation}
In our case, it is sufficient to enforcing this condition for $n=0$,
which implies $\sum_{k=0}^{k_{\rm max}} c_k = 0$ meaning that for a
given $k_{\rm max}$, instead of $k_{\rm max}+1$ fit parameters we have
$k_{\rm max}$, with one being fixed by this condition.  We follow our
approach used in Ref.~\cite{Alexandrou:2023qbg} for the axial form
factors, which  was adopted from
Refs.~\cite{Meyer:2016oeg,Hill:2010yb}, namely, we use Gaussian priors for
the parameters not directly related to the charge and radius, i.e. for
$k\ge 2$,
\begin{equation}\label{eq:zexp_priors}
    c_{k,0}\sim 0(w/k),\quad c_{k,2}\sim 0(20w/k)\quad\text{for}\quad k\ge2.
\end{equation}
 $w\le5$ is a hyper-parameter controlling the width of the priors
and which we vary together with the $k_{\max}$.

\subsection{Proton electric form factor}
We first present our analysis of the proton electric form factor $
G^p_E(Q^2)$.
\subsubsection{Dipole analysis}
Since we use the lattice conserved current, the form factor is equal
to 1 at $Q^2=0$ for any lattice spacing by symmetry, and we can
therefore  set $g=1$ in Eq.~(\ref{eq:dipole}) and $g_0=1$, $g_2$=0 in
Eq.~(\ref{eq:dipole_a}). 

In Table~\ref{tab:r2_protonge}, we summarize the values of $\langle
r^2_{\rm E} \rangle^p$ obtained when fitting the individual ensembles
and at the continuum limit when using either the one- or two-step
approaches. In Fig.~\ref{fig:dipole_proton_ge_ff}, we plot the data
for each ensemble and the bands corresponding to the dipole fits. We
vary the largest momentum transfer ($Q^2_{\rm cut}$) that we include
in the fit and monitor the
$\tilde{\chi}^2\equiv\chi^2/N_\mathrm{dof}$, where $N_\mathrm{dof}$
are the number of degrees of freedom. We find that a value of
$Q^2_{\rm cut}=0.4$~GeV$^2$ describes all three ensembles well, as can
be seen in Fig.~\ref{fig:dipole_proton_ge_ff}, as well as via the
values of $\tilde{\chi}^2$ quoted in Table~\ref{tab:r2_protonge}.
\begin{table}
    \centering
    \caption{Results on the proton electric charge radius $\langle
      r^2_{\rm E} \rangle^p$ and
      $\tilde{\chi}^2\def\chi^2/{N_{dof}}$ obtained when using dipole
      fits. The first three rows show the values obtained from fits to
      the individual ensembles. The continuum result (marked $a=0$) is
      obtained either by single fit to the combined $Q^2$- and
      $a^2$-dependence (1-step approach) or by linear extrapolation of
      the values of the radii obtained at the individual ensembles
      (2-step approach). }
    \label{tab:r2_protonge}
    \begin{tabular}{ccc}
    \hline\hline Ensemble & $\langle r^2_{\rm E} \rangle^p$ [fm$^2$] & $\tilde{\chi}^2$ \\
    \hline \texttt{cB211.72.64} & 0.619(31) & 0.518\\
    \texttt{cC211.60.80} & 0.609(17) & 0.635\\
    \texttt{cD211.54.96} & 0.635(20) & 1.969\\
    \hline 
    $a=0$, 1-step & 0.650(52) & 1.042\\
    $a=0$, 2-step & 0.650(52) & 0.552\\
    \hline
    \end{tabular}
\end{table}

\begin{figure}
    \centering
    \includegraphics[width=1\linewidth]{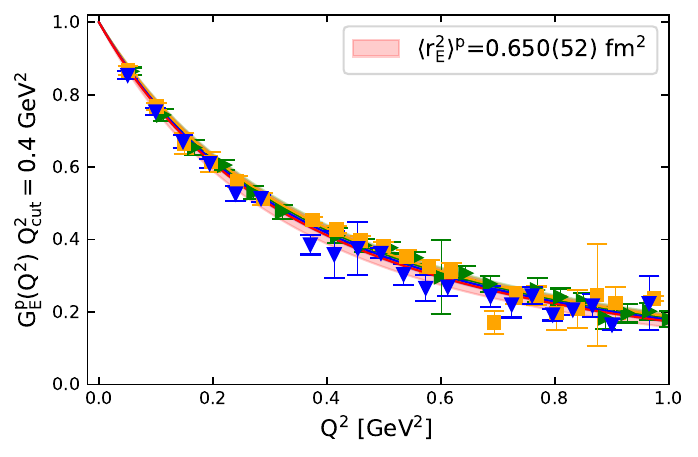}
    \caption{ Results on the proton electric form factor $G^p_E(Q^2)$
      for each ensemble denoted by the same symbols as in
      Fig.~\ref{fig:GEMvs}. The bands show the results obtained by
      fitting the dipole form to each ensemble (green band for
      \texttt{cB211.72.64}, orange band for \texttt{cC211.60.80} and
      blue band for \texttt{cD211.54.96}), and at the continuum limit
      (red band) using the ``one-step'' approach to fit the combined
      ($Q^2$, $a^2$)-dependence using $Q^2_{\rm cut} =
      0.4$~GeV\textsuperscript{2}.  We find $\langle r_E^2
      \rangle^p=0.650(52)$~fm$^2$ at the continuum obtained from the
      ``one-step'' approach. The fit $\tilde{\chi}^2= 1.042$.}
    \label{fig:dipole_proton_ge_ff}
\end{figure}

\begin{figure}
    \centering
    \includegraphics[width=1\linewidth]{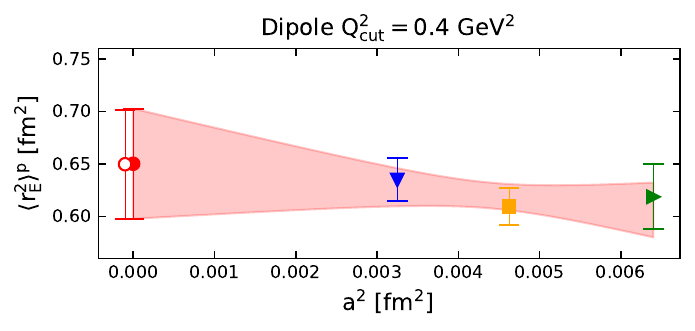}
    \caption{$\langle r^2_E\rangle ^p$ obtained using dipole fits to
      the \texttt{cB211.72.64} (green right-pointing triangle),
      \texttt{cC211.60.80} (orange square), and \texttt{cD211.54.96}
      (blue downward-pointing triangle) ensembles. A linear fit to the
      three values of the radii is also shown (red band) yielding the
      result at $a^2=0$ (filled red circle), which corresponds to the
      two-step approach. The one-step approach result is also shown at
      $a^2=0$, slightly shifted for clarity (open red circle).}
    \label{fig:dipole_proton_ge_continuum}
\end{figure}

A comparison of the one- and two-step approaches is shown in
Fig.~\ref{fig:dipole_proton_ge_continuum}, where the linear
extrapolation of the radius to the continuum limit is shown for the
case of the two-step approach and compared against the value obtained
from the combined fit in the one-step approach. As can be seen, the
two approaches yield the same result.

The analysis using the dipole fits has shown that the two- and
one-step approaches yield the same results at the continuum limit. We
note that an advantage of the one-step approach is that it provides
for a single $\tilde{\chi}^2$ value to the combined continuum
extrapolation and $Q^2$-dependence fit. The one-step approach is thus
more easily incorporated within a model-averaging procedure rather
than the two-step approach.

\subsubsection{$z$-expansion results}

For the $z$-expansion analysis we restrict to presenting results using
the one-step approach only.

\begin{figure}
    \centering
    \includegraphics[width=\linewidth]{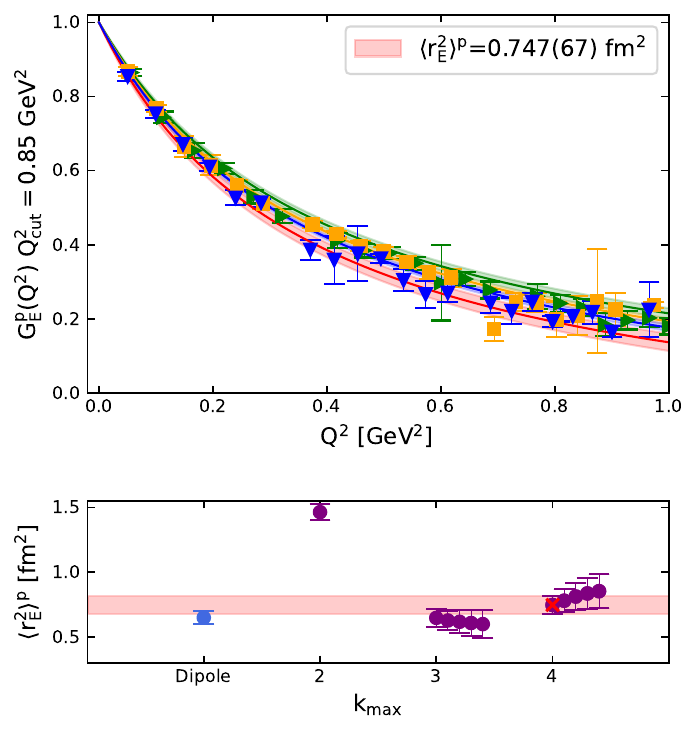}
    \caption{The $z$-expansion fit to $G^p_E(Q^2)$ with
      $Q_{cut}^2=0.85$~GeV$^2$ (top panel) and the dependence of the
      extracted radius on the $z$-expansion order ($k_{\rm max}$) and
      prior widths ($w$) (bottom panel). In the top panel, the green,
      orange, and blue bands show the result of the one-step
      $z$-expansion fit for $a=0.080$, 0.068, and 0.057~fm,
      respectively and the red band for $a=0$. In the bottom panel,
      for $k_{\rm max}=3$ and 4, $w$=1 for the leftmost point and
      increases by 1 up to $w=5$. For $k_{\rm max}=2$ no priors are
      employed, as explained in the text. In the top panel, we use
      $k_{max}=4$ and $w=1$, indicated by the red cross and the red band spanning horizontally the bottom
      panel. The blue point is the result from the one-step dipole fit, shown for
      comparison.}
    \label{fig:zexp_proton_ge_continuum}
\end{figure}

For the case of the proton electric form factor, we set $c_{0,0}=1$
and $c_{0,2}=0$ so that $G(0, a^2)=1$ for any value of $a^2$. For the
$z$-expansion fits, we vary $Q^2_{\rm cut}$, the order of the
expansion $k_{\rm max}$, and the width of the priors $w$. In
Fig.~(\ref{fig:zexp_proton_ge_continuum}), we show an example fit with
$Q^2_{\rm cut} = 0.85$~GeV$^2$, as well as the value obtained for the
radius when varying $k_{\rm max}$ and $w$. For comparison, we also
show the result of the dipole fit for the radius. Note that for
$k_{\rm max}=2$, no prior is used since $c_0$ is fixed, $c_1$ is free
with no prior and $c_2 = -(c_0+c_1)$, making this effectively a single
parameter fit. For $k_{\max} = 3$ and 4, we employ priors as in
Eq.~(\ref{eq:zexp_priors}) starting with $w=1$ and increasing by 1
from left to right until $w=5$. We see agreement between the
$z$-expansion fits for $k_{\max} = 3$ and 4 and relatively weak
dependence on $w$. The bands showing the form factor $Q^2$-dependence
in Fig.~\ref{fig:zexp_proton_ge_continuum} are from the fits with
$k_{\rm max}$=4 and $w=1$.

Once the choice for the order of $z$-expansion and the width of the
prior is set, we vary the maximum momentum transfer used for the fits
($Q^2_{\rm cut}$), obtaining the results shown in
Table~\ref{tab:protonge_zexp}.

\begin{table}
    \caption{The proton electric charge radius $\langle r^2_{\rm E}
      \rangle^p$ obtained using the one-step approach and the
      $z$-expansion with $k_{\rm max}=4$ and $w=1$ as we  increase
      $Q^2_{\rm cut}$. The values of  $\tilde{\chi}^2$ are also provided.}
    \label{tab:protonge_zexp}
    \begin{tabular}{ccc}
    \hline\hline $Q^2_{\rm cut} [\rm GeV^2]$ & $\langle r^2_{\rm E} \rangle^p$ [fm$^2$] & $\tilde{\chi}^2$ \\
    \hline 0.40 & 0.700(76) & 0.770\\
           0.50 & 0.713(72) & 0.638\\
           0.70 & 0.716(69) & 1.016\\
           0.85 & 0.747(67) & 1.032\\
           1.00 & 0.816(64) & 1.261\\
    \hline
    \end{tabular}
\end{table}

\subsection{Proton magnetic form factor}
\subsubsection{Dipole results}
\begin{figure}
    \centering
    \includegraphics[width=\linewidth]{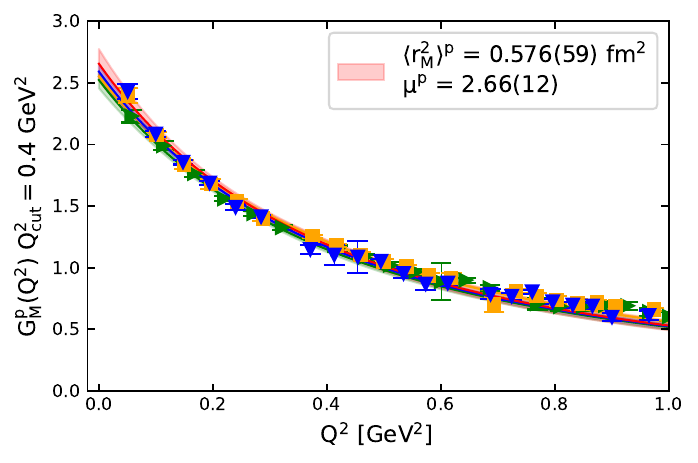}
    \caption{The same as in Fig.~\ref{fig:dipole_proton_ge_ff} but for
      $G^p_M(Q^2)$.}
    \label{fig:dipole_proton_gm_ff}
\end{figure}

\begin{figure}
    \centering
    \includegraphics[width=\linewidth]{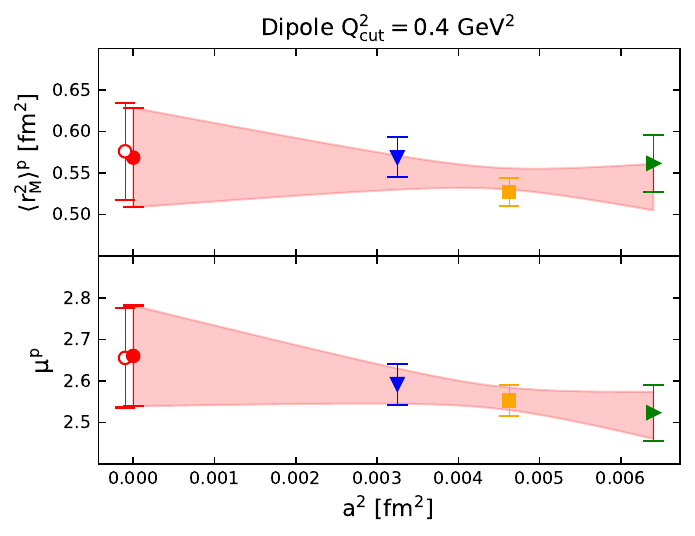}
    \caption{Results on $\langle r^2_M\rangle ^p$ (top panel) and
      $\mu^p$ (bottom panel) obtained using dipole fits to the three
      ensembles. The notation is the same as that of
      Fig.~\ref{fig:dipole_proton_ge_continuum}.}
    \label{fig:dipole_proton_gm_continuum}
\end{figure}

The same analysis is carried out for the proton magnetic form factor,
starting with the dipole fits. In this case, the proton magnetic
moment ($\mu^p$) is a fit parameter, in addition to the magnetic
radius $\langle r_M^2 \rangle^p$. In
Fig.~\ref{fig:dipole_proton_gm_ff}, we show the fit results using the
same convention as used for the proton electric form factor. For
completeness we carry out both one- and two-step approaches as in the
case of the dipole fits to the electric form factor, with results
summarized in Table~\ref{tab:mu_r2_protongm}.
A continuum extrapolation in $a^2$ results in consistent values for
the radius and the magnetic moment when using either one- or two-step
approaches, as shown in Fig.~\ref{fig:dipole_proton_gm_continuum}.

\begin{table}
    \centering
    \caption{The proton magnetic moment ($\mu^p$), magnetic radius
      ($\langle r^2_{\rm M} \rangle^p$), and $\tilde{\chi}^2$
      extracted from dipole fits to the proton magnetic form factor
      for each ensemble, along with continuum extrapolation results
      for both one- and two-step approaches, as in
      Table~\ref{tab:r2_protonge}. For the two-step approach, the
      $\tilde{\chi}^2$ is given as $(\tilde{\chi}^2_{\langle r^2
        \rangle}, \tilde{\chi}^2_{\mu})$ where the former corresponds
      to the fit for $\langle r^2_{\rm M} \rangle^p$ and the latter
      for $\mu^p$.}
    \label{tab:mu_r2_protongm}
    \begin{tabular}{cccc}
    \hline\hline Ensemble & $\mu^p$ & $\langle r^2_{\rm M} \rangle^p$ [fm$^2$] & $\tilde{\chi}^2$ \\
    \hline
    \texttt{cB211.72.64} & 2.524(67) & 0.562(34) & 1.016\\
    \texttt{cC211.60.80} & 2.553(37) & 0.527(17) & 2.230\\
    \texttt{cD211.54.96} & 2.592(49) & 0.569(24) & 2.732\\
    \hline 
    $a=0$, 1-step & 2.66(12) & 0.576(59) & 2.326\\
    $a=0$, 2-step & 2.66(12) & 0.569(60) & (2.182, 0.031)\\
    \hline
    \end{tabular}
\end{table}

\subsubsection{$z$-expansion results}
\begin{figure}
    \centering
    \includegraphics[width=\linewidth]{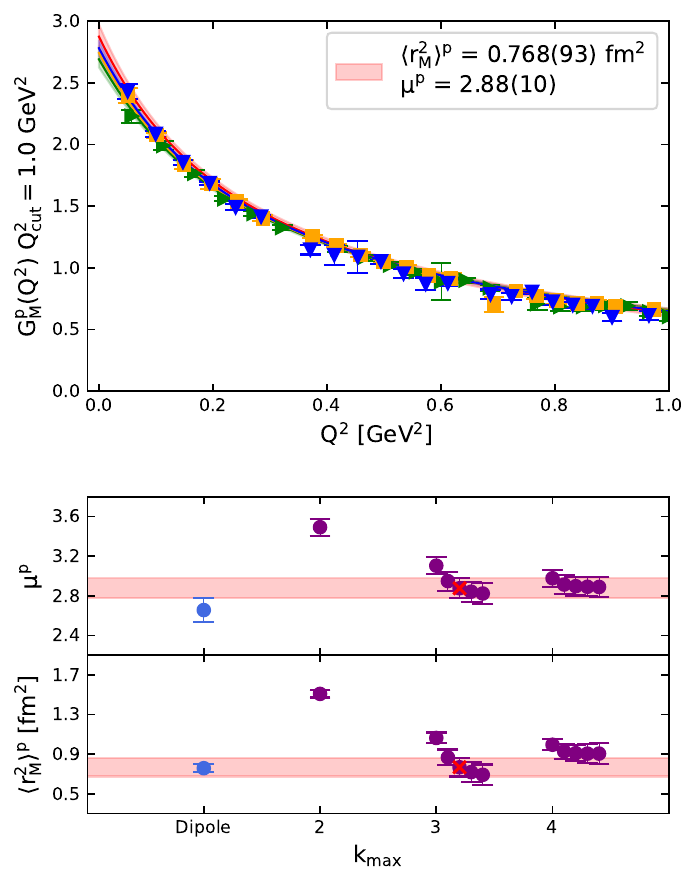}
    \caption{Results on $G^p_M(Q^2)$ using the $z$-expansion with
      $Q_\mathrm{cut}^2=1.0$~GeV$^2$ and $k_{max}=4$ and $w=1$, (top
      panel) and the dependence of the extracted magnetic moment and
      radius on the $z$-expansion order ($k_{\rm max}$) and prior widths
      ($w$) (bottom panel). In the top panel, the green, orange, and
      blue bands show the result of the one-step $z$-expansion fit for
      the B, C and D ensembles, respectively and the red band
      corresponds to the continuum limit. In the bottom panel, we show
      results on the magnetic moment $\mu^p$ and square radius when
      using the $z$-expansion with $k_{\rm max}=2$, 3, and 4, and with
      $w$=1 for the leftmost points and increasing by 1 up to $w=5$.
      The blue point is the result from the
      one-step dipole fit, shown for comparison.}
    \label{fig:zexp_proton_gm_continuum}
\end{figure}

The analysis of the proton magnetic form factors using the $z$-expansion
follows that of the proton electric form factor, where now the $k=0$
coefficients, $c_{0,0}$ and $c_{0,2}$ are fit parameters. The
investigation of $k_{\rm max}$ and prior widths $w$ is shown in
Fig.~\ref{fig:zexp_proton_gm_continuum}. A stronger dependence on $w$
is observed compared to the electric case, with convergence achieved
for $w=3$. At this value of $w$, we observe agreement between $k_{\rm
  max}=3$ and 4. We take $k_{\rm max}=3$, which results in the same
number of fit parameters as the electric case for $k_{\rm max}=4$.

\begin{table}
  \caption{The proton magnetic moment ($\mu^p$), magnetic radius
    ($\langle r^2_{\rm M} \rangle^p$), and  $\tilde{\chi}^2$
    obtained using the one-step approach and the $z$-expansion with
    $k_{\rm max}=3$ and $w=3$ as  we increase $Q^2_{\rm cut}$.}
    \label{tab:protongm_zexp}
    \centering
    \begin{tabular}{cccc}
      \hline\hline $Q^2_{\rm cut} [\rm GeV^2]$ &  $\mu^p$ & $\langle r^2_{\rm M} \rangle^p$ [fm$^2$] & $\tilde{\chi}^2$ \\
    \hline 0.40 & 2.97(12) & 0.98(12) & 1.172\\
           0.50 & 2.92(12) & 0.91(11) & 1.007\\
           0.70 & 2.89(10) & 0.80(10) & 1.311\\
           0.85 & 2.86(10) & 0.79(10) & 1.338\\
           1.00 & 2.88(10) & 0.768(93) & 1.282\\
    \hline
    \end{tabular}
\end{table}

\subsection{Neutron electric form factor}
\begin{figure}
    \centering
    \includegraphics[width=\linewidth]{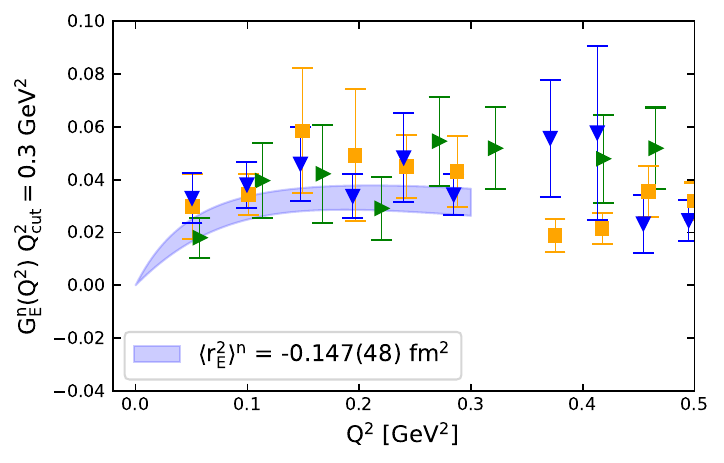}
    \caption{Galster-like fit to $G^n_E(Q^2)$ with
      $Q_{cut}^2=0.3$~GeV$^2$. The blue band shows the resulting fit to
      all ensembles neglecting any $a^2$-dependence. We find $\langle r_E^2 \rangle^n = -0.147(48)$~fm$^2$ and $\tilde{\chi}^2=0.656$.}
    \label{fig:dipole_neutron_ge_ff}
\end{figure}

The neutron electric form factor at $Q^2=0$ is zero by current
conservation. At finite $Q^2$, due to the small absolute magnitude of
the signal arising from the subtraction in Eq.~(\ref{eq:vstopn}), our
data have large relative statistical errors, as shown in
Fig.~\ref{fig:dipole_neutron_ge_ff}. Within these errors, we cannot
discern a dependence on the lattice spacing, and we thus fit all data
using the Galster-like parameterization without including any
$a^2$-dependence. Our fit is thus constrained by the data with the
smaller statistical errors, and we find that taking $Q^2_{\rm
  cut}$=0.3 GeV$^2$ yields a fit with $\tilde{\chi}^2=0.656$. The
value of the neutron electric radius extracted form this fit is
$\langle r_E^2 \rangle^n = -0.147(48)$~fm$^2$.

\subsection{Neutron magnetic form factor}
\subsubsection{Dipole results}
\begin{figure}
    \centering
    \includegraphics[width=1\linewidth]{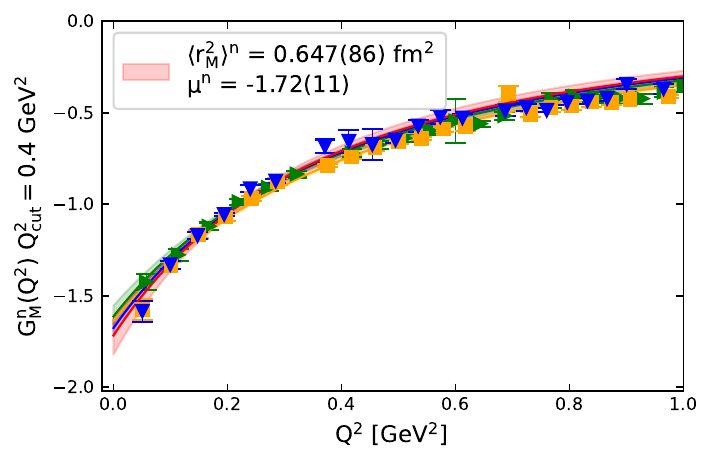}
    \caption{The same as in Fig.~\ref{fig:dipole_proton_gm_ff} but for
      $G^n_M(Q^2)$.}
    \label{fig:dipole_neutron_gm_ff}
\end{figure}

\begin{figure}
    \centering
    \includegraphics[width=\linewidth]{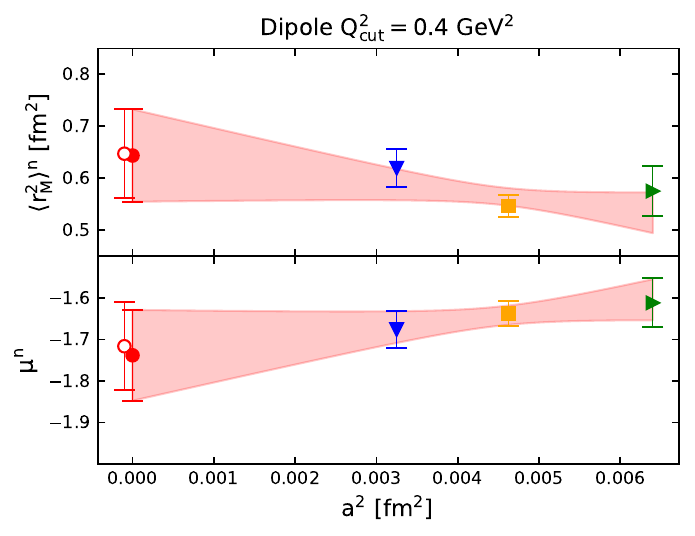}
    \caption{The same as in Fig.~\ref{fig:dipole_proton_gm_continuum}
      but for the neutron magnetic radius $\langle r_M^2 \rangle^n$
      (top) and the neutron magnetic moment $\mu^n$ (bottom).}
    \label{fig:dipole_neutron_gm_continuum}
\end{figure}

Our analysis of the neutron magnetic form factor follows that of the
proton. The dipole fits to the form factor are
shown in Fig.~\ref{fig:dipole_neutron_gm_ff} and in
Fig.~\ref{fig:dipole_neutron_gm_continuum}, we compare the continuum
extrapolation between the one- and two-step approaches, showing that
also in this case the two approaches yield consistent results.  In
Table~\ref{tab:mu_r2_neutrongm}, we summarize the values of the
neutron magnetic radius and the neutron magnetic moment obtained using
the one- and two-step approaches.
\begin{table}
  \centering
  \caption{The neutron magnetic moment ($\mu^n$), magnetic radius
    ($\langle r^2_{\rm M} \rangle^n$), and $\tilde{\chi}^2$ extracted
    from dipole fits to the neutron magnetic form factor for each
    ensemble, along with continuum extrapolation results for both one-
    and two-step approaches, as in Table~\ref{tab:mu_r2_protongm}.
  }
  \label{tab:mu_r2_neutrongm}
  \begin{tabular}{cccc}
    \hline\hline Ensemble & $\mu^n$ & $\langle r^2_{\rm M} \rangle^n$ [fm$^2$] & $\tilde{\chi}^2$ \\
    \hline
    \texttt{cB211.72.64} & -1.612(59) & 0.575(48) & 0.770\\
    \texttt{cC211.60.80} & -1.637(30) & 0.547(21) & 1.883\\
    \texttt{cD211.54.96} & -1.676(45) & 0.619(37) & 2.182\\
    \hline 
    $a=0$, 1-step & -1.72(11) & 0.647(86) & 2.072\\
    $a=0$, 2-step & -1.74(11) & 0.644(89) & (2.207, 0.052)\\
    \hline
  \end{tabular}
\end{table}

\subsubsection{$z$-expansion results}
\begin{figure}[h]
    \centering
    \includegraphics[width=\linewidth]{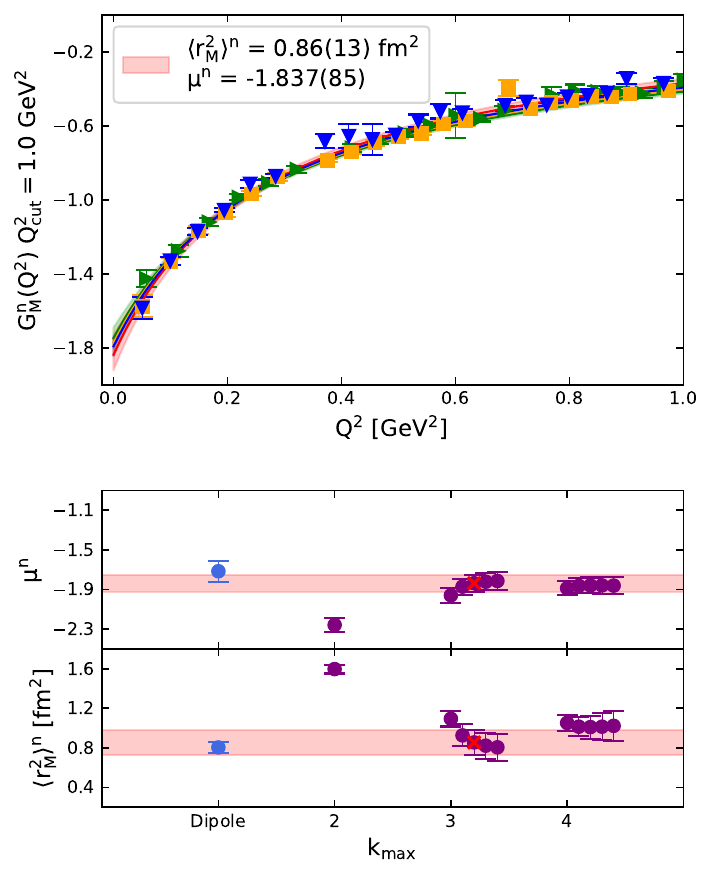}
    \caption{The notation is the same as  the one in Fig.~\ref{fig:zexp_proton_gm_continuum}
      but for $G^n_M(Q^2)$.}
    \label{fig:zexp_neutron_gm_continuum}
\end{figure}

We proceed to carry out the analysis using the $z$-expansion following
the analysis  presented for the proton case. In
Fig.~\ref{fig:zexp_neutron_gm_continuum}, we show the $z$-expansion
fits, with the analysis of $k_{\rm max}$ and $w$ in the bottom
panel. We see that as in the proton case, the fits with $k_{\rm
  max}=3$ and $w=3$ achieve convergence for both the radius and the
moment. Fixing these parameters, we vary $Q^2_{\rm cut}$ and give the
results  in Table~\ref{tab:neutrongm_zexp}.

\begin{table}
    \centering
    \caption{The neutron magnetic moment ($\mu^n$), magnetic radius
      ($\langle r^2_{\rm M} \rangle^n$), and   $\tilde{\chi}^2$
      obtained using the one-step approach and the $z$-expansion with
      $k_{\rm max}=3$ and $w=3$ as we  increase $Q^2_{\rm cut}$.}    
    \label{tab:neutrongm_zexp}    \begin{tabular}{cccc}
      \hline\hline $Q^2_{\rm cut} [\rm GeV^2]$ &  $\mu^n$ & $\langle r^2_{\rm M} \rangle^n$ [fm$^2$] & $\tilde{\chi}^2$ \\
      \hline 0.40 & -1.95(12) & 1.11(17) & 0.987\\
      0.50 & -1.88(11) & 1.02(16) & 0.890\\
      0.70 & -1.856(92) & 0.90(14) & 1.113\\
      0.85 & -1.838(91) & 0.88(14) & 1.024\\
      1.00 & -1.837(85) & 0.86(13) & 1.012\\
      \hline    \end{tabular}
\end{table}

\section{Results and comparison to other studies}
\label{sec:final_results}
\subsection{Final results on the electric and magnetic radii and moments}
For our final results we use the one-step approach with either a
linear or no $a^2$-dependence since, as already explained, compared to
the two-step approach it allows for a single value of $\tilde{\chi}^2$
for the combined $Q^2$-dependence and continuum extrapolation fit.
The models used are given in Table~\ref{tab:final_results}, and
include one-step dipole and $z$-expansion fits with varying values of
$Q^2_{\rm cut}$.  We note the exception of the neutron electric form
factor, for which we use a single, Galster-like fit for the
description of the $Q^2$-dependence. The probability associated with
each fit is provided using the AIC.

A comparison of these fits based on the fit probability shows that for
the electric and magnetic proton form factors and for the magnetic
neutron form factor, the one-step $z$-expansion parameterization with
linear $a^2$-dependence has the highest weight, with the dipole fit
effectively not contributing to the model average. Furthermore, all
three cases favor a higher $Q^2_{\rm cut}$, with the proton electric
favoring $Q^2_{\rm cut} = 0.85$~GeV$^2$ and the magnetic proton and
neutron form factors favoring $Q^2_{\rm cut} = 1$~GeV$^2$.

For the final row of Table~\ref{tab:final_results}, we quote our
statistical and systematic errors separately, the latter derived from
the $Q^2_{\rm cut}$- and model-dependence. Namely, if $p_j$ is the
probability of the combination of $Q^2_{\rm cut}$ and model $j$, with
observable given by $\bar{\mathcal{O}}_j(\sigma_j)$ we define,
\begin{align}
\langle \mathcal{O} \rangle = \bar{\mathcal{O}} \; (\sigma^{\rm stat}) \; (\sigma^{\rm sys}) 
\quad \text{where,} \quad
\bar{\mathcal{O}} = \sum_j \bar{\mathcal{O}}_j p_j \;,\nonumber\\
(\sigma^{\rm stat})^2 = \sum_j \sigma_j^2 p_j 
\quad \text{and,} \quad
(\sigma^{\rm sys})^2 = \sum_j \bar{\mathcal{O}}_j^2 p_j - \bar{\mathcal{O}}^2.
\end{align}

\begin{table*}
  \centering
    \caption{Results for the proton and neutron electric and magnetic
      radii squared and for their magnetic moments. We use the
      one-step approach and a $z$-expansion or dipole with a linear
      $a^2$ dependence (``z-exp($a^2$)'' and ``Dipole($a^2$)'' in the
      first column, respectively) or with no $a^2$ dependence
      (``z-exp'' and ``Dipole'' in the first column,
      respectively). For each combination of model and $Q^2_{\rm cut}$
      we indicate the probability from the model-averaging procedure
      in the columns labeled ``prob''.  For the neutron electric case,
      we use the Galster-like parameterization with a single $Q^2_{\rm
        cut}$ and no $a^2$ dependence as explained in the text.}
    \label{tab:final_results}
    \begin{tabular}{cc|cc|ccc|ccc|ccc|c}
    \hline
    \hline  Model &\makecell[c]{$Q^2_{\rm cut}$\\$[\rm GeV^2]$} & \makecell[c]{$\langle r^2_{\rm E} \rangle^p$\\ $[$fm$^2]$}  & \makecell[c]{prob\\$[\%]$} & $\mu^p$ & \makecell[c]{$\langle r^2_{\rm M} \rangle^p$\\ $[$fm$^2]$} & \makecell[c]{prob\\ $[\%]$} & $\mu^n$ & \makecell[c]{$\langle r^2_{\rm M} \rangle^n$\\ $[$fm$^2]$} & \makecell[c]{prob\\ $[\%]$} & \makecell[c]{$\langle r^2_{\rm E} \rangle^n$\\ $[$fm$^2]$} \\
    \hline
       
    z-exp($a^2$) &0.40 & 0.700(76) & 0.000 & 2.97(12) & 0.98(12) & 0.000 & -1.95(12) & 1.11(17) & 0.000 & - \\
    z-exp($a^2$) &0.50 & 0.713(72) & 0.186 & 2.92(12) & 0.91(11) & 0.027 & -1.88(11) & 1.02(16) & 0.000 & - \\
    z-exp($a^2$) &0.70 & 0.716(69) & 0.565 & 2.89(10) & 0.80(10) & 0.033 & -1.856(92) & 0.90(14) & 0.000 & - \\
    z-exp($a^2$) &0.85 & 0.747(67) & 82.151 & 2.86(10) & 0.79(10) & 0.735 & -1.838(91) & 0.88(14) & 0.613 & - \\
    z-exp($a^2$) &1.00 & 0.816(64) & 5.983 & 2.88(10) & 0.768(93) & 79.124 & -1.837(85) & 0.86(13) & 70.807 & - \\
    z-exp &0.40 & 0.652(49) & 0.000 & 2.849(68) & 0.93(11) & 0.000 & -1.855(65) & 1.03(16) & 0.000 & - \\
    z-exp &0.50 & 0.658(48) & 0.301 & 2.828(66) & 0.88(10) & 0.010 & -1.840(63) & 0.98(15) & 0.000 & - \\
    z-exp &0.70 & 0.651(47) & 0.644 & 2.763(63) & 0.75(10) & 0.010 & -1.795(59) & 0.85(14) & 0.000 & - \\
    z-exp &0.85 & 0.636(46) & 10.158 & 2.757(61) & 0.739(94) & 0.277 & -1.775(55) & 0.79(13) & 0.302 & - \\
    z-exp &1.00 & 0.667(46) & 0.009 & 2.744(58) & 0.705(87) & 19.784 & -1.772(52) & 0.78(12) & 28.278 & - \\
    
    \hline 
   Dipole($a^2$) & 0.4 & 0.650(52) & 0.000 & 2.66(12) & 0.576(59) & 0.000 & -1.72(11) & 0.647(86) & 0.000 & - \\
   Dipole & 0.4 & 0.620(12) & 0.000 & 2.551(27) & 0.545(13) & 0.000 & -1.626(22) & 0.563(17) & 0.000 & - \\

    \hline Galster-like & 0.3 & - & - & - & - & - & - & - & - &  -0.147(48)\\
    \hline \multicolumn{2}{c}{Final results} & 0.739(64)(39) & & 2.849(92)(52) & 0.756(92)(25) & & -1.819(76)(29) & 0.83(12)(03) & & -0.147(48)\\
    \hline
    \end{tabular}
\end{table*}

\begin{figure}
  \centering
    \includegraphics[width=\linewidth]{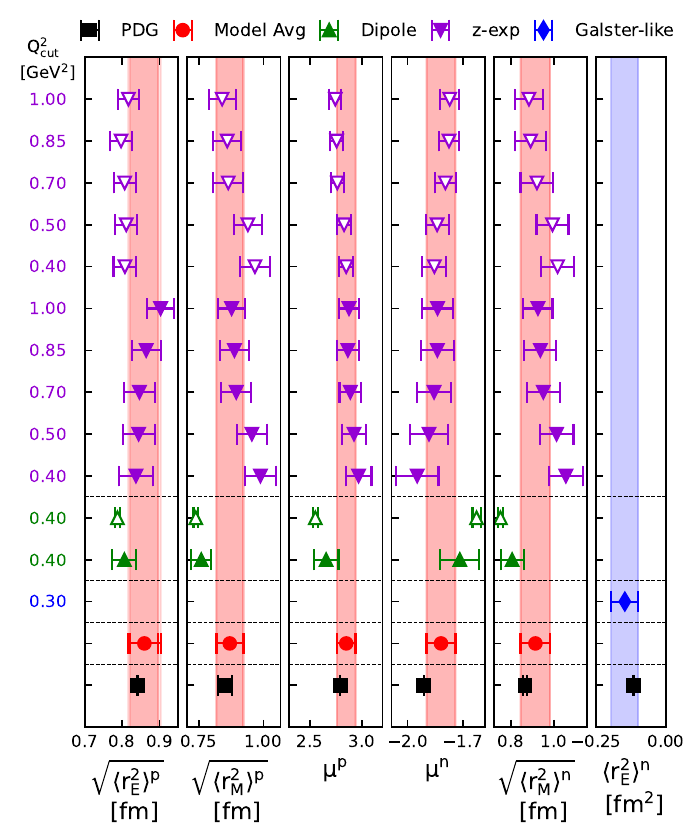}
    \caption{Electric and magnetic radii and magnetic moments of the
      proton and neutron for all combinations of $Q^2_{\rm cut}$ and
      models considered in Table~\ref{tab:final_results}. Namely, we
      show the $z$-expansion (purple downward-pointing triangles) and
      dipole fits (green triangles) using the one-step approach with
      (filled symbols) or without (open symbols) the
      $a^2$-dependence. The blue diamond and blue band corresponds to
      the Galster-like fit to $G_E^n(Q^2)$. The red point and band
      denoted ``Model average'' is obtained by weighting according to
      the AIC as explained in Sec.~\ref{sec:extraction}.}
    \label{fig:Results}
\end{figure}

Our results for the radii and magnetic moments are shown in
Fig.~\ref{fig:Results} for varying $Q^2_{\rm cut}$ and for the
one-step approach with and without an $a^2$-dependence. We overall
observe consistent results among the variations.  The model-averaged
result, also shown, is consistent with the PDG
values~\cite{ParticleDataGroup:2024cfk} for these quantities.
\begin{figure*}
    \includegraphics[width=1\linewidth]{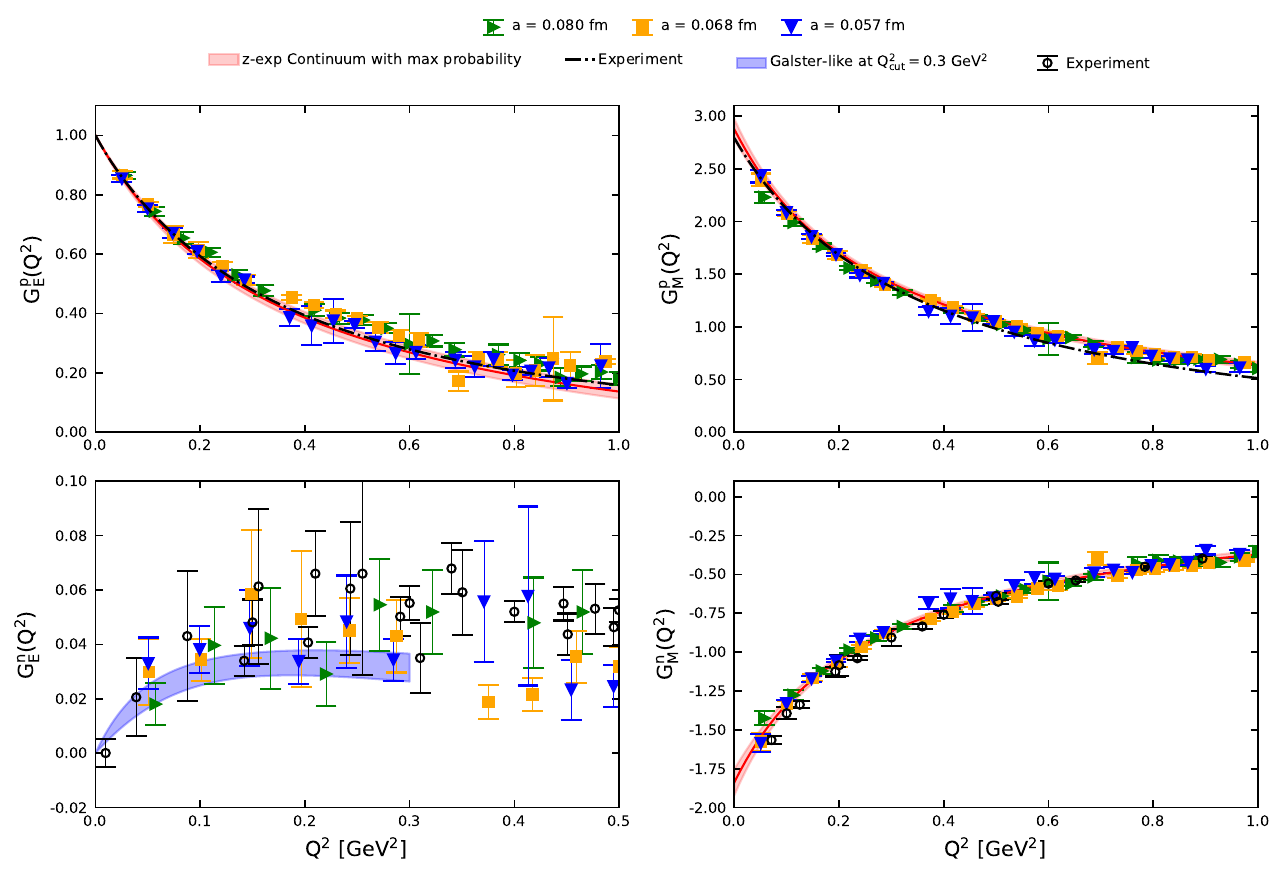}
    \caption{$G_E(Q^2)$ (left) and $G_M(Q^2)$ (right), proton (top)
      and neutron (bottom) form factors as a function of $Q^2$ for the
      three ensembles analyzed. The red bands indicate the continuum
      limit using the $z$-expansion with $Q^2_{\rm cut} =
      0.85$~GeV$^2$ for the case of proton electric and $Q^2_{\rm cut}
      = 1$~GeV$^2$ for the proton and neutron magnetic form
      factors. The light blue band corresponds to the Galster-like fit
      to the neutron electric form factor with $Q^2_{\rm cut} =
      0.3$~GeV$^2$. The black dashed lines for the proton case are
      $z$-expansion fits to experimental
      results~\cite{Ye:2017gyb}. The black circles for the neutron
      case are from a collection of experimental
      results~\cite{Golak:2000nt,Becker:1999tw,Eden:1994ji,Meyerhoff:1994ev,Passchier:1999cj,JeffersonLabE93-026:2003tty,E93026:2001css,E93-038:2003ixb,
        E93-038:2003ixb,
        Rohe:1999sh,Bermuth:2003qh,Glazier:2004ny,Herberg:1999ud,Schiavilla:2001qe,Ostrick:1999xa,JeffersonLabE95-001:2006dax,Gao:1994ud,Anklin:1994ae,Anklin:1998ae,
        Kubon:2001rj,Alarcon:2007zza}.}
    \label{fig:GEMpn}
\end{figure*}

Our final results for the proton and neutron electromagnetic form
factors are shown in Fig.~\ref{fig:GEMpn}, where we also compare with
experimental fits and data. For our continuum limit band, we show the
highest probability fit for each case, namely the $z$-expansion with
$Q_{\rm cut}^2 = 0.85$~GeV$^2$ for $G_E^p$ and $Q_{\rm cut}^2 =
1$~GeV$^2$ for $G_M^p$ and $G_M^n$ and using the Galster-like
parameterization for $G_E^n$. For comparison with experiment, for the
proton form factors we use a $z$-expansion fit to experimental data
from Ref.~\cite{Ye:2017gyb} and for the neutron form factors we show
experimental data from
Refs.~\cite{Golak:2000nt,Becker:1999tw,Eden:1994ji,Meyerhoff:1994ev,Passchier:1999cj,JeffersonLabE93-026:2003tty,E93026:2001css,E93-038:2003ixb,
  E93-038:2003ixb,
  Rohe:1999sh,Bermuth:2003qh,Glazier:2004ny,Herberg:1999ud,Schiavilla:2001qe,Ostrick:1999xa,JeffersonLabE95-001:2006dax,Gao:1994ud,Anklin:1994ae,Anklin:1998ae,
  Kubon:2001rj,Alarcon:2007zza}. The values for the radii and magnetic
moments that correspond to the final results of the form factors are
included in Fig.~\ref{fig:Results}.

For completeness we summarize below our final result for electric and
magnetic radii and magnetic moment of the proton,
\begin{align}
  \langle r^2_E \rangle^p     &= 0.739(64)(39)~{\rm fm}^2, \nonumber\\
  \langle r^2_M \rangle^p     &= 0.756(92)(25)~{\rm fm}^2, \nonumber\\
  \mu^p                       &= 2.849(92)(25),
\end{align}
the neutron,
\begin{align}
  \langle r^2_E \rangle^n     &= -0.147(48)~{\rm fm}^2, \nonumber\\
  \langle r^2_M \rangle^n     &= 0.83(12)(03)~{\rm fm}^2, \nonumber\\
  \mu^n                       &= -1.819(76)(29),
\end{align}
and the isovector combinations,
\begin{align}
  \langle r^2_E \rangle^{u-d}     &= 0.886(80)(39)~{\rm fm}^2, \nonumber\\
  \langle r^2_M \rangle^{u-d}     &= 0.786(73)(20)~{\rm fm}^2, \nonumber\\
  \mu^{u-d}                       &= 4.67(12)(6).
\end{align}

\subsection{Comparison of electric and magnetic radii and moments with other lattice QCD studies}
\begin{figure*}
  \centering
  \includegraphics[width=\linewidth]{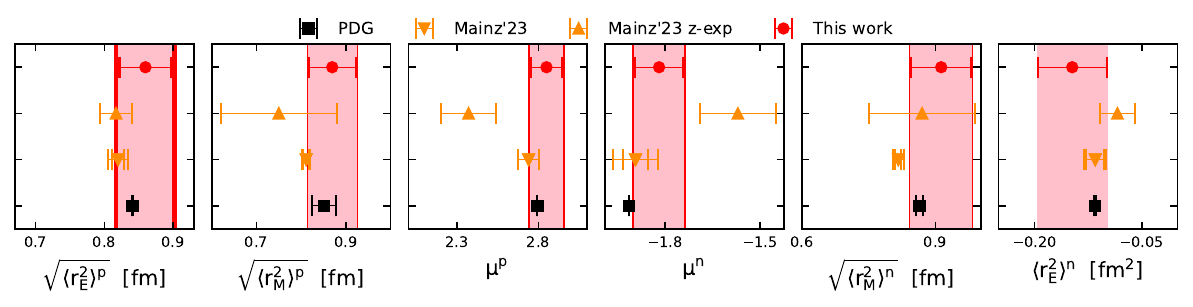}
    \caption{Results for the electric and magnetic radii and the
      magnetic moments of the proton and neutron obtained within this
      work (red circles). The light vertical bands correspond to
      statistical error and the dark red bands corresponds to the
      total error, obtained by adding statistical and systematic
      errors in quadrature. We compare to results by the Mainz
      collaboration (Mainz'23 and Mainz'23
      z-exp.~\cite{Djukanovic:2023beb}) as well as results from the
      Particle Data Group (PDG~\cite{ParticleDataGroup:2024cfk})}
    \label{fig:Results_compare}
\end{figure*}

The electromagnetic form factors of the nucleon were among the first
nucleon structure quantities studied within lattice QCD, and therefore
results, especially for the isovector case in which disconnected
contributions cancel, go back several decades. For the purposes of
comparing with similar results in the literature, in
Fig.~\ref{fig:Results_compare}, we compare our results to a recent
calculation by the Mainz collaboration which includes a continuum
extrapolation using one physical pion mass ensemble and 10 additional
ensembles at heavier than physical pion
mass~\cite{Djukanovic:2023beb}. This calculation employs \Nf{2}{1} CLS
ensembles and includes disconnected contributions. The summation
method is used for the treatment of their excited states and they
perform a simultaneous chiral, continuum, and infinite volume fit
using an expression from chiral perturbation theory. Their final
result is obtained after a cut in the pion mass of $M_{\pi}\le
0.3$~GeV that leaves six of the 11 ensembles contributing. Beyond
their quoted final result, we include their result obtained using the
$z$-expansion (denoted as Mainz'23 z-exp. in
Fig.~\ref{fig:Results_compare}) for a direct comparison to our
results. We observe good agreement with the results by the Mainz group
within 1.5$\sigma$.

Other recent results in the literature which quote proton and neutron
form factors and that include physical point ensembles but lack a
continuum limit include two previous studies by the
ETMC~\cite{Alexandrou:2017ypw,Alexandrou:2018sjm} and
PACS~\cite{Shintani:2018ozy,Tsuji:2023llh}.

\begin{itemize}
    \item ETMC results include one study using a single ensemble of
      \Nf{2} twisted mass fermions with lattice spacing $a=0.094$~fm
      \cite{Alexandrou:2017ypw} and physical pion mass, as well as a
      more recent result obtained using only the B ensemble used in
      this work~\cite{Alexandrou:2018sjm}. Our current work, which
      includes a continuum extrapolation, yields larger values for the
      radii compared to our previous results. Beyond the continuum
      limit extrapolation, this work includes a more thorough
      excited-state analysis with systematic errors obtained from
      model-averaging.

    \item The PACS Collaboration has two results using \Nf{2}{1}
      stout-smeared, $\mathcal{O}$(a)-improved Wilson-clover fermions,
      with $a$=0.084~fm and lattice volume
      $128^3\times128$~\cite{Shintani:2018ozy} and with $a$=0.063~fm
      and lattice volume $160^3\times160$~\cite{Tsuji:2023llh}. They
      perform plateau fits to obtain their final results using source
      sink separations ranging between 0.82~fm to 1.35~fm. Their
      results on proton and neutron electromagnetic form factors are
      obtained by combining with only the connected contribution to
      the isoscalar form factor since they do not perform the
      calculation of disconnected contributions. There is qualitative
      agreement between our results and those of the PACS
      Collaboration which suggests that finite volume effects are
      likely within our combined statistical and systematic errors.
\end{itemize}

\subsection{Zemach and Friar radii}
With our results for the form factors at the continuum limit, we can
further analyze the $Q^2$ dependence to study moments of the charge
and magnetization distribution for both proton and neutron, known as
the Zemach moments. Experimentally, the Zemach moments enter the
determination of the proton radius from Lamb shifts, such as for
example the high-precision determination using muonic
hydrogen~\cite{Antognini:2013txn} which led to the proton radius
puzzle. The Zemach moments can be related to the electromagnetic form
factors~\cite{PhysRevD.102.074012, Distler:2010zq,
  PhysRevLett.128.052002}, for example the Zemach radius is given by,

\begin{align}
  r_Z &= -\frac{2}{\pi} \int_0^\infty \frac{dQ^2}{(Q^2)^{3/2}} \left[ \frac{G_E(Q^2) G_M(Q^2)}{\mu_M} - 1 \right].
  \label{eq:zemach}
\end{align}
while the third Zemach moment is given by,
\begin{align}
\left\langle r_F^3 \right\rangle 
&= \frac{24}{\pi} \int_0^\infty \frac{dQ^2}{(Q^2)^{5/2}} \left[ \left(G_E(Q^2)\right)^2 + \eta(Q^2) \right]\label{eq:friar},
\end{align}
where $\eta(Q^2) = \left(- 1 + \frac{1}{3} \langle r_E^2 \rangle^p
Q^2\right)$ for proton and $0$ for neutron. $\left\langle r_F^3
\right\rangle^{1/3}$ corresponds to the Friar radius.

\begin{figure*}
  \centering
    \includegraphics[width=0.9\linewidth]{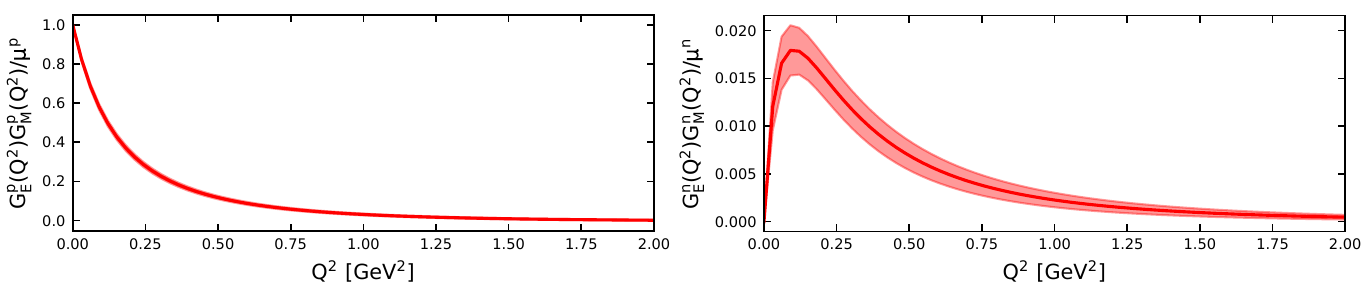}
    \caption{Product of proton (left) and neutron (right) electric and
      magnetic form factors as a function of $Q^2$, normalized by the
      respective magnetic moment.}
    \label{fig:Integrand_z}
\end{figure*}

We use our final result for the form factors at the continuum limit as
presented in Fig.~\ref{fig:GEMpn} to compute the Zemach and Friar
radii. Namely, using the parameters given in Table~\ref{tab:fitp},
which correspond to the most probable $z$-expansion fits to $G_E^p$,
$G_M^p$, and $G_M^n$, and the Galster-like fit to $G_E^n$, we carry
out numerically the two integrals in Eqs.~(\ref{eq:zemach})
and~(\ref{eq:friar}). In Fig.~\ref{fig:Integrand_z}, we plot the
product of the electric and magnetic form factors for the proton and
neutron, normalized by the respective magnetic moment, which forms the
main part of the integrand of Eq.~(\ref{eq:zemach}) which yields the
Zemach radius.

\begin{figure}[!h]
  \centering
    \includegraphics[width=\linewidth]{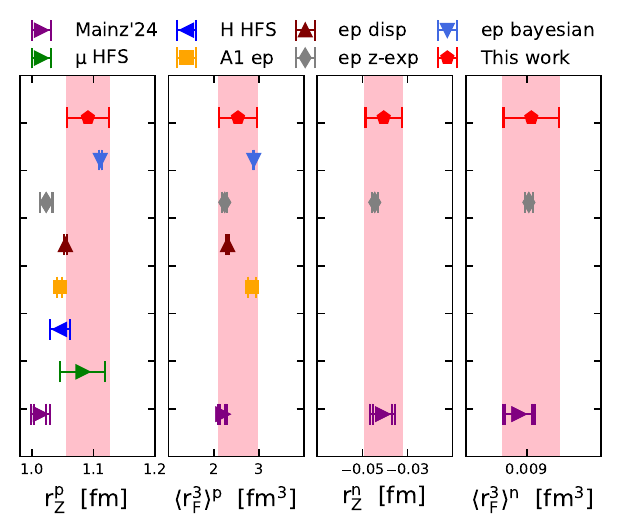}
    \caption{Results for the Zemach and Friar proton (left two
      columns) and neutron (right two columns) radii as obtained
      within this work (red circles and vertical bar) and compared to
      lattice results by the Mainz collaboration (Mainz'24,
      \cite{PhysRevD.110.L011503}), and experimental results
      (\cite{PhysRevD.102.074012, Distler:2010zq,
        PhysRevLett.128.052002, Graczyk:2015kka, Volotka:2004zu, Antognini:2013txn}). }
    \label{fig:Results_compare_zf}
\end{figure}

Our results for the Zemach and Friar radii are given in
Table~\ref{tab:zemach_friar} and plotted in
Fig.~\ref{fig:Results_compare_zf} where we also compare with other
works in the literature. Errors are obtained by carrying out the
numerical integration over bootstrap samples drawn according to the
errors and correlations in the $z$-expansion and Galster-like fit
parameters. The only other available lattice study for these
quantities is by the Mainz collaboration~\cite{PhysRevD.110.L011503},
where we observe agreement within $1\sigma$ for all quantities except
proton Zemach radius, where we agree within $2\sigma$. Our results are
also in agreement with experimental results based on $ep$-scattering
and electronic and muonic hydrogen hyperfine splitting (HFS)
experiments, which we also compare to~\cite{PhysRevD.102.074012,
  Distler:2010zq, PhysRevLett.128.052002, Graczyk:2015kka,
  Graczyk:2015kka, Volotka:2004zu, Antognini:2013txn}.

\begin{table}
\caption{Results for the Zemach and Friar radii for proton and
  neutron}
\label{tab:zemach_friar}
\begin{tabular}{llll}
\hline
\hline
$r_Z^{p}$  [fm] & $\langle r_F^3 \rangle^{p}$  [fm$^3$] & $r_Z^{n}$  [fm] & $\langle r_F^3 \rangle^{n}$  [fm$^3$] \\
\hline
1.091(34) & 2.54(43) & -0.0405(81) & 0.0097(41) \\
\hline
\end{tabular}
\end{table}

\section{Conclusions}
\label{sec:conclusions}
An analysis of the electromagnetic form factors of the nucleon is
presented. We use three ensembles of \Nf{2}{1}{1} twisted mass
fermions at three lattice spacings simulated at approximately physical
pion mass. We carry out an extended excited state analysis which
employs multi-state fits and data with high statistics. This allows us
to use different excited states when fitting the two- and three-point
functions. We carry out the calculation of the disconnected
contributions to the isoscalar form factors, which allows us to obtain
the proton and neutron electromagnetic form factors.

The $Q^2$-dependence of the proton and neutron magnetic form factors
are fitted with both dipole and $z$-expansion forms to extract the
related radii and magnetic moments. The neutron electric form factor
carries larger errors, and we use a Galster-like parameterization.
Our final results include statistical and systematic errors, where the
latter are obtained from a model-average when considering the
different fits to extract the ground state matrix elements and using
the dipole and $z$-expansion Ans\"atze and varying
$Q^2_\mathrm{cut}$. Our results on the form factors agree with
experimental results. For the neutron electric form factor, our
results are more precise than the experimental ones.  For the radii
and magnetic moments we extract the following values: $\langle
\sqrt{r_E^2\rangle^p} = 0.860(38)(23)$~fm and $\langle r_E^2\rangle^n
= -0.147(48)$~fm$^2$ for the proton and neutron electric radius,
respectively, $\mu^p=2.849(92)(52)$ and $\mu^n=-1.819(76)(29)$ for the
proton and neutron magnetic moments, respectively, and $\sqrt{\langle
  r_M^2\rangle^p} = 0.870(53)(15)$~fm and $\sqrt{\langle
  r_M^2\rangle^n} = 0.913(67)(19)$~fm for the proton and neutron
magnetic radii, respectively. We also proceed to compute the Zemach
and Friar radii, obtaining $r_Z^{p}=1.091(34)$~fm and $\langle
r_F^3\rangle^p$=2.54(43)~fm$^3$ for the proton and
$r_Z^{n}=-0.0405(81)$~fm and $\langle
r_F^3\rangle^p$=0.0097(41)~fm$^3$ for the neutron. Our results for the
electric and magnetic radii and magnetic moments are consistent with
the PDG values for these quantities within our combined statistical
and systematic errors. Our result for the proton electric radius
carries an error that cannot distinguish between the larger value
extracted from historical electron–proton scattering
results~\cite{Mohr:2015ccw} and the smaller radius measured in muonic
hydrogen spectroscopy experiments~\cite{Antognini:2013txn}.
Increasing the accuracy of the results for each ensemble, but even
more importantly, including more gauge ensembles with smaller lattice
spacing will help reduce this error. Furthermore, an analysis
including gauge ensembles with larger physical volumes simulated at
physical pion mass recently by ETMC~\cite{Alexandrou:2025bkm} will
allow for more lattice momenta close to $Q^2$=0, which can potentially
lead to more robust fits of the $Q^2$-dependence in turn reducing
systematic errors in the determination of the radii and magnetic
moments.

\section*{Acknowledgements}
We thank Matteo Di Carlo, Antonio Evangelista, Roberto Frezzotti,
Giuseppe Gagliardi and Vittorio Lubicz, for useful discussions and
crosschecks on the analysis of the renormalization factors.  C.A.,
S.B, G.K., and G.S.  acknowledge partial support by the projects PulseQCD,
3D-nucleon, NiceQuarks, Image-n, MuonHVP, Baryon8 and HyperON 
(EXCELLENCE/0524/0269, EXCELLENCE/0421/0043, EXCELLENCE/0421/0195, 
EXCELLENCE/0524/0459, EXCELLENCE/0524/0017, POSTDOC/0524/0001, VISION ERC-PATH 2/0524/0001) co-financed by the European Regional Development
Fund and co-funded by the EU within the framework of the Cohesion Policy Programme “THALIA 2021-2027” and the Republic of Cyprus through the Research and Innovation
Foundation as well as AQTIVATE that received funding from the European
Union’s research and innovation program under the Marie
Sklodowska-Curie Doctoral Networks action, Grant Agreement No
101072344. C.A acknowledges support by the University of Cyprus
projects ``Nucleon-GPDs'' and ``PDFs-LQCD''. S.B. is supported by
Inno4scale, which received funding from the European High-Performance
Computing Joint Undertaking (JU) GA No.~101118139. B.P. is supported
by ENGAGE which received funding from the EU's Horizon 2020 Research
and Innovation Programme under the Marie Skłodowska-Curie GA
No. 101034267. This work was supported by grants from the Swiss
National Supercomputing Centre (CSCS) under projects with ids s702 and
s1174. The authors gratefully acknowledge the Gauss Centre for
Supercomputing e.V. (www.gauss-centre.eu) for funding this project by
providing computing time through the John von Neumann Institute for
Computing (NIC) on the GCS Supercomputers JUWELS~\cite{JUWELS} and
JUWELS Booster~\cite{JUWELS-BOOSTER} at J\"ulich Supercomputing Centre
(JSC). The authors also acknowledge the Texas Advanced Computing
Center (TACC) at University of Texas at Austin for providing HPC
resources.

 \bibliography{refs}

\begin{thebibliography}{76}%
\makeatletter
\providecommand \@ifxundefined [1]{%
 \@ifx{#1\undefined}
}%
\providecommand \@ifnum [1]{%
 \ifnum #1\expandafter \@firstoftwo
 \else \expandafter \@secondoftwo
 \fi
}%
\providecommand \@ifx [1]{%
 \ifx #1\expandafter \@firstoftwo
 \else \expandafter \@secondoftwo
 \fi
}%
\providecommand \natexlab [1]{#1}%
\providecommand \enquote  [1]{``#1''}%
\providecommand \bibnamefont  [1]{#1}%
\providecommand \bibfnamefont [1]{#1}%
\providecommand \citenamefont [1]{#1}%
\providecommand \href@noop [0]{\@secondoftwo}%
\providecommand \href [0]{\begingroup \@sanitize@url \@href}%
\providecommand \@href[1]{\@@startlink{#1}\@@href}%
\providecommand \@@href[1]{\endgroup#1\@@endlink}%
\providecommand \@sanitize@url [0]{\catcode `\\12\catcode `\$12\catcode
  `\&12\catcode `\#12\catcode `\^12\catcode `\_12\catcode `\%12\relax}%
\providecommand \@@startlink[1]{}%
\providecommand \@@endlink[0]{}%
\providecommand \url  [0]{\begingroup\@sanitize@url \@url }%
\providecommand \@url [1]{\endgroup\@href {#1}{\urlprefix }}%
\providecommand \urlprefix  [0]{URL }%
\providecommand \Eprint [0]{\href }%
\providecommand \doibase [0]{https://doi.org/}%
\providecommand \selectlanguage [0]{\@gobble}%
\providecommand \bibinfo  [0]{\@secondoftwo}%
\providecommand \bibfield  [0]{\@secondoftwo}%
\providecommand \translation [1]{[#1]}%
\providecommand \BibitemOpen [0]{}%
\providecommand \bibitemStop [0]{}%
\providecommand \bibitemNoStop [0]{.\EOS\space}%
\providecommand \EOS [0]{\spacefactor3000\relax}%
\providecommand \BibitemShut  [1]{\csname bibitem#1\endcsname}%
\let\auto@bib@innerbib\@empty
\bibitem [{\citenamefont {Bernauer}\ \emph {et~al.}(2014)\citenamefont
  {Bernauer} \emph {et~al.}}]{A1:2013fsc}%
  \BibitemOpen
  \bibfield  {author} {\bibinfo {author} {\bibfnamefont {J.~C.}\ \bibnamefont
  {Bernauer}} \emph {et~al.} (\bibinfo {collaboration} {A1}),\ }\bibfield
  {title} {\bibinfo {title} {{Electric and magnetic form factors of the
  proton}},\ }\href {https://doi.org/10.1103/PhysRevC.90.015206} {\bibfield
  {journal} {\bibinfo  {journal} {Phys. Rev. C}\ }\textbf {\bibinfo {volume}
  {90}},\ \bibinfo {pages} {015206} (\bibinfo {year} {2014})},\ \Eprint
  {https://arxiv.org/abs/1307.6227} {arXiv:1307.6227 [nucl-ex]} \BibitemShut
  {NoStop}%
\bibitem [{\citenamefont {Punjabi}\ \emph {et~al.}(2015)\citenamefont
  {Punjabi}, \citenamefont {Perdrisat}, \citenamefont {Jones}, \citenamefont
  {Brash},\ and\ \citenamefont {Carlson}}]{Punjabi:2015bba}%
  \BibitemOpen
  \bibfield  {author} {\bibinfo {author} {\bibfnamefont {V.}~\bibnamefont
  {Punjabi}}, \bibinfo {author} {\bibfnamefont {C.~F.}\ \bibnamefont
  {Perdrisat}}, \bibinfo {author} {\bibfnamefont {M.~K.}\ \bibnamefont
  {Jones}}, \bibinfo {author} {\bibfnamefont {E.~J.}\ \bibnamefont {Brash}},\
  and\ \bibinfo {author} {\bibfnamefont {C.~E.}\ \bibnamefont {Carlson}},\
  }\bibfield  {title} {\bibinfo {title} {{The Structure of the Nucleon: Elastic
  Electromagnetic Form Factors}},\ }\href
  {https://doi.org/10.1140/epja/i2015-15079-x} {\bibfield  {journal} {\bibinfo
  {journal} {Eur. Phys. J. A}\ }\textbf {\bibinfo {volume} {51}},\ \bibinfo
  {pages} {79} (\bibinfo {year} {2015})},\ \Eprint
  {https://arxiv.org/abs/1503.01452} {arXiv:1503.01452 [nucl-ex]} \BibitemShut
  {NoStop}%
\bibitem [{\citenamefont {Pohl}\ \emph {et~al.}(2010)\citenamefont {Pohl} \emph
  {et~al.}}]{Pohl:2010zza}%
  \BibitemOpen
  \bibfield  {author} {\bibinfo {author} {\bibfnamefont {R.}~\bibnamefont
  {Pohl}} \emph {et~al.},\ }\bibfield  {title} {\bibinfo {title} {{The size of
  the proton}},\ }\href {https://doi.org/10.1038/nature09250} {\bibfield
  {journal} {\bibinfo  {journal} {Nature}\ }\textbf {\bibinfo {volume} {466}},\
  \bibinfo {pages} {213} (\bibinfo {year} {2010})}\BibitemShut {NoStop}%
\bibitem [{\citenamefont {Golak}\ \emph {et~al.}(2001)\citenamefont {Golak},
  \citenamefont {Ziemer}, \citenamefont {Kamada}, \citenamefont {Witala},\ and\
  \citenamefont {Gloeckle}}]{Golak:2000nt}%
  \BibitemOpen
  \bibfield  {author} {\bibinfo {author} {\bibfnamefont {J.}~\bibnamefont
  {Golak}}, \bibinfo {author} {\bibfnamefont {G.}~\bibnamefont {Ziemer}},
  \bibinfo {author} {\bibfnamefont {H.}~\bibnamefont {Kamada}}, \bibinfo
  {author} {\bibfnamefont {H.}~\bibnamefont {Witala}},\ and\ \bibinfo {author}
  {\bibfnamefont {W.}~\bibnamefont {Gloeckle}},\ }\bibfield  {title} {\bibinfo
  {title} {{Extraction of electromagnetic neutron form-factors through
  inclusive and exclusive polarized electron scattering on polarized He-3
  target}},\ }\href {https://doi.org/10.1103/PhysRevC.63.034006} {\bibfield
  {journal} {\bibinfo  {journal} {Phys. Rev. C}\ }\textbf {\bibinfo {volume}
  {63}},\ \bibinfo {pages} {034006} (\bibinfo {year} {2001})},\ \Eprint
  {https://arxiv.org/abs/nucl-th/0008008} {arXiv:nucl-th/0008008} \BibitemShut
  {NoStop}%
\bibitem [{\citenamefont {Xiong}\ \emph {et~al.}(2019)\citenamefont {Xiong}
  \emph {et~al.}}]{Xiong:2019umf}%
  \BibitemOpen
  \bibfield  {author} {\bibinfo {author} {\bibfnamefont {W.}~\bibnamefont
  {Xiong}} \emph {et~al.},\ }\bibfield  {title} {\bibinfo {title} {{A small
  proton charge radius from an electron\textendash{}proton scattering
  experiment}},\ }\href {https://doi.org/10.1038/s41586-019-1721-2} {\bibfield
  {journal} {\bibinfo  {journal} {Nature}\ }\textbf {\bibinfo {volume} {575}},\
  \bibinfo {pages} {147} (\bibinfo {year} {2019})}\BibitemShut {NoStop}%
\bibitem [{\citenamefont {Djukanovic}\ \emph
  {et~al.}(2024{\natexlab{a}})\citenamefont {Djukanovic}, \citenamefont {von
  Hippel}, \citenamefont {Meyer}, \citenamefont {Ottnad}, \citenamefont
  {Salg},\ and\ \citenamefont {Wittig}}]{Djukanovic:2023beb}%
  \BibitemOpen
  \bibfield  {author} {\bibinfo {author} {\bibfnamefont {D.}~\bibnamefont
  {Djukanovic}}, \bibinfo {author} {\bibfnamefont {G.}~\bibnamefont {von
  Hippel}}, \bibinfo {author} {\bibfnamefont {H.~B.}\ \bibnamefont {Meyer}},
  \bibinfo {author} {\bibfnamefont {K.}~\bibnamefont {Ottnad}}, \bibinfo
  {author} {\bibfnamefont {M.}~\bibnamefont {Salg}},\ and\ \bibinfo {author}
  {\bibfnamefont {H.}~\bibnamefont {Wittig}},\ }\bibfield  {title} {\bibinfo
  {title} {{Electromagnetic form factors of the nucleon from Nf=2+1 lattice
  QCD}},\ }\href {https://doi.org/10.1103/PhysRevD.109.094510} {\bibfield
  {journal} {\bibinfo  {journal} {Phys. Rev. D}\ }\textbf {\bibinfo {volume}
  {109}},\ \bibinfo {pages} {094510} (\bibinfo {year} {2024}{\natexlab{a}})},\
  \Eprint {https://arxiv.org/abs/2309.06590} {arXiv:2309.06590 [hep-lat]}
  \BibitemShut {NoStop}%
\bibitem [{\citenamefont {Djukanovic}\ \emph
  {et~al.}(2024{\natexlab{b}})\citenamefont {Djukanovic}, \citenamefont {von
  Hippel}, \citenamefont {Meyer}, \citenamefont {Ottnad}, \citenamefont
  {Salg},\ and\ \citenamefont {Wittig}}]{Djukanovic:2023jag}%
  \BibitemOpen
  \bibfield  {author} {\bibinfo {author} {\bibfnamefont {D.}~\bibnamefont
  {Djukanovic}}, \bibinfo {author} {\bibfnamefont {G.}~\bibnamefont {von
  Hippel}}, \bibinfo {author} {\bibfnamefont {H.~B.}\ \bibnamefont {Meyer}},
  \bibinfo {author} {\bibfnamefont {K.}~\bibnamefont {Ottnad}}, \bibinfo
  {author} {\bibfnamefont {M.}~\bibnamefont {Salg}},\ and\ \bibinfo {author}
  {\bibfnamefont {H.}~\bibnamefont {Wittig}},\ }\bibfield  {title} {\bibinfo
  {title} {{Precision Calculation of the Electromagnetic Radii of the Proton
  and Neutron from Lattice QCD}},\ }\href
  {https://doi.org/10.1103/PhysRevLett.132.211901} {\bibfield  {journal}
  {\bibinfo  {journal} {Phys. Rev. Lett.}\ }\textbf {\bibinfo {volume} {132}},\
  \bibinfo {pages} {211901} (\bibinfo {year} {2024}{\natexlab{b}})},\ \Eprint
  {https://arxiv.org/abs/2309.07491} {arXiv:2309.07491 [hep-lat]} \BibitemShut
  {NoStop}%
\bibitem [{\citenamefont {Alexandrou}\ \emph {et~al.}(2019)\citenamefont
  {Alexandrou}, \citenamefont {Bacchio}, \citenamefont {Constantinou},
  \citenamefont {Finkenrath}, \citenamefont {Hadjiyiannakou}, \citenamefont
  {Jansen}, \citenamefont {Koutsou},\ and\ \citenamefont {Vaquero
  Aviles-Casco}}]{Alexandrou:2018sjm}%
  \BibitemOpen
  \bibfield  {author} {\bibinfo {author} {\bibfnamefont {C.}~\bibnamefont
  {Alexandrou}}, \bibinfo {author} {\bibfnamefont {S.}~\bibnamefont {Bacchio}},
  \bibinfo {author} {\bibfnamefont {M.}~\bibnamefont {Constantinou}}, \bibinfo
  {author} {\bibfnamefont {J.}~\bibnamefont {Finkenrath}}, \bibinfo {author}
  {\bibfnamefont {K.}~\bibnamefont {Hadjiyiannakou}}, \bibinfo {author}
  {\bibfnamefont {K.}~\bibnamefont {Jansen}}, \bibinfo {author} {\bibfnamefont
  {G.}~\bibnamefont {Koutsou}},\ and\ \bibinfo {author} {\bibfnamefont
  {A.}~\bibnamefont {Vaquero Aviles-Casco}},\ }\bibfield  {title} {\bibinfo
  {title} {{Proton and neutron electromagnetic form factors from lattice
  QCD}},\ }\href {https://doi.org/10.1103/PhysRevD.100.014509} {\bibfield
  {journal} {\bibinfo  {journal} {Phys. Rev. D}\ }\textbf {\bibinfo {volume}
  {100}},\ \bibinfo {pages} {014509} (\bibinfo {year} {2019})},\ \Eprint
  {https://arxiv.org/abs/1812.10311} {arXiv:1812.10311 [hep-lat]} \BibitemShut
  {NoStop}%
\bibitem [{\citenamefont {Tsuji}\ \emph {et~al.}(2024)\citenamefont {Tsuji},
  \citenamefont {Aoki}, \citenamefont {Ishikawa}, \citenamefont {Kuramashi},
  \citenamefont {Sasaki}, \citenamefont {Sato}, \citenamefont {Shintani},
  \citenamefont {Watanabe},\ and\ \citenamefont {Yamazaki}}]{Tsuji:2023llh}%
  \BibitemOpen
  \bibfield  {author} {\bibinfo {author} {\bibfnamefont {R.}~\bibnamefont
  {Tsuji}}, \bibinfo {author} {\bibfnamefont {Y.}~\bibnamefont {Aoki}},
  \bibinfo {author} {\bibfnamefont {K.-I.}\ \bibnamefont {Ishikawa}}, \bibinfo
  {author} {\bibfnamefont {Y.}~\bibnamefont {Kuramashi}}, \bibinfo {author}
  {\bibfnamefont {S.}~\bibnamefont {Sasaki}}, \bibinfo {author} {\bibfnamefont
  {K.}~\bibnamefont {Sato}}, \bibinfo {author} {\bibfnamefont {E.}~\bibnamefont
  {Shintani}}, \bibinfo {author} {\bibfnamefont {H.}~\bibnamefont {Watanabe}},\
  and\ \bibinfo {author} {\bibfnamefont {T.}~\bibnamefont {Yamazaki}} (\bibinfo
  {collaboration} {PACS}),\ }\bibfield  {title} {\bibinfo {title} {{Nucleon
  form factors in Nf=2+1 lattice QCD at the physical point: Finite lattice
  spacing effect on the root-mean-square radii}},\ }\href
  {https://doi.org/10.1103/PhysRevD.109.094505} {\bibfield  {journal} {\bibinfo
   {journal} {Phys. Rev. D}\ }\textbf {\bibinfo {volume} {109}},\ \bibinfo
  {pages} {094505} (\bibinfo {year} {2024})},\ \Eprint
  {https://arxiv.org/abs/2311.10345} {arXiv:2311.10345 [hep-lat]} \BibitemShut
  {NoStop}%
\bibitem [{\citenamefont {Jang}\ \emph {et~al.}(2020)\citenamefont {Jang},
  \citenamefont {Gupta}, \citenamefont {Lin}, \citenamefont {Yoon},\ and\
  \citenamefont {Bhattacharya}}]{Jang:2019jkn}%
  \BibitemOpen
  \bibfield  {author} {\bibinfo {author} {\bibfnamefont {Y.-C.}\ \bibnamefont
  {Jang}}, \bibinfo {author} {\bibfnamefont {R.}~\bibnamefont {Gupta}},
  \bibinfo {author} {\bibfnamefont {H.-W.}\ \bibnamefont {Lin}}, \bibinfo
  {author} {\bibfnamefont {B.}~\bibnamefont {Yoon}},\ and\ \bibinfo {author}
  {\bibfnamefont {T.}~\bibnamefont {Bhattacharya}},\ }\bibfield  {title}
  {\bibinfo {title} {{Nucleon electromagnetic form factors in the continuum
  limit from ( 2+1+1 )-flavor lattice QCD}},\ }\href
  {https://doi.org/10.1103/PhysRevD.101.014507} {\bibfield  {journal} {\bibinfo
   {journal} {Phys. Rev. D}\ }\textbf {\bibinfo {volume} {101}},\ \bibinfo
  {pages} {014507} (\bibinfo {year} {2020})},\ \Eprint
  {https://arxiv.org/abs/1906.07217} {arXiv:1906.07217 [hep-lat]} \BibitemShut
  {NoStop}%
\bibitem [{\citenamefont {Frezzotti}\ and\ \citenamefont
  {Rossi}(2004)}]{Frezzotti:2003ni}%
  \BibitemOpen
  \bibfield  {author} {\bibinfo {author} {\bibfnamefont {R.}~\bibnamefont
  {Frezzotti}}\ and\ \bibinfo {author} {\bibfnamefont {G.~C.}\ \bibnamefont
  {Rossi}},\ }\bibfield  {title} {\bibinfo {title} {{Chirally improving Wilson
  fermions. 1. O(a) improvement}},\ }\href
  {https://doi.org/10.1088/1126-6708/2004/08/007} {\bibfield  {journal}
  {\bibinfo  {journal} {JHEP}\ }\textbf {\bibinfo {volume} {08}},\ \bibinfo
  {pages} {007}},\ \Eprint {https://arxiv.org/abs/hep-lat/0306014}
  {arXiv:hep-lat/0306014} \BibitemShut {NoStop}%
\bibitem [{\citenamefont {Frezzotti}\ \emph {et~al.}(2001)\citenamefont
  {Frezzotti}, \citenamefont {Grassi}, \citenamefont {Sint},\ and\
  \citenamefont {Weisz}}]{Frezzotti:2000nk}%
  \BibitemOpen
  \bibfield  {author} {\bibinfo {author} {\bibfnamefont {R.}~\bibnamefont
  {Frezzotti}}, \bibinfo {author} {\bibfnamefont {P.~A.}\ \bibnamefont
  {Grassi}}, \bibinfo {author} {\bibfnamefont {S.}~\bibnamefont {Sint}},\ and\
  \bibinfo {author} {\bibfnamefont {P.}~\bibnamefont {Weisz}} (\bibinfo
  {collaboration} {Alpha}),\ }\bibfield  {title} {\bibinfo {title} {{Lattice
  QCD with a chirally twisted mass term}},\ }\href
  {https://doi.org/10.1088/1126-6708/2001/08/058} {\bibfield  {journal}
  {\bibinfo  {journal} {JHEP}\ }\textbf {\bibinfo {volume} {08}},\ \bibinfo
  {pages} {058}},\ \Eprint {https://arxiv.org/abs/hep-lat/0101001}
  {arXiv:hep-lat/0101001} \BibitemShut {NoStop}%
\bibitem [{\citenamefont {Constantinou}\ \emph {et~al.}(2010)\citenamefont
  {Constantinou} \emph {et~al.}}]{ETM:2010iwh}%
  \BibitemOpen
  \bibfield  {author} {\bibinfo {author} {\bibfnamefont {M.}~\bibnamefont
  {Constantinou}} \emph {et~al.} (\bibinfo {collaboration} {ETM}),\ }\bibfield
  {title} {\bibinfo {title} {{Non-perturbative renormalization of quark
  bilinear operators with $N_f$ = 2 (tmQCD) Wilson fermions and the tree-level
  improved gauge action}},\ }\href {https://doi.org/10.1007/JHEP08(2010)068}
  {\bibfield  {journal} {\bibinfo  {journal} {JHEP}\ }\textbf {\bibinfo
  {volume} {08}},\ \bibinfo {pages} {068}},\ \Eprint
  {https://arxiv.org/abs/1004.1115} {arXiv:1004.1115 [hep-lat]} \BibitemShut
  {NoStop}%
\bibitem [{\citenamefont {Alexandrou}\ \emph {et~al.}(2018)\citenamefont
  {Alexandrou} \emph {et~al.}}]{Alexandrou:2018egz}%
  \BibitemOpen
  \bibfield  {author} {\bibinfo {author} {\bibfnamefont {C.}~\bibnamefont
  {Alexandrou}} \emph {et~al.},\ }\bibfield  {title} {\bibinfo {title}
  {{Simulating twisted mass fermions at physical light, strange and charm quark
  masses}},\ }\href {https://doi.org/10.1103/PhysRevD.98.054518} {\bibfield
  {journal} {\bibinfo  {journal} {Phys. Rev. D}\ }\textbf {\bibinfo {volume}
  {98}},\ \bibinfo {pages} {054518} (\bibinfo {year} {2018})},\ \Eprint
  {https://arxiv.org/abs/1807.00495} {arXiv:1807.00495 [hep-lat]} \BibitemShut
  {NoStop}%
\bibitem [{\citenamefont {Finkenrath}\ \emph {et~al.}(2022)\citenamefont
  {Finkenrath} \emph {et~al.}}]{Finkenrath:2022eon}%
  \BibitemOpen
  \bibfield  {author} {\bibinfo {author} {\bibfnamefont {J.}~\bibnamefont
  {Finkenrath}} \emph {et~al.},\ }\bibfield  {title} {\bibinfo {title}
  {{Twisted mass gauge ensembles at physical values of the light, strange and
  charm quark masses}},\ }\href {https://doi.org/10.22323/1.396.0284}
  {\bibfield  {journal} {\bibinfo  {journal} {PoS}\ }\textbf {\bibinfo {volume}
  {LATTICE2021}},\ \bibinfo {pages} {284} (\bibinfo {year} {2022})},\ \Eprint
  {https://arxiv.org/abs/2201.02551} {arXiv:2201.02551 [hep-lat]} \BibitemShut
  {NoStop}%
\bibitem [{\citenamefont {Alexandrou}\ \emph
  {et~al.}(2023{\natexlab{a}})\citenamefont {Alexandrou} \emph
  {et~al.}}]{ExtendedTwistedMass:2022jpw}%
  \BibitemOpen
  \bibfield  {author} {\bibinfo {author} {\bibfnamefont {C.}~\bibnamefont
  {Alexandrou}} \emph {et~al.} (\bibinfo {collaboration} {Extended Twisted
  Mass}),\ }\bibfield  {title} {\bibinfo {title} {{Lattice calculation of the
  short and intermediate time-distance hadronic vacuum polarization
  contributions to the muon magnetic moment using twisted-mass fermions}},\
  }\href {https://doi.org/10.1103/PhysRevD.107.074506} {\bibfield  {journal}
  {\bibinfo  {journal} {Phys. Rev. D}\ }\textbf {\bibinfo {volume} {107}},\
  \bibinfo {pages} {074506} (\bibinfo {year} {2023}{\natexlab{a}})},\ \Eprint
  {https://arxiv.org/abs/2206.15084} {arXiv:2206.15084 [hep-lat]} \BibitemShut
  {NoStop}%
\bibitem [{\citenamefont {Alexandrou}\ \emph {et~al.}(2021)\citenamefont
  {Alexandrou} \emph {et~al.}}]{ExtendedTwistedMass:2021gbo}%
  \BibitemOpen
  \bibfield  {author} {\bibinfo {author} {\bibfnamefont {C.}~\bibnamefont
  {Alexandrou}} \emph {et~al.} (\bibinfo {collaboration} {Extended Twisted
  Mass}),\ }\bibfield  {title} {\bibinfo {title} {{Quark masses using
  twisted-mass fermion gauge ensembles}},\ }\href
  {https://doi.org/10.1103/PhysRevD.104.074515} {\bibfield  {journal} {\bibinfo
   {journal} {Phys. Rev. D}\ }\textbf {\bibinfo {volume} {104}},\ \bibinfo
  {pages} {074515} (\bibinfo {year} {2021})},\ \Eprint
  {https://arxiv.org/abs/2104.13408} {arXiv:2104.13408 [hep-lat]} \BibitemShut
  {NoStop}%
\bibitem [{\citenamefont {Gusken}(1990)}]{Gusken:1989qx}%
  \BibitemOpen
  \bibfield  {author} {\bibinfo {author} {\bibfnamefont {S.}~\bibnamefont
  {Gusken}},\ }\bibfield  {title} {\bibinfo {title} {{A Study of smearing
  techniques for hadron correlation functions}},\ }\href
  {https://doi.org/10.1016/0920-5632(90)90273-W} {\bibfield  {journal}
  {\bibinfo  {journal} {Nucl. Phys. B Proc. Suppl.}\ }\textbf {\bibinfo
  {volume} {17}},\ \bibinfo {pages} {361} (\bibinfo {year} {1990})}\BibitemShut
  {NoStop}%
\bibitem [{\citenamefont {Alexandrou}\ \emph {et~al.}(1994)\citenamefont
  {Alexandrou}, \citenamefont {Gusken}, \citenamefont {Jegerlehner},
  \citenamefont {Schilling},\ and\ \citenamefont {Sommer}}]{Alexandrou:1992ti}%
  \BibitemOpen
  \bibfield  {author} {\bibinfo {author} {\bibfnamefont {C.}~\bibnamefont
  {Alexandrou}}, \bibinfo {author} {\bibfnamefont {S.}~\bibnamefont {Gusken}},
  \bibinfo {author} {\bibfnamefont {F.}~\bibnamefont {Jegerlehner}}, \bibinfo
  {author} {\bibfnamefont {K.}~\bibnamefont {Schilling}},\ and\ \bibinfo
  {author} {\bibfnamefont {R.}~\bibnamefont {Sommer}},\ }\bibfield  {title}
  {\bibinfo {title} {{The Static approximation of heavy - light quark systems:
  A Systematic lattice study}},\ }\href
  {https://doi.org/10.1016/0550-3213(94)90262-3} {\bibfield  {journal}
  {\bibinfo  {journal} {Nucl. Phys. B}\ }\textbf {\bibinfo {volume} {414}},\
  \bibinfo {pages} {815} (\bibinfo {year} {1994})},\ \Eprint
  {https://arxiv.org/abs/hep-lat/9211042} {arXiv:hep-lat/9211042} \BibitemShut
  {NoStop}%
\bibitem [{\citenamefont {Alexandrou}\ \emph {et~al.}(2020)\citenamefont
  {Alexandrou} \emph {et~al.}}]{Alexandrou:2019ali}%
  \BibitemOpen
  \bibfield  {author} {\bibinfo {author} {\bibfnamefont {C.}~\bibnamefont
  {Alexandrou}} \emph {et~al.},\ }\bibfield  {title} {\bibinfo {title}
  {{Moments of nucleon generalized parton distributions from lattice QCD
  simulations at physical pion mass}},\ }\href
  {https://doi.org/10.1103/PhysRevD.101.034519} {\bibfield  {journal} {\bibinfo
   {journal} {Phys. Rev. D}\ }\textbf {\bibinfo {volume} {101}},\ \bibinfo
  {pages} {034519} (\bibinfo {year} {2020})},\ \Eprint
  {https://arxiv.org/abs/1908.10706} {arXiv:1908.10706 [hep-lat]} \BibitemShut
  {NoStop}%
\bibitem [{\citenamefont {Albanese}\ \emph {et~al.}(1987)\citenamefont
  {Albanese} \emph {et~al.}}]{APE:1987ehd}%
  \BibitemOpen
  \bibfield  {author} {\bibinfo {author} {\bibfnamefont {M.}~\bibnamefont
  {Albanese}} \emph {et~al.} (\bibinfo {collaboration} {APE}),\ }\bibfield
  {title} {\bibinfo {title} {{Glueball Masses and String Tension in Lattice
  QCD}},\ }\href {https://doi.org/10.1016/0370-2693(87)91160-9} {\bibfield
  {journal} {\bibinfo  {journal} {Phys. Lett. B}\ }\textbf {\bibinfo {volume}
  {192}},\ \bibinfo {pages} {163} (\bibinfo {year} {1987})}\BibitemShut
  {NoStop}%
\bibitem [{\citenamefont {Alexandrou}\ \emph
  {et~al.}(2023{\natexlab{b}})\citenamefont {Alexandrou} \emph
  {et~al.}}]{Alexandrou:2022dtc}%
  \BibitemOpen
  \bibfield  {author} {\bibinfo {author} {\bibfnamefont {C.}~\bibnamefont
  {Alexandrou}} \emph {et~al.},\ }\bibfield  {title} {\bibinfo {title}
  {{Moments of the nucleon transverse quark spin densities using lattice
  QCD}},\ }\href {https://doi.org/10.1103/PhysRevD.107.054504} {\bibfield
  {journal} {\bibinfo  {journal} {Phys. Rev. D}\ }\textbf {\bibinfo {volume}
  {107}},\ \bibinfo {pages} {054504} (\bibinfo {year} {2023}{\natexlab{b}})},\
  \Eprint {https://arxiv.org/abs/2202.09871} {arXiv:2202.09871 [hep-lat]}
  \BibitemShut {NoStop}%
\bibitem [{\citenamefont {McNeile}\ and\ \citenamefont
  {Michael}(2006)}]{McNeile:2006bz}%
  \BibitemOpen
  \bibfield  {author} {\bibinfo {author} {\bibfnamefont {C.}~\bibnamefont
  {McNeile}}\ and\ \bibinfo {author} {\bibfnamefont {C.}~\bibnamefont
  {Michael}} (\bibinfo {collaboration} {UKQCD}),\ }\bibfield  {title} {\bibinfo
  {title} {{Decay width of light quark hybrid meson from the lattice}},\ }\href
  {https://doi.org/10.1103/PhysRevD.73.074506} {\bibfield  {journal} {\bibinfo
  {journal} {Phys. Rev. D}\ }\textbf {\bibinfo {volume} {73}},\ \bibinfo
  {pages} {074506} (\bibinfo {year} {2006})},\ \Eprint
  {https://arxiv.org/abs/hep-lat/0603007} {arXiv:hep-lat/0603007} \BibitemShut
  {NoStop}%
\bibitem [{\citenamefont {Alexandrou}\ \emph {et~al.}(2014)\citenamefont
  {Alexandrou}, \citenamefont {Constantinou}, \citenamefont {Drach},
  \citenamefont {Hadjiyiannakou}, \citenamefont {Jansen}, \citenamefont
  {Koutsou}, \citenamefont {Strelchenko},\ and\ \citenamefont
  {Vaquero}}]{Alexandrou:2013wca}%
  \BibitemOpen
  \bibfield  {author} {\bibinfo {author} {\bibfnamefont {C.}~\bibnamefont
  {Alexandrou}}, \bibinfo {author} {\bibfnamefont {M.}~\bibnamefont
  {Constantinou}}, \bibinfo {author} {\bibfnamefont {V.}~\bibnamefont {Drach}},
  \bibinfo {author} {\bibfnamefont {K.}~\bibnamefont {Hadjiyiannakou}},
  \bibinfo {author} {\bibfnamefont {K.}~\bibnamefont {Jansen}}, \bibinfo
  {author} {\bibfnamefont {G.}~\bibnamefont {Koutsou}}, \bibinfo {author}
  {\bibfnamefont {A.}~\bibnamefont {Strelchenko}},\ and\ \bibinfo {author}
  {\bibfnamefont {A.}~\bibnamefont {Vaquero}},\ }\bibfield  {title} {\bibinfo
  {title} {{Evaluation of disconnected quark loops for hadron structure using
  GPUs}},\ }\href {https://doi.org/10.1016/j.cpc.2014.01.009} {\bibfield
  {journal} {\bibinfo  {journal} {Comput. Phys. Commun.}\ }\textbf {\bibinfo
  {volume} {185}},\ \bibinfo {pages} {1370} (\bibinfo {year} {2014})},\ \Eprint
  {https://arxiv.org/abs/1309.2256} {arXiv:1309.2256 [hep-lat]} \BibitemShut
  {NoStop}%
\bibitem [{\citenamefont {Alexandrou}\ \emph
  {et~al.}(2017{\natexlab{a}})\citenamefont {Alexandrou}, \citenamefont
  {Constantinou}, \citenamefont {Hadjiyiannakou}, \citenamefont {Jansen},
  \citenamefont {Kallidonis}, \citenamefont {Koutsou},\ and\ \citenamefont
  {Vaquero Aviles-Casco}}]{Alexandrou:2017hac}%
  \BibitemOpen
  \bibfield  {author} {\bibinfo {author} {\bibfnamefont {C.}~\bibnamefont
  {Alexandrou}}, \bibinfo {author} {\bibfnamefont {M.}~\bibnamefont
  {Constantinou}}, \bibinfo {author} {\bibfnamefont {K.}~\bibnamefont
  {Hadjiyiannakou}}, \bibinfo {author} {\bibfnamefont {K.}~\bibnamefont
  {Jansen}}, \bibinfo {author} {\bibfnamefont {C.}~\bibnamefont {Kallidonis}},
  \bibinfo {author} {\bibfnamefont {G.}~\bibnamefont {Koutsou}},\ and\ \bibinfo
  {author} {\bibfnamefont {A.}~\bibnamefont {Vaquero Aviles-Casco}},\
  }\bibfield  {title} {\bibinfo {title} {{Nucleon axial form factors using
  $N_f$ = 2 twisted mass fermions with a physical value of the pion mass}},\
  }\href {https://doi.org/10.1103/PhysRevD.96.054507} {\bibfield  {journal}
  {\bibinfo  {journal} {Phys. Rev. D}\ }\textbf {\bibinfo {volume} {96}},\
  \bibinfo {pages} {054507} (\bibinfo {year} {2017}{\natexlab{a}})},\ \Eprint
  {https://arxiv.org/abs/1705.03399} {arXiv:1705.03399 [hep-lat]} \BibitemShut
  {NoStop}%
\bibitem [{\citenamefont {Alexandrou}\ \emph
  {et~al.}(2017{\natexlab{b}})\citenamefont {Alexandrou} \emph
  {et~al.}}]{Alexandrou:2017qyt}%
  \BibitemOpen
  \bibfield  {author} {\bibinfo {author} {\bibfnamefont {C.}~\bibnamefont
  {Alexandrou}} \emph {et~al.},\ }\bibfield  {title} {\bibinfo {title}
  {{Nucleon scalar and tensor charges using lattice QCD simulations at the
  physical value of the pion mass}},\ }\href
  {https://doi.org/10.1103/PhysRevD.95.114514} {\bibfield  {journal} {\bibinfo
  {journal} {Phys. Rev. D}\ }\textbf {\bibinfo {volume} {95}},\ \bibinfo
  {pages} {114514} (\bibinfo {year} {2017}{\natexlab{b}})},\ \bibinfo {note}
  {[Erratum: Phys.Rev.D 96, 099906 (2017)]},\ \Eprint
  {https://arxiv.org/abs/1703.08788} {arXiv:1703.08788 [hep-lat]} \BibitemShut
  {NoStop}%
\bibitem [{\citenamefont {Alexandrou}\ \emph
  {et~al.}(2017{\natexlab{c}})\citenamefont {Alexandrou}, \citenamefont
  {Constantinou}, \citenamefont {Hadjiyiannakou}, \citenamefont {Jansen},
  \citenamefont {Kallidonis}, \citenamefont {Koutsou}, \citenamefont {Vaquero
  Avil\'es-Casco},\ and\ \citenamefont {Wiese}}]{Alexandrou:2017oeh}%
  \BibitemOpen
  \bibfield  {author} {\bibinfo {author} {\bibfnamefont {C.}~\bibnamefont
  {Alexandrou}}, \bibinfo {author} {\bibfnamefont {M.}~\bibnamefont
  {Constantinou}}, \bibinfo {author} {\bibfnamefont {K.}~\bibnamefont
  {Hadjiyiannakou}}, \bibinfo {author} {\bibfnamefont {K.}~\bibnamefont
  {Jansen}}, \bibinfo {author} {\bibfnamefont {C.}~\bibnamefont {Kallidonis}},
  \bibinfo {author} {\bibfnamefont {G.}~\bibnamefont {Koutsou}}, \bibinfo
  {author} {\bibfnamefont {A.}~\bibnamefont {Vaquero Avil\'es-Casco}},\ and\
  \bibinfo {author} {\bibfnamefont {C.}~\bibnamefont {Wiese}},\ }\bibfield
  {title} {\bibinfo {title} {{Nucleon Spin and Momentum Decomposition Using
  Lattice QCD Simulations}},\ }\href
  {https://doi.org/10.1103/PhysRevLett.119.142002} {\bibfield  {journal}
  {\bibinfo  {journal} {Phys. Rev. Lett.}\ }\textbf {\bibinfo {volume} {119}},\
  \bibinfo {pages} {142002} (\bibinfo {year} {2017}{\natexlab{c}})},\ \Eprint
  {https://arxiv.org/abs/1706.02973} {arXiv:1706.02973 [hep-lat]} \BibitemShut
  {NoStop}%
\bibitem [{\citenamefont {Stathopoulos}\ \emph {et~al.}(2013)\citenamefont
  {Stathopoulos}, \citenamefont {Laeuchli},\ and\ \citenamefont
  {Orginos}}]{Stathopoulos:2013aci}%
  \BibitemOpen
  \bibfield  {author} {\bibinfo {author} {\bibfnamefont {A.}~\bibnamefont
  {Stathopoulos}}, \bibinfo {author} {\bibfnamefont {J.}~\bibnamefont
  {Laeuchli}},\ and\ \bibinfo {author} {\bibfnamefont {K.}~\bibnamefont
  {Orginos}},\ }\bibfield  {title} {\bibinfo {title} {{Hierarchical Probing for
  Estimating the Trace of the Matrix Inverse on Toroidal Lattices}},\ }\href
  {https://doi.org/10.1137/120881452} {\bibfield  {journal} {\bibinfo
  {journal} {SIAM J. Sci. Comput.}\ }\textbf {\bibinfo {volume} {35}},\
  \bibinfo {pages} {S299} (\bibinfo {year} {2013})},\ \Eprint
  {https://arxiv.org/abs/1302.4018} {arXiv:1302.4018 [hep-lat]} \BibitemShut
  {NoStop}%
\bibitem [{\citenamefont {Gambhir}\ \emph {et~al.}(2017)\citenamefont
  {Gambhir}, \citenamefont {Stathopoulos},\ and\ \citenamefont
  {Orginos}}]{Gambhir:2016uwp}%
  \BibitemOpen
  \bibfield  {author} {\bibinfo {author} {\bibfnamefont {A.~S.}\ \bibnamefont
  {Gambhir}}, \bibinfo {author} {\bibfnamefont {A.}~\bibnamefont
  {Stathopoulos}},\ and\ \bibinfo {author} {\bibfnamefont {K.}~\bibnamefont
  {Orginos}},\ }\bibfield  {title} {\bibinfo {title} {{Deflation as a Method of
  Variance Reduction for Estimating the Trace of a Matrix Inverse}},\ }\href
  {https://doi.org/10.1137/16M1066361} {\bibfield  {journal} {\bibinfo
  {journal} {SIAM J. Sci. Comput.}\ }\textbf {\bibinfo {volume} {39}},\
  \bibinfo {pages} {A532} (\bibinfo {year} {2017})},\ \Eprint
  {https://arxiv.org/abs/1603.05988} {arXiv:1603.05988 [hep-lat]} \BibitemShut
  {NoStop}%
\bibitem [{\citenamefont {Martinelli}\ \emph {et~al.}(1995)\citenamefont
  {Martinelli}, \citenamefont {Pittori}, \citenamefont {Sachrajda},
  \citenamefont {Testa},\ and\ \citenamefont {Vladikas}}]{Martinelli:1994ty}%
  \BibitemOpen
  \bibfield  {author} {\bibinfo {author} {\bibfnamefont {G.}~\bibnamefont
  {Martinelli}}, \bibinfo {author} {\bibfnamefont {C.}~\bibnamefont {Pittori}},
  \bibinfo {author} {\bibfnamefont {C.~T.}\ \bibnamefont {Sachrajda}}, \bibinfo
  {author} {\bibfnamefont {M.}~\bibnamefont {Testa}},\ and\ \bibinfo {author}
  {\bibfnamefont {A.}~\bibnamefont {Vladikas}},\ }\bibfield  {title} {\bibinfo
  {title} {{A General method for nonperturbative renormalization of lattice
  operators}},\ }\href {https://doi.org/10.1016/0550-3213(95)00126-D}
  {\bibfield  {journal} {\bibinfo  {journal} {Nucl. Phys. B}\ }\textbf
  {\bibinfo {volume} {445}},\ \bibinfo {pages} {81} (\bibinfo {year} {1995})},\
  \Eprint {https://arxiv.org/abs/hep-lat/9411010} {arXiv:hep-lat/9411010}
  \BibitemShut {NoStop}%
\bibitem [{\citenamefont {{Extended Twisted Mass
  Collaboration}}()}]{ETMC-ren:2025}%
  \BibitemOpen
  \bibfield  {author} {\bibinfo {author} {\bibnamefont {{Extended Twisted Mass
  Collaboration}}},\ }\bibfield  {title} {\bibinfo {title} {{Non-perturbative
  renormalisation of quark bilinear operators with $N_f=2+1+1$ Wilson-clover
  twisted mass fermions}},\ }\href@noop {} {\bibinfo  {journal} {In
  preparation}\ }\BibitemShut {NoStop}%
\bibitem [{\citenamefont {Gockeler}\ \emph {et~al.}(1999)\citenamefont
  {Gockeler}, \citenamefont {Horsley}, \citenamefont {Oelrich}, \citenamefont
  {Perlt}, \citenamefont {Petters}, \citenamefont {Rakow}, \citenamefont
  {Schafer}, \citenamefont {Schierholz},\ and\ \citenamefont
  {Schiller}}]{Gockeler:1998ye}%
  \BibitemOpen
\bibfield  {journal} {  }\bibfield  {author} {\bibinfo {author} {\bibfnamefont
  {M.}~\bibnamefont {Gockeler}}, \bibinfo {author} {\bibfnamefont
  {R.}~\bibnamefont {Horsley}}, \bibinfo {author} {\bibfnamefont
  {H.}~\bibnamefont {Oelrich}}, \bibinfo {author} {\bibfnamefont
  {H.}~\bibnamefont {Perlt}}, \bibinfo {author} {\bibfnamefont
  {D.}~\bibnamefont {Petters}}, \bibinfo {author} {\bibfnamefont {P.~E.~L.}\
  \bibnamefont {Rakow}}, \bibinfo {author} {\bibfnamefont {A.}~\bibnamefont
  {Schafer}}, \bibinfo {author} {\bibfnamefont {G.}~\bibnamefont
  {Schierholz}},\ and\ \bibinfo {author} {\bibfnamefont {A.}~\bibnamefont
  {Schiller}},\ }\bibfield  {title} {\bibinfo {title} {{Nonperturbative
  renormalization of composite operators in lattice QCD}},\ }\href
  {https://doi.org/10.1016/S0550-3213(99)00036-X} {\bibfield  {journal}
  {\bibinfo  {journal} {Nucl. Phys. B}\ }\textbf {\bibinfo {volume} {544}},\
  \bibinfo {pages} {699} (\bibinfo {year} {1999})},\ \Eprint
  {https://arxiv.org/abs/hep-lat/9807044} {arXiv:hep-lat/9807044} \BibitemShut
  {NoStop}%
\bibitem [{\citenamefont {Alexandrou}\ \emph
  {et~al.}(2025{\natexlab{a}})\citenamefont {Alexandrou}, \citenamefont
  {Bacchio}, \citenamefont {Finkenrath}, \citenamefont {Iona}, \citenamefont
  {Koutsou}, \citenamefont {Li},\ and\ \citenamefont
  {Spanoudes}}]{Alexandrou:2024ozj}%
  \BibitemOpen
  \bibfield  {author} {\bibinfo {author} {\bibfnamefont {C.}~\bibnamefont
  {Alexandrou}}, \bibinfo {author} {\bibfnamefont {S.}~\bibnamefont {Bacchio}},
  \bibinfo {author} {\bibfnamefont {J.}~\bibnamefont {Finkenrath}}, \bibinfo
  {author} {\bibfnamefont {C.}~\bibnamefont {Iona}}, \bibinfo {author}
  {\bibfnamefont {G.}~\bibnamefont {Koutsou}}, \bibinfo {author} {\bibfnamefont
  {Y.}~\bibnamefont {Li}},\ and\ \bibinfo {author} {\bibfnamefont
  {G.}~\bibnamefont {Spanoudes}},\ }\bibfield  {title} {\bibinfo {title}
  {{Nucleon charges and \ensuremath{\sigma}-terms in lattice QCD}},\ }\href
  {https://doi.org/10.1103/PhysRevD.111.054505} {\bibfield  {journal} {\bibinfo
   {journal} {Phys. Rev. D}\ }\textbf {\bibinfo {volume} {111}},\ \bibinfo
  {pages} {054505} (\bibinfo {year} {2025}{\natexlab{a}})},\ \Eprint
  {https://arxiv.org/abs/2412.01535} {arXiv:2412.01535 [hep-lat]} \BibitemShut
  {NoStop}%
\bibitem [{\citenamefont {Alexandrou}\ \emph
  {et~al.}(2017{\natexlab{d}})\citenamefont {Alexandrou}, \citenamefont
  {Constantinou},\ and\ \citenamefont {Panagopoulos}}]{Alexandrou:2015sea}%
  \BibitemOpen
  \bibfield  {author} {\bibinfo {author} {\bibfnamefont {C.}~\bibnamefont
  {Alexandrou}}, \bibinfo {author} {\bibfnamefont {M.}~\bibnamefont
  {Constantinou}},\ and\ \bibinfo {author} {\bibfnamefont {H.}~\bibnamefont
  {Panagopoulos}} (\bibinfo {collaboration} {ETM}),\ }\bibfield  {title}
  {\bibinfo {title} {{Renormalization functions for Nf=2 and Nf=4 twisted mass
  fermions}},\ }\href {https://doi.org/10.1103/PhysRevD.95.034505} {\bibfield
  {journal} {\bibinfo  {journal} {Phys. Rev. D}\ }\textbf {\bibinfo {volume}
  {95}},\ \bibinfo {pages} {034505} (\bibinfo {year} {2017}{\natexlab{d}})},\
  \Eprint {https://arxiv.org/abs/1509.00213} {arXiv:1509.00213 [hep-lat]}
  \BibitemShut {NoStop}%
\bibitem [{\citenamefont {Bar}\ and\ \citenamefont
  {Colic}(2021)}]{Bar:2021crj}%
  \BibitemOpen
  \bibfield  {author} {\bibinfo {author} {\bibfnamefont {O.}~\bibnamefont
  {Bar}}\ and\ \bibinfo {author} {\bibfnamefont {H.}~\bibnamefont {Colic}},\
  }\bibfield  {title} {\bibinfo {title} {{N\ensuremath{\pi}-state contamination
  in lattice calculations of the nucleon electromagnetic form factors}},\
  }\href {https://doi.org/10.1103/PhysRevD.103.114514} {\bibfield  {journal}
  {\bibinfo  {journal} {Phys. Rev. D}\ }\textbf {\bibinfo {volume} {103}},\
  \bibinfo {pages} {114514} (\bibinfo {year} {2021})},\ \Eprint
  {https://arxiv.org/abs/2104.00329} {arXiv:2104.00329 [hep-lat]} \BibitemShut
  {NoStop}%
\bibitem [{\citenamefont {Jay}\ and\ \citenamefont {Neil}(2021)}]{Jay:2020jkz}%
  \BibitemOpen
  \bibfield  {author} {\bibinfo {author} {\bibfnamefont {W.~I.}\ \bibnamefont
  {Jay}}\ and\ \bibinfo {author} {\bibfnamefont {E.~T.}\ \bibnamefont {Neil}},\
  }\bibfield  {title} {\bibinfo {title} {{Bayesian model averaging for analysis
  of lattice field theory results}},\ }\href
  {https://doi.org/10.1103/PhysRevD.103.114502} {\bibfield  {journal} {\bibinfo
   {journal} {Phys. Rev. D}\ }\textbf {\bibinfo {volume} {103}},\ \bibinfo
  {pages} {114502} (\bibinfo {year} {2021})},\ \Eprint
  {https://arxiv.org/abs/2008.01069} {arXiv:2008.01069 [stat.ME]} \BibitemShut
  {NoStop}%
\bibitem [{\citenamefont {Neil}\ and\ \citenamefont
  {Sitison}(2024)}]{Neil:2022joj}%
  \BibitemOpen
  \bibfield  {author} {\bibinfo {author} {\bibfnamefont {E.~T.}\ \bibnamefont
  {Neil}}\ and\ \bibinfo {author} {\bibfnamefont {J.~W.}\ \bibnamefont
  {Sitison}},\ }\bibfield  {title} {\bibinfo {title} {{Improved information
  criteria for Bayesian model averaging in lattice field theory}},\ }\href
  {https://doi.org/10.1103/PhysRevD.109.014510} {\bibfield  {journal} {\bibinfo
   {journal} {Phys. Rev. D}\ }\textbf {\bibinfo {volume} {109}},\ \bibinfo
  {pages} {014510} (\bibinfo {year} {2024})},\ \Eprint
  {https://arxiv.org/abs/2208.14983} {arXiv:2208.14983 [stat.ME]} \BibitemShut
  {NoStop}%
\bibitem [{\citenamefont {Alexandrou}\ \emph {et~al.}(2024)\citenamefont
  {Alexandrou}, \citenamefont {Bacchio}, \citenamefont {Constantinou},
  \citenamefont {Finkenrath}, \citenamefont {Frezzotti}, \citenamefont
  {Kostrzewa}, \citenamefont {Koutsou}, \citenamefont {Spanoudes},\ and\
  \citenamefont {Urbach}}]{Alexandrou:2023qbg}%
  \BibitemOpen
  \bibfield  {author} {\bibinfo {author} {\bibfnamefont {C.}~\bibnamefont
  {Alexandrou}}, \bibinfo {author} {\bibfnamefont {S.}~\bibnamefont {Bacchio}},
  \bibinfo {author} {\bibfnamefont {M.}~\bibnamefont {Constantinou}}, \bibinfo
  {author} {\bibfnamefont {J.}~\bibnamefont {Finkenrath}}, \bibinfo {author}
  {\bibfnamefont {R.}~\bibnamefont {Frezzotti}}, \bibinfo {author}
  {\bibfnamefont {B.}~\bibnamefont {Kostrzewa}}, \bibinfo {author}
  {\bibfnamefont {G.}~\bibnamefont {Koutsou}}, \bibinfo {author} {\bibfnamefont
  {G.}~\bibnamefont {Spanoudes}},\ and\ \bibinfo {author} {\bibfnamefont
  {C.}~\bibnamefont {Urbach}} (\bibinfo {collaboration} {Extended Twisted
  Mass}),\ }\bibfield  {title} {\bibinfo {title} {{Nucleon axial and
  pseudoscalar form factors using twisted-mass fermion ensembles at the
  physical point}},\ }\href {https://doi.org/10.1103/PhysRevD.109.034503}
  {\bibfield  {journal} {\bibinfo  {journal} {Phys. Rev. D}\ }\textbf {\bibinfo
  {volume} {109}},\ \bibinfo {pages} {034503} (\bibinfo {year} {2024})},\
  \Eprint {https://arxiv.org/abs/2309.05774} {arXiv:2309.05774 [hep-lat]}
  \BibitemShut {NoStop}%
\bibitem [{\citenamefont {Galster}\ \emph {et~al.}(1971)\citenamefont
  {Galster}, \citenamefont {Klein}, \citenamefont {Moritz}, \citenamefont
  {Schmidt}, \citenamefont {Wegener},\ and\ \citenamefont
  {Bleckwenn}}]{Galster:1971kv}%
  \BibitemOpen
  \bibfield  {author} {\bibinfo {author} {\bibfnamefont {S.}~\bibnamefont
  {Galster}}, \bibinfo {author} {\bibfnamefont {H.}~\bibnamefont {Klein}},
  \bibinfo {author} {\bibfnamefont {J.}~\bibnamefont {Moritz}}, \bibinfo
  {author} {\bibfnamefont {K.~H.}\ \bibnamefont {Schmidt}}, \bibinfo {author}
  {\bibfnamefont {D.}~\bibnamefont {Wegener}},\ and\ \bibinfo {author}
  {\bibfnamefont {J.}~\bibnamefont {Bleckwenn}},\ }\bibfield  {title} {\bibinfo
  {title} {{Elastic electron-deuteron scattering and the electric neutron form
  factor at four-momentum transfers 5fm$^{-2} < q^2 < 14$fm$^{-2}$}},\ }\href
  {https://doi.org/10.1016/0550-3213(71)90068-X} {\bibfield  {journal}
  {\bibinfo  {journal} {Nucl. Phys.}\ }\textbf {\bibinfo {volume} {B32}},\
  \bibinfo {pages} {221} (\bibinfo {year} {1971})}\BibitemShut {NoStop}%
\bibitem [{\citenamefont {Lee}\ \emph {et~al.}(2015)\citenamefont {Lee},
  \citenamefont {Arrington},\ and\ \citenamefont {Hill}}]{Lee_2015}%
  \BibitemOpen
  \bibfield  {author} {\bibinfo {author} {\bibfnamefont {G.}~\bibnamefont
  {Lee}}, \bibinfo {author} {\bibfnamefont {J.~R.}\ \bibnamefont {Arrington}},\
  and\ \bibinfo {author} {\bibfnamefont {R.~J.}\ \bibnamefont {Hill}},\
  }\bibfield  {title} {\bibinfo {title} {Extraction of the proton radius from
  electron-proton scattering data},\ }\bibfield  {journal} {\bibinfo  {journal}
  {Physical Review D}\ }\textbf {\bibinfo {volume} {92}},\ \href
  {https://doi.org/10.1103/physrevd.92.013013} {10.1103/physrevd.92.013013}
  (\bibinfo {year} {2015})\BibitemShut {NoStop}%
\bibitem [{\citenamefont {Meyer}\ \emph {et~al.}(2016)\citenamefont {Meyer},
  \citenamefont {Betancourt}, \citenamefont {Gran},\ and\ \citenamefont
  {Hill}}]{Meyer:2016oeg}%
  \BibitemOpen
  \bibfield  {author} {\bibinfo {author} {\bibfnamefont {A.~S.}\ \bibnamefont
  {Meyer}}, \bibinfo {author} {\bibfnamefont {M.}~\bibnamefont {Betancourt}},
  \bibinfo {author} {\bibfnamefont {R.}~\bibnamefont {Gran}},\ and\ \bibinfo
  {author} {\bibfnamefont {R.~J.}\ \bibnamefont {Hill}},\ }\bibfield  {title}
  {\bibinfo {title} {{Deuterium target data for precision neutrino-nucleus
  cross sections}},\ }\href {https://doi.org/10.1103/PhysRevD.93.113015}
  {\bibfield  {journal} {\bibinfo  {journal} {Phys. Rev. D}\ }\textbf {\bibinfo
  {volume} {93}},\ \bibinfo {pages} {113015} (\bibinfo {year} {2016})},\
  \Eprint {https://arxiv.org/abs/1603.03048} {arXiv:1603.03048 [hep-ph]}
  \BibitemShut {NoStop}%
\bibitem [{\citenamefont {Hill}\ and\ \citenamefont {Paz}(2010)}]{Hill:2010yb}%
  \BibitemOpen
  \bibfield  {author} {\bibinfo {author} {\bibfnamefont {R.~J.}\ \bibnamefont
  {Hill}}\ and\ \bibinfo {author} {\bibfnamefont {G.}~\bibnamefont {Paz}},\
  }\bibfield  {title} {\bibinfo {title} {{Model independent extraction of the
  proton charge radius from electron scattering}},\ }\href
  {https://doi.org/10.1103/PhysRevD.82.113005} {\bibfield  {journal} {\bibinfo
  {journal} {Phys. Rev. D}\ }\textbf {\bibinfo {volume} {82}},\ \bibinfo
  {pages} {113005} (\bibinfo {year} {2010})},\ \Eprint
  {https://arxiv.org/abs/1008.4619} {arXiv:1008.4619 [hep-ph]} \BibitemShut
  {NoStop}%
\bibitem [{\citenamefont {Navas}\ \emph {et~al.}(2024)\citenamefont {Navas}
  \emph {et~al.}}]{ParticleDataGroup:2024cfk}%
  \BibitemOpen
  \bibfield  {author} {\bibinfo {author} {\bibfnamefont {S.}~\bibnamefont
  {Navas}} \emph {et~al.} (\bibinfo {collaboration} {Particle Data Group}),\
  }\bibfield  {title} {\bibinfo {title} {{Review of particle physics}},\ }\href
  {https://doi.org/10.1103/PhysRevD.110.030001} {\bibfield  {journal} {\bibinfo
   {journal} {Phys. Rev. D}\ }\textbf {\bibinfo {volume} {110}},\ \bibinfo
  {pages} {030001} (\bibinfo {year} {2024})}\BibitemShut {NoStop}%
\bibitem [{\citenamefont {Ye}\ \emph {et~al.}(2018)\citenamefont {Ye},
  \citenamefont {Arrington}, \citenamefont {Hill},\ and\ \citenamefont
  {Lee}}]{Ye:2017gyb}%
  \BibitemOpen
  \bibfield  {author} {\bibinfo {author} {\bibfnamefont {Z.}~\bibnamefont
  {Ye}}, \bibinfo {author} {\bibfnamefont {J.}~\bibnamefont {Arrington}},
  \bibinfo {author} {\bibfnamefont {R.~J.}\ \bibnamefont {Hill}},\ and\
  \bibinfo {author} {\bibfnamefont {G.}~\bibnamefont {Lee}},\ }\bibfield
  {title} {\bibinfo {title} {{Proton and Neutron Electromagnetic Form Factors
  and Uncertainties}},\ }\href {https://doi.org/10.1016/j.physletb.2017.11.023}
  {\bibfield  {journal} {\bibinfo  {journal} {Phys. Lett. B}\ }\textbf
  {\bibinfo {volume} {777}},\ \bibinfo {pages} {8} (\bibinfo {year} {2018})},\
  \Eprint {https://arxiv.org/abs/1707.09063} {arXiv:1707.09063 [nucl-ex]}
  \BibitemShut {NoStop}%
\bibitem [{\citenamefont {Becker}\ \emph {et~al.}(1999)\citenamefont {Becker}
  \emph {et~al.}}]{Becker:1999tw}%
  \BibitemOpen
  \bibfield  {author} {\bibinfo {author} {\bibfnamefont {J.}~\bibnamefont
  {Becker}} \emph {et~al.},\ }\bibfield  {title} {\bibinfo {title}
  {{Determination of the neutron electric form-factor from the reaction
  He-3(e,e' n) at medium momentum transfer}},\ }\href
  {https://doi.org/10.1007/s100500050351} {\bibfield  {journal} {\bibinfo
  {journal} {Eur. Phys. J. A}\ }\textbf {\bibinfo {volume} {6}},\ \bibinfo
  {pages} {329} (\bibinfo {year} {1999})}\BibitemShut {NoStop}%
\bibitem [{\citenamefont {Eden}\ \emph {et~al.}(1994)\citenamefont {Eden} \emph
  {et~al.}}]{Eden:1994ji}%
  \BibitemOpen
  \bibfield  {author} {\bibinfo {author} {\bibfnamefont {T.}~\bibnamefont
  {Eden}} \emph {et~al.},\ }\bibfield  {title} {\bibinfo {title} {{Electric
  form factor of the neutron from the
  $^{2}H(\overrightarrow{e},e’\overrightarrow{n})^{1}H$ reaction at $Q^{2} =$
  0.255 (GeV/c)$^2$}},\ }\href {https://doi.org/10.1103/PhysRevC.50.R1749}
  {\bibfield  {journal} {\bibinfo  {journal} {Phys. Rev. C}\ }\textbf {\bibinfo
  {volume} {50}},\ \bibinfo {pages} {R1749} (\bibinfo {year}
  {1994})}\BibitemShut {NoStop}%
\bibitem [{\citenamefont {Meyerhoff}\ \emph {et~al.}(1994)\citenamefont
  {Meyerhoff} \emph {et~al.}}]{Meyerhoff:1994ev}%
  \BibitemOpen
  \bibfield  {author} {\bibinfo {author} {\bibfnamefont {M.}~\bibnamefont
  {Meyerhoff}} \emph {et~al.},\ }\bibfield  {title} {\bibinfo {title} {{First
  measurement of the electric form-factor of the neutron in the exclusive
  quasielastic scattering of polarized electrons from polarized He-3}},\ }\href
  {https://doi.org/10.1016/0370-2693(94)90718-8} {\bibfield  {journal}
  {\bibinfo  {journal} {Phys. Lett. B}\ }\textbf {\bibinfo {volume} {327}},\
  \bibinfo {pages} {201} (\bibinfo {year} {1994})}\BibitemShut {NoStop}%
\bibitem [{\citenamefont {Passchier}\ \emph {et~al.}(1999)\citenamefont
  {Passchier} \emph {et~al.}}]{Passchier:1999cj}%
  \BibitemOpen
  \bibfield  {author} {\bibinfo {author} {\bibfnamefont {I.}~\bibnamefont
  {Passchier}} \emph {et~al.},\ }\bibfield  {title} {\bibinfo {title} {{The
  Charge form-factor of the neutron from the reaction polarized H-2(polarized
  e, e-prime n) p}},\ }\href {https://doi.org/10.1103/PhysRevLett.82.4988}
  {\bibfield  {journal} {\bibinfo  {journal} {Phys. Rev. Lett.}\ }\textbf
  {\bibinfo {volume} {82}},\ \bibinfo {pages} {4988} (\bibinfo {year}
  {1999})},\ \Eprint {https://arxiv.org/abs/nucl-ex/9907012}
  {arXiv:nucl-ex/9907012} \BibitemShut {NoStop}%
\bibitem [{\citenamefont {Warren}\ \emph {et~al.}(2004)\citenamefont {Warren}
  \emph {et~al.}}]{JeffersonLabE93-026:2003tty}%
  \BibitemOpen
  \bibfield  {author} {\bibinfo {author} {\bibfnamefont {G.}~\bibnamefont
  {Warren}} \emph {et~al.} (\bibinfo {collaboration} {Jefferson Lab E93-026}),\
  }\bibfield  {title} {\bibinfo {title} {{Measurement of the electric
  form-factor of the neutron at $Q^2$ = 0.5 and 1.0 $GeV^2/c^2$}},\ }\href
  {https://doi.org/10.1103/PhysRevLett.92.042301} {\bibfield  {journal}
  {\bibinfo  {journal} {Phys. Rev. Lett.}\ }\textbf {\bibinfo {volume} {92}},\
  \bibinfo {pages} {042301} (\bibinfo {year} {2004})},\ \Eprint
  {https://arxiv.org/abs/nucl-ex/0308021} {arXiv:nucl-ex/0308021} \BibitemShut
  {NoStop}%
\bibitem [{\citenamefont {Zhu}\ \emph {et~al.}(2001)\citenamefont {Zhu} \emph
  {et~al.}}]{E93026:2001css}%
  \BibitemOpen
  \bibfield  {author} {\bibinfo {author} {\bibfnamefont {H.}~\bibnamefont
  {Zhu}} \emph {et~al.} (\bibinfo {collaboration} {E93026}),\ }\bibfield
  {title} {\bibinfo {title} {{A Measurement of the electric form-factor of the
  neutron through polarized-d (polarized-e, e-prime n)p at Q**2 =
  0.5-(GeV/c)**2}},\ }\href {https://doi.org/10.1103/PhysRevLett.87.081801}
  {\bibfield  {journal} {\bibinfo  {journal} {Phys. Rev. Lett.}\ }\textbf
  {\bibinfo {volume} {87}},\ \bibinfo {pages} {081801} (\bibinfo {year}
  {2001})},\ \Eprint {https://arxiv.org/abs/nucl-ex/0105001}
  {arXiv:nucl-ex/0105001} \BibitemShut {NoStop}%
\bibitem [{\citenamefont {Madey}\ \emph {et~al.}(2003)\citenamefont {Madey}
  \emph {et~al.}}]{E93-038:2003ixb}%
  \BibitemOpen
  \bibfield  {author} {\bibinfo {author} {\bibfnamefont {R.}~\bibnamefont
  {Madey}} \emph {et~al.} (\bibinfo {collaboration} {E93-038}),\ }\bibfield
  {title} {\bibinfo {title} {{Measurements of G(E)n / G(M)n from the
  H-2(polarized-e,e-prime polarized-n) reaction to Q**2 = 1.45 (GeV/c)**2}},\
  }\href {https://doi.org/10.1103/PhysRevLett.91.122002} {\bibfield  {journal}
  {\bibinfo  {journal} {Phys. Rev. Lett.}\ }\textbf {\bibinfo {volume} {91}},\
  \bibinfo {pages} {122002} (\bibinfo {year} {2003})},\ \Eprint
  {https://arxiv.org/abs/nucl-ex/0308007} {arXiv:nucl-ex/0308007} \BibitemShut
  {NoStop}%
\bibitem [{\citenamefont {Rohe}\ \emph {et~al.}(1999)\citenamefont {Rohe} \emph
  {et~al.}}]{Rohe:1999sh}%
  \BibitemOpen
  \bibfield  {author} {\bibinfo {author} {\bibfnamefont {D.}~\bibnamefont
  {Rohe}} \emph {et~al.},\ }\bibfield  {title} {\bibinfo {title} {{Measurement
  of the neutron electric form-factor G(en) at 0.67-(GeV/c)**2 via
  He-3(pol.)(e(pol.),e' n)}},\ }\href
  {https://doi.org/10.1103/PhysRevLett.83.4257} {\bibfield  {journal} {\bibinfo
   {journal} {Phys. Rev. Lett.}\ }\textbf {\bibinfo {volume} {83}},\ \bibinfo
  {pages} {4257} (\bibinfo {year} {1999})}\BibitemShut {NoStop}%
\bibitem [{\citenamefont {Bermuth}\ \emph {et~al.}(2003)\citenamefont {Bermuth}
  \emph {et~al.}}]{Bermuth:2003qh}%
  \BibitemOpen
  \bibfield  {author} {\bibinfo {author} {\bibfnamefont {J.}~\bibnamefont
  {Bermuth}} \emph {et~al.},\ }\bibfield  {title} {\bibinfo {title} {{The
  Neutron charge form-factor and target analyzing powers from polarized-He-3
  (polarized-e,e-prime n) scattering}},\ }\href
  {https://doi.org/10.1016/S0370-2693(03)00725-1} {\bibfield  {journal}
  {\bibinfo  {journal} {Phys. Lett. B}\ }\textbf {\bibinfo {volume} {564}},\
  \bibinfo {pages} {199} (\bibinfo {year} {2003})},\ \Eprint
  {https://arxiv.org/abs/nucl-ex/0303015} {arXiv:nucl-ex/0303015} \BibitemShut
  {NoStop}%
\bibitem [{\citenamefont {Glazier}\ \emph {et~al.}(2005)\citenamefont {Glazier}
  \emph {et~al.}}]{Glazier:2004ny}%
  \BibitemOpen
  \bibfield  {author} {\bibinfo {author} {\bibfnamefont {D.~I.}\ \bibnamefont
  {Glazier}} \emph {et~al.},\ }\bibfield  {title} {\bibinfo {title}
  {{Measurement of the electric form-factor of the neutron at Q**2 =
  0.3-(GeV/c)**2 to 0.8-(GeV/c)**2}},\ }\href
  {https://doi.org/10.1140/epja/i2004-10115-8} {\bibfield  {journal} {\bibinfo
  {journal} {Eur. Phys. J. A}\ }\textbf {\bibinfo {volume} {24}},\ \bibinfo
  {pages} {101} (\bibinfo {year} {2005})},\ \Eprint
  {https://arxiv.org/abs/nucl-ex/0410026} {arXiv:nucl-ex/0410026} \BibitemShut
  {NoStop}%
\bibitem [{\citenamefont {Herberg}\ \emph {et~al.}(1999)\citenamefont {Herberg}
  \emph {et~al.}}]{Herberg:1999ud}%
  \BibitemOpen
  \bibfield  {author} {\bibinfo {author} {\bibfnamefont {C.}~\bibnamefont
  {Herberg}} \emph {et~al.},\ }\bibfield  {title} {\bibinfo {title}
  {{Determination of the neutron electric form-factor in the D(e,e' n)p
  reaction and the influence of nuclear binding}},\ }\href
  {https://doi.org/10.1007/s100500050268} {\bibfield  {journal} {\bibinfo
  {journal} {Eur. Phys. J. A}\ }\textbf {\bibinfo {volume} {5}},\ \bibinfo
  {pages} {131} (\bibinfo {year} {1999})}\BibitemShut {NoStop}%
\bibitem [{\citenamefont {Schiavilla}\ and\ \citenamefont
  {Sick}(2001)}]{Schiavilla:2001qe}%
  \BibitemOpen
  \bibfield  {author} {\bibinfo {author} {\bibfnamefont {R.}~\bibnamefont
  {Schiavilla}}\ and\ \bibinfo {author} {\bibfnamefont {I.}~\bibnamefont
  {Sick}},\ }\bibfield  {title} {\bibinfo {title} {{Neutron charge form-factor
  at large q**2}},\ }\href {https://doi.org/10.1103/PhysRevC.64.041002}
  {\bibfield  {journal} {\bibinfo  {journal} {Phys. Rev. C}\ }\textbf {\bibinfo
  {volume} {64}},\ \bibinfo {pages} {041002} (\bibinfo {year} {2001})},\
  \Eprint {https://arxiv.org/abs/nucl-ex/0107004} {arXiv:nucl-ex/0107004}
  \BibitemShut {NoStop}%
\bibitem [{\citenamefont {Ostrick}\ \emph {et~al.}(1999)\citenamefont {Ostrick}
  \emph {et~al.}}]{Ostrick:1999xa}%
  \BibitemOpen
  \bibfield  {author} {\bibinfo {author} {\bibfnamefont {M.}~\bibnamefont
  {Ostrick}} \emph {et~al.},\ }\bibfield  {title} {\bibinfo {title}
  {{Measurement of the neutron electric form-factor G(E,n) in the quasifree
  H-2(e(pol.),e' n(pol.))p reaction}},\ }\href
  {https://doi.org/10.1103/PhysRevLett.83.276} {\bibfield  {journal} {\bibinfo
  {journal} {Phys. Rev. Lett.}\ }\textbf {\bibinfo {volume} {83}},\ \bibinfo
  {pages} {276} (\bibinfo {year} {1999})}\BibitemShut {NoStop}%
\bibitem [{\citenamefont {Anderson}\ \emph {et~al.}(2007)\citenamefont
  {Anderson} \emph {et~al.}}]{JeffersonLabE95-001:2006dax}%
  \BibitemOpen
  \bibfield  {author} {\bibinfo {author} {\bibfnamefont {B.}~\bibnamefont
  {Anderson}} \emph {et~al.} (\bibinfo {collaboration} {Jefferson Lab
  E95-001}),\ }\bibfield  {title} {\bibinfo {title} {{Extraction of the Neutron
  Magnetic Form Factor from Quasi-elastic $^{3}\vec{He}(\vec{e},e')$ at Q$^2$ =
  0.1 - 0.6 (GeV/c)$^2$}},\ }\href {https://doi.org/10.1103/PhysRevC.75.034003}
  {\bibfield  {journal} {\bibinfo  {journal} {Phys. Rev. C}\ }\textbf {\bibinfo
  {volume} {75}},\ \bibinfo {pages} {034003} (\bibinfo {year} {2007})},\
  \Eprint {https://arxiv.org/abs/nucl-ex/0605006} {arXiv:nucl-ex/0605006}
  \BibitemShut {NoStop}%
\bibitem [{\citenamefont {Gao}\ \emph {et~al.}(1994)\citenamefont {Gao} \emph
  {et~al.}}]{Gao:1994ud}%
  \BibitemOpen
  \bibfield  {author} {\bibinfo {author} {\bibfnamefont {H.}~\bibnamefont
  {Gao}} \emph {et~al.},\ }\bibfield  {title} {\bibinfo {title} {{Measurement
  of the neutron magnetic form-factor from inclusive quasielastic scattering of
  polarized electrons from polarized He-3}},\ }\href
  {https://doi.org/10.1103/PhysRevC.50.R546} {\bibfield  {journal} {\bibinfo
  {journal} {Phys. Rev. C}\ }\textbf {\bibinfo {volume} {50}},\ \bibinfo
  {pages} {R546} (\bibinfo {year} {1994})}\BibitemShut {NoStop}%
\bibitem [{\citenamefont {Anklin}\ \emph {et~al.}(1994)\citenamefont {Anklin}
  \emph {et~al.}}]{Anklin:1994ae}%
  \BibitemOpen
  \bibfield  {author} {\bibinfo {author} {\bibfnamefont {H.}~\bibnamefont
  {Anklin}} \emph {et~al.},\ }\bibfield  {title} {\bibinfo {title} {{Precision
  measurement of the neutron magnetic form-factor}},\ }\href
  {https://doi.org/10.1016/0370-2693(94)90538-X} {\bibfield  {journal}
  {\bibinfo  {journal} {Phys. Lett. B}\ }\textbf {\bibinfo {volume} {336}},\
  \bibinfo {pages} {313} (\bibinfo {year} {1994})}\BibitemShut {NoStop}%
\bibitem [{\citenamefont {Anklin}\ \emph {et~al.}(1998)\citenamefont {Anklin}
  \emph {et~al.}}]{Anklin:1998ae}%
  \BibitemOpen
  \bibfield  {author} {\bibinfo {author} {\bibfnamefont {H.}~\bibnamefont
  {Anklin}} \emph {et~al.},\ }\bibfield  {title} {\bibinfo {title} {{Precise
  measurements of the neutron magnetic form-factor}},\ }\href
  {https://doi.org/10.1016/S0370-2693(98)00442-0} {\bibfield  {journal}
  {\bibinfo  {journal} {Phys. Lett. B}\ }\textbf {\bibinfo {volume} {428}},\
  \bibinfo {pages} {248} (\bibinfo {year} {1998})}\BibitemShut {NoStop}%
\bibitem [{\citenamefont {Kubon}\ \emph {et~al.}(2002)\citenamefont {Kubon}
  \emph {et~al.}}]{Kubon:2001rj}%
  \BibitemOpen
  \bibfield  {author} {\bibinfo {author} {\bibfnamefont {G.}~\bibnamefont
  {Kubon}} \emph {et~al.},\ }\bibfield  {title} {\bibinfo {title} {{Precise
  neutron magnetic form-factors}},\ }\href
  {https://doi.org/10.1016/S0370-2693(01)01386-7} {\bibfield  {journal}
  {\bibinfo  {journal} {Phys. Lett. B}\ }\textbf {\bibinfo {volume} {524}},\
  \bibinfo {pages} {26} (\bibinfo {year} {2002})},\ \Eprint
  {https://arxiv.org/abs/nucl-ex/0107016} {arXiv:nucl-ex/0107016} \BibitemShut
  {NoStop}%
\bibitem [{\citenamefont {Alarcon}(2007)}]{Alarcon:2007zza}%
  \BibitemOpen
  \bibfield  {author} {\bibinfo {author} {\bibfnamefont {R.}~\bibnamefont
  {Alarcon}} (\bibinfo {collaboration} {BLAST}),\ }\bibfield  {title} {\bibinfo
  {title} {{Nucleon form factors and the BLAST experiment}},\ }\href
  {https://doi.org/10.1140/epja/i2006-10395-x} {\bibfield  {journal} {\bibinfo
  {journal} {Eur. Phys. J. A}\ }\textbf {\bibinfo {volume} {32}},\ \bibinfo
  {pages} {477} (\bibinfo {year} {2007})}\BibitemShut {NoStop}%
\bibitem [{\citenamefont {Alexandrou}\ \emph
  {et~al.}(2017{\natexlab{e}})\citenamefont {Alexandrou}, \citenamefont
  {Constantinou}, \citenamefont {Hadjiyiannakou}, \citenamefont {Jansen},
  \citenamefont {Kallidonis}, \citenamefont {Koutsou},\ and\ \citenamefont
  {Vaquero Aviles-Casco}}]{Alexandrou:2017ypw}%
  \BibitemOpen
  \bibfield  {author} {\bibinfo {author} {\bibfnamefont {C.}~\bibnamefont
  {Alexandrou}}, \bibinfo {author} {\bibfnamefont {M.}~\bibnamefont
  {Constantinou}}, \bibinfo {author} {\bibfnamefont {K.}~\bibnamefont
  {Hadjiyiannakou}}, \bibinfo {author} {\bibfnamefont {K.}~\bibnamefont
  {Jansen}}, \bibinfo {author} {\bibfnamefont {C.}~\bibnamefont {Kallidonis}},
  \bibinfo {author} {\bibfnamefont {G.}~\bibnamefont {Koutsou}},\ and\ \bibinfo
  {author} {\bibfnamefont {A.}~\bibnamefont {Vaquero Aviles-Casco}},\
  }\bibfield  {title} {\bibinfo {title} {{Nucleon electromagnetic form factors
  using lattice simulations at the physical point}},\ }\href
  {https://doi.org/10.1103/PhysRevD.96.034503} {\bibfield  {journal} {\bibinfo
  {journal} {Phys. Rev. D}\ }\textbf {\bibinfo {volume} {96}},\ \bibinfo
  {pages} {034503} (\bibinfo {year} {2017}{\natexlab{e}})},\ \Eprint
  {https://arxiv.org/abs/1706.00469} {arXiv:1706.00469 [hep-lat]} \BibitemShut
  {NoStop}%
\bibitem [{\citenamefont {Shintani}\ \emph {et~al.}(2019)\citenamefont
  {Shintani}, \citenamefont {Ishikawa}, \citenamefont {Kuramashi},
  \citenamefont {Sasaki},\ and\ \citenamefont {Yamazaki}}]{Shintani:2018ozy}%
  \BibitemOpen
  \bibfield  {author} {\bibinfo {author} {\bibfnamefont {E.}~\bibnamefont
  {Shintani}}, \bibinfo {author} {\bibfnamefont {K.-I.}\ \bibnamefont
  {Ishikawa}}, \bibinfo {author} {\bibfnamefont {Y.}~\bibnamefont {Kuramashi}},
  \bibinfo {author} {\bibfnamefont {S.}~\bibnamefont {Sasaki}},\ and\ \bibinfo
  {author} {\bibfnamefont {T.}~\bibnamefont {Yamazaki}},\ }\bibfield  {title}
  {\bibinfo {title} {{Nucleon form factors and root-mean-square radii on a
  (10.8 fm)$^4$ lattice at the physical point}},\ }\href
  {https://doi.org/10.1103/PhysRevD.99.014510} {\bibfield  {journal} {\bibinfo
  {journal} {Phys. Rev. D}\ }\textbf {\bibinfo {volume} {99}},\ \bibinfo
  {pages} {014510} (\bibinfo {year} {2019})},\ \bibinfo {note} {[Erratum:
  Phys.Rev.D 102, 019902 (2020)]},\ \Eprint {https://arxiv.org/abs/1811.07292}
  {arXiv:1811.07292 [hep-lat]} \BibitemShut {NoStop}%
\bibitem [{\citenamefont {Antognini}\ \emph {et~al.}(2013)\citenamefont
  {Antognini} \emph {et~al.}}]{Antognini:2013txn}%
  \BibitemOpen
  \bibfield  {author} {\bibinfo {author} {\bibfnamefont {A.}~\bibnamefont
  {Antognini}} \emph {et~al.},\ }\bibfield  {title} {\bibinfo {title} {{Proton
  Structure from the Measurement of $2S-2P$ Transition Frequencies of Muonic
  Hydrogen}},\ }\href {https://doi.org/10.1126/science.1230016} {\bibfield
  {journal} {\bibinfo  {journal} {Science}\ }\textbf {\bibinfo {volume}
  {339}},\ \bibinfo {pages} {417} (\bibinfo {year} {2013})}\BibitemShut
  {NoStop}%
\bibitem [{\citenamefont {Borah}\ \emph {et~al.}(2020)\citenamefont {Borah},
  \citenamefont {Hill}, \citenamefont {Lee},\ and\ \citenamefont
  {Tomalak}}]{PhysRevD.102.074012}%
  \BibitemOpen
  \bibfield  {author} {\bibinfo {author} {\bibfnamefont {K.}~\bibnamefont
  {Borah}}, \bibinfo {author} {\bibfnamefont {R.~J.}\ \bibnamefont {Hill}},
  \bibinfo {author} {\bibfnamefont {G.}~\bibnamefont {Lee}},\ and\ \bibinfo
  {author} {\bibfnamefont {O.}~\bibnamefont {Tomalak}},\ }\bibfield  {title}
  {\bibinfo {title} {Parametrization and applications of the low-${Q}^{2}$
  nucleon vector form factors},\ }\href
  {https://doi.org/10.1103/PhysRevD.102.074012} {\bibfield  {journal} {\bibinfo
   {journal} {Phys. Rev. D}\ }\textbf {\bibinfo {volume} {102}},\ \bibinfo
  {pages} {074012} (\bibinfo {year} {2020})}\BibitemShut {NoStop}%
\bibitem [{\citenamefont {Distler}\ \emph {et~al.}(2011)\citenamefont
  {Distler}, \citenamefont {Bernauer},\ and\ \citenamefont
  {Walcher}}]{Distler:2010zq}%
  \BibitemOpen
  \bibfield  {author} {\bibinfo {author} {\bibfnamefont {M.~O.}\ \bibnamefont
  {Distler}}, \bibinfo {author} {\bibfnamefont {J.~C.}\ \bibnamefont
  {Bernauer}},\ and\ \bibinfo {author} {\bibfnamefont {T.}~\bibnamefont
  {Walcher}},\ }\bibfield  {title} {\bibinfo {title} {{The RMS Charge Radius of
  the Proton and Zemach Moments}},\ }\href
  {https://doi.org/10.1016/j.physletb.2010.12.067} {\bibfield  {journal}
  {\bibinfo  {journal} {Phys. Lett. B}\ }\textbf {\bibinfo {volume} {696}},\
  \bibinfo {pages} {343} (\bibinfo {year} {2011})},\ \Eprint
  {https://arxiv.org/abs/1011.1861} {arXiv:1011.1861 [nucl-th]} \BibitemShut
  {NoStop}%
\bibitem [{\citenamefont {Lin}\ \emph {et~al.}(2022)\citenamefont {Lin},
  \citenamefont {Hammer},\ and\ \citenamefont
  {Mei\ss{}ner}}]{PhysRevLett.128.052002}%
  \BibitemOpen
  \bibfield  {author} {\bibinfo {author} {\bibfnamefont {Y.-H.}\ \bibnamefont
  {Lin}}, \bibinfo {author} {\bibfnamefont {H.-W.}\ \bibnamefont {Hammer}},\
  and\ \bibinfo {author} {\bibfnamefont {U.-G.}\ \bibnamefont {Mei\ss{}ner}},\
  }\bibfield  {title} {\bibinfo {title} {New insights into the nucleon's
  electromagnetic structure},\ }\href
  {https://doi.org/10.1103/PhysRevLett.128.052002} {\bibfield  {journal}
  {\bibinfo  {journal} {Phys. Rev. Lett.}\ }\textbf {\bibinfo {volume} {128}},\
  \bibinfo {pages} {052002} (\bibinfo {year} {2022})}\BibitemShut {NoStop}%
\bibitem [{\citenamefont {Djukanovic}\ \emph
  {et~al.}(2024{\natexlab{c}})\citenamefont {Djukanovic}, \citenamefont {von
  Hippel}, \citenamefont {Meyer}, \citenamefont {Ottnad}, \citenamefont
  {Salg},\ and\ \citenamefont {Wittig}}]{PhysRevD.110.L011503}%
  \BibitemOpen
  \bibfield  {author} {\bibinfo {author} {\bibfnamefont {D.}~\bibnamefont
  {Djukanovic}}, \bibinfo {author} {\bibfnamefont {G.}~\bibnamefont {von
  Hippel}}, \bibinfo {author} {\bibfnamefont {H.~B.}\ \bibnamefont {Meyer}},
  \bibinfo {author} {\bibfnamefont {K.}~\bibnamefont {Ottnad}}, \bibinfo
  {author} {\bibfnamefont {M.}~\bibnamefont {Salg}},\ and\ \bibinfo {author}
  {\bibfnamefont {H.}~\bibnamefont {Wittig}},\ }\bibfield  {title} {\bibinfo
  {title} {Zemach and friar radii of the proton and neutron from lattice qcd},\
  }\href {https://doi.org/10.1103/PhysRevD.110.L011503} {\bibfield  {journal}
  {\bibinfo  {journal} {Phys. Rev. D}\ }\textbf {\bibinfo {volume} {110}},\
  \bibinfo {pages} {L011503} (\bibinfo {year}
  {2024}{\natexlab{c}})}\BibitemShut {NoStop}%
\bibitem [{\citenamefont {Graczyk}\ and\ \citenamefont
  {Juszczak}(2015)}]{Graczyk:2015kka}%
  \BibitemOpen
  \bibfield  {author} {\bibinfo {author} {\bibfnamefont {K.~M.}\ \bibnamefont
  {Graczyk}}\ and\ \bibinfo {author} {\bibfnamefont {C.}~\bibnamefont
  {Juszczak}},\ }\bibfield  {title} {\bibinfo {title} {{Zemach moments of the
  proton from Bayesian inference}},\ }\href
  {https://doi.org/10.1103/PhysRevC.91.045205} {\bibfield  {journal} {\bibinfo
  {journal} {Phys. Rev. C}\ }\textbf {\bibinfo {volume} {91}},\ \bibinfo
  {pages} {045205} (\bibinfo {year} {2015})}\BibitemShut {NoStop}%
\bibitem [{\citenamefont {Volotka}\ \emph {et~al.}(2005)\citenamefont
  {Volotka}, \citenamefont {Shabaev}, \citenamefont {Plunien},\ and\
  \citenamefont {Soff}}]{Volotka:2004zu}%
  \BibitemOpen
  \bibfield  {author} {\bibinfo {author} {\bibfnamefont {A.~V.}\ \bibnamefont
  {Volotka}}, \bibinfo {author} {\bibfnamefont {V.~M.}\ \bibnamefont
  {Shabaev}}, \bibinfo {author} {\bibfnamefont {G.}~\bibnamefont {Plunien}},\
  and\ \bibinfo {author} {\bibfnamefont {G.}~\bibnamefont {Soff}},\ }\bibfield
  {title} {\bibinfo {title} {{Zemach and magnetic radius of the proton from the
  hyperfine splitting in hydrogen}},\ }\href
  {https://doi.org/10.1140/epjd/e2005-00025-9} {\bibfield  {journal} {\bibinfo
  {journal} {Eur. Phys. J. D}\ }\textbf {\bibinfo {volume} {33}},\ \bibinfo
  {pages} {23} (\bibinfo {year} {2005})},\ \Eprint
  {https://arxiv.org/abs/physics/0405118} {arXiv:physics/0405118} \BibitemShut
  {NoStop}%
\bibitem [{\citenamefont {Mohr}\ \emph {et~al.}(2016)\citenamefont {Mohr},
  \citenamefont {Newell},\ and\ \citenamefont {Taylor}}]{Mohr:2015ccw}%
  \BibitemOpen
  \bibfield  {author} {\bibinfo {author} {\bibfnamefont {P.~J.}\ \bibnamefont
  {Mohr}}, \bibinfo {author} {\bibfnamefont {D.~B.}\ \bibnamefont {Newell}},\
  and\ \bibinfo {author} {\bibfnamefont {B.~N.}\ \bibnamefont {Taylor}},\
  }\bibfield  {title} {\bibinfo {title} {{CODATA Recommended Values of the
  Fundamental Physical Constants: 2014}},\ }\href
  {https://doi.org/10.1103/RevModPhys.88.035009} {\bibfield  {journal}
  {\bibinfo  {journal} {Rev. Mod. Phys.}\ }\textbf {\bibinfo {volume} {88}},\
  \bibinfo {pages} {035009} (\bibinfo {year} {2016})},\ \Eprint
  {https://arxiv.org/abs/1507.07956} {arXiv:1507.07956 [physics.atom-ph]}
  \BibitemShut {NoStop}%
\bibitem [{\citenamefont {Alexandrou}\ \emph
  {et~al.}(2025{\natexlab{b}})\citenamefont {Alexandrou} \emph
  {et~al.}}]{Alexandrou:2025bkm}%
  \BibitemOpen
  \bibfield  {author} {\bibinfo {author} {\bibfnamefont {C.}~\bibnamefont
  {Alexandrou}} \emph {et~al.},\ }\bibfield  {title} {\bibinfo {title}
  {{Large-scale simulations of lattice QCD for nucleon structure using Nf=2+1+1
  flavors of twisted mass fermions}},\ }\href
  {https://doi.org/10.1016/j.procs.2025.08.236} {\bibfield  {journal} {\bibinfo
   {journal} {Procedia Comput. Sci.}\ }\textbf {\bibinfo {volume} {267}},\
  \bibinfo {pages} {92} (\bibinfo {year} {2025}{\natexlab{b}})}\BibitemShut
  {NoStop}%
\bibitem [{\citenamefont {{J\"{u}lich Supercomputing Centre}}(2019)}]{JUWELS}%
  \BibitemOpen
  \bibfield  {author} {\bibinfo {author} {\bibnamefont {{J\"{u}lich
  Supercomputing Centre}}},\ }\bibfield  {title} {\bibinfo {title} {{JUWELS:
  Modular Tier-0/1 Supercomputer at the J\"{u}lich Supercomputing Centre}},\
  }\bibfield  {journal} {\bibinfo  {journal} {Journal of large-scale research
  facilities}\ }\textbf {\bibinfo {volume} {5}},\ \href
  {https://doi.org/10.17815/jlsrf-5-171} {10.17815/jlsrf-5-171} (\bibinfo
  {year} {2019})\BibitemShut {NoStop}%
\bibitem [{\citenamefont {{J\"{u}lich Supercomputing
  Centre}}(2021)}]{JUWELS-BOOSTER}%
  \BibitemOpen
  \bibfield  {author} {\bibinfo {author} {\bibnamefont {{J\"{u}lich
  Supercomputing Centre}}},\ }\bibfield  {title} {\bibinfo {title} {{JUWELS
  Cluster and Booster: Exascale Pathfinder with Modular Supercomputing
  Architecture at Juelich Supercomputing Centre}},\ }\bibfield  {journal}
  {\bibinfo  {journal} {Journal of large-scale research facilities}\ }\textbf
  {\bibinfo {volume} {7}},\ \href {https://doi.org/10.17815/jlsrf-7-183}
  {10.17815/jlsrf-7-183} (\bibinfo {year} {2021})\BibitemShut {NoStop}%
\end{thebibliography}%

\appendix

\section{Excited state analysis of connected and disconnected contributions}
\label{sec:appendix_excited}
We compare the ground state element obtained for the isoscalar case
when i) the connected and disconnected contributions are fitted
separately, as done in the main text and ii) the connected and
disconnected contributions are added first and then fitted. For the
purposes of this comparison, we use the same data for both cases,
i.e. in both i) and ii) we use the disconnected contributions with
$\vec{p}\,'=0$ and only use the sink-source separations available for
the connected case.

\begin{figure}[h]
  \includegraphics[width=1\linewidth]{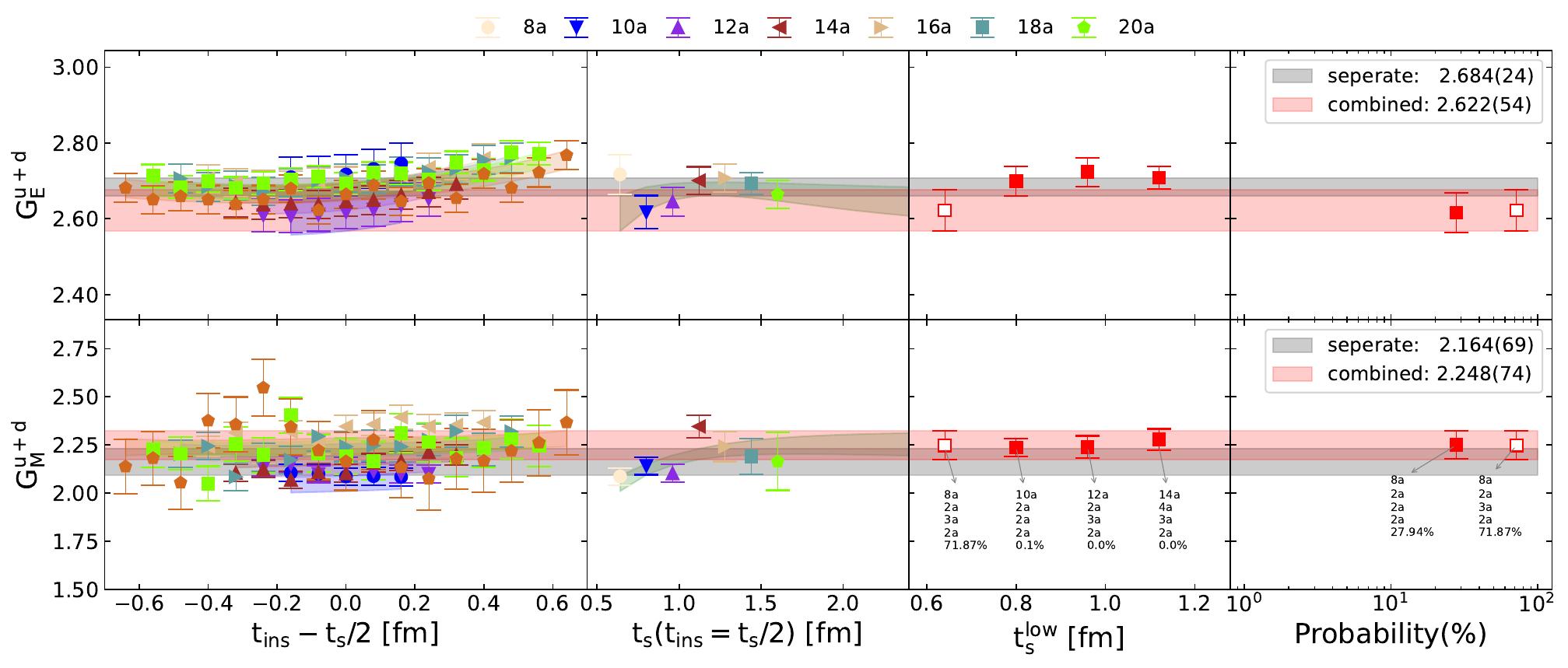}
  \caption{Excited state analysis of the isoscalar matrix elements for
    the B ensemble for the first non-zero momentum transfer. The
    procedure and notation is the same as for the connected case,
    shown in Fig.~\ref{fig:cD96_Q2_2_s}, but after adding the
    connected and disconnected three-point function data. The gray
    band is obtained by adding the connected and disconnected
    contributions to the ground-state matrix element after fitting
    them separately.}\label{fig:appendix_excited}
\end{figure}

The comparison is shown in Fig.~\ref{fig:appendix_excited}. The data
plotted are the sum of connected and disconnected contributions for
each case for the B ensemble. Otherwise, the procedure followed is the
same as that followed for the connected contribution in the main text
(see Fig.~\ref{fig:cD96_Q2_2_s} and related discussion). As can be
seen, the ground state matrix element of the total contribution is
consistent between the two approaches within statistical errors.

\begin{figure}[h]
  \includegraphics[width=1\linewidth]{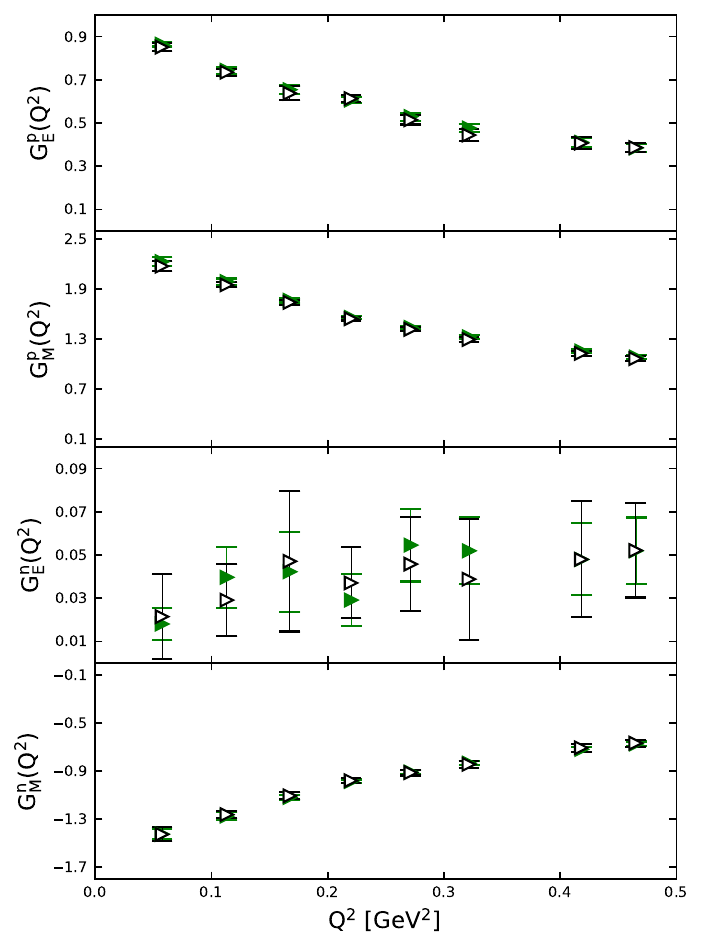}
  \caption{Comparison of the electric and magnetic proton and neutron
    form factors when the ground state analysis is carried out for the
    connected and disconnected contributions separately (solid
    triangles) and when the connected and disconnected contributions are
    added first and then fitted for the ground state (open
    triangles). We show the B ensemble up to
    $Q^2$=0.5~GeV$^2$.}\label{fig:appendix_excited_Q2}
\end{figure}
The comparison is carried out for additional values of $Q^2$, as shown
in Fig.~\ref{fig:appendix_excited_Q2}, where we show that the two
approaches yield consistent results for the B ensemble for all form
factors up to $Q^2$=0.5~GeV$^2$.

\section{Decomposition of lattice matrix elements}
\label{sec:appendix_equations}

We provide the general equations, in Euclidean space, corresponding to
the ratio in Eq.~\ref{eq:full_ratio} when considered for the case of
$p'\neq 0$. For compact representation, we take $G_E \equiv G_E(Q^2)$,
$G_M \equiv G_M(Q^2)$, $\Pi_{\mu,0} \equiv
\Pi^\mu(\Gamma_0;\vec{p}\,', \vec{p}) $, $\Pi_{\mu,k} \equiv
\Pi^\mu(\Gamma_k;\vec{p}\,', \vec{p})$ where ${\vec p}$ (${\vec p}'$)
corresponds to the initial (final) momentum. The four momenta are then
given by $\{E,\vec{p}\}$ ($\{E',\vec{p'}\}$), where $E$ ($E'$) is the
initial (final) energy.

\begin{align}
\Pi_{\mu,0} &=  
-iCG_E\Big[ \left(p'_\mu + p_\mu\right) \left(m\left( E'+ E + m \right) - p'_\rho p_\rho \right) \Big]
\nonumber \\&\quad
+ \frac{CG_M}{2m} \Big[\delta_{\mu 0}( 4m^2 + Q^2 )(m^2 + p'_\rho p_\rho) - i E Q^2 p'_\mu  
\nonumber \\&\quad
+ 2 i m^2 (E' - E)(p'_\mu - p_\mu) - i E' Q^2p_\mu 
\nonumber \\&\quad
 - i m Q^2 (p'_\mu + p_\mu) (2m^2 + Q^2 + 2p'_\rho p_\rho) \Big]
\label{eq:pi_full_0}
\end{align}

\begin{align}
\Pi_{\mu,k} &= CG_E\Big[ \epsilon_{\mu k 0 \rho} (p_\rho' - p_\rho) (m^2 - p_\sigma' p_\sigma)
\nonumber \\&\quad 
- i\epsilon_{\mu k \rho \sigma} p_\rho' p_\sigma (E'  + E)+\epsilon_{\mu 0 \rho \sigma}p_\rho' p_\sigma (p_k' + p_k) \Big]
\nonumber \\&\quad 
+ \frac{ CG_M}{2m} \Big[ m\epsilon_{\mu k 0 \rho} (p_\rho' - p_\rho) (2m^2 + Q^2 + 2p'_\sigma p_\sigma)
\nonumber \\&\quad
+ 2im\epsilon_{\mu k \rho \sigma} p'_\rho p_\sigma(2m+E' + E +\frac{Q^2}{2m})
\nonumber \\&\quad 
-2m\epsilon_{\mu 0 \rho \sigma}p'_\rho p_\sigma (p_k' + p_k) \Big],
\label{eq:eq:pi_full_k}
\end{align}
where $C$ is a kinematic factor given by
\begin{equation}
  C = \frac{m(4m^2+Q^2)^{-1}}{E (E'+m)} \sqrt{\frac{E (E'+m)}{E'(E+m)}}
\end{equation}

\section{{Comparison with previous work}}
\label{sec:appendix_comparison}

\begin{figure}
    \centering
    \includegraphics[width=\linewidth]{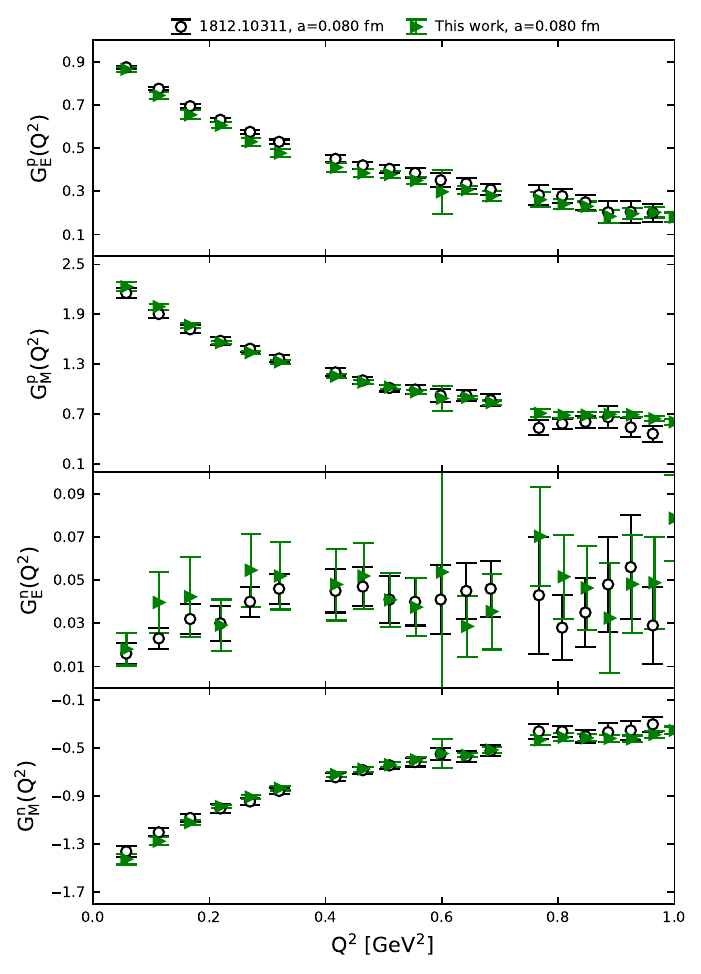}
    \caption{Comparison of proton and neutron electromagnetic form
      factors for the ensemble \texttt{cB211.072.64}, in
      \cite{Alexandrou:2018sjm} (open black circles) and current work
      (green right-pointing triangles).}
    \label{fig:appendix_compare_old}
\end{figure}
Here we compare the results obtained using the current method with our
previous work presented in \cite{Alexandrou:2018sjm}. A major
difference between the analysis procedure of the previous work and the
current work is the treatment of excited states, where previous
results were obtained using plateau fits. Our current work includes
all systematics due to presence of excited states. While most of the
results are compatible within $2\sigma$, the minor deviations at lower
$Q^2$ region, expecially for the magnetic form factors explains the
shift in the magnetic moments obtained for the given ensemble.

\section{Tabulated results}
\label{sec:appendix_results}
We provide the results of the form factors obtained for each
ensemble. In Table~\ref{tab:results_B}, we provide our values for the
electric and magnetic form factors of the proton, neutron, isoscalar
connected, and isovector combinations for the \texttt{cB211.072.64}
ensemble, where the results for the proton and neutron have the
disconnected contributions included. Similarly,
Table~\ref{tab:results_C} and Table~\ref{tab:results_D} tabulate the
results obtained for ensembles \texttt{cC211.060.80} and
\texttt{cD211.054.96}, respectively. The disconnected contributions
are also provided in Table~\ref{tab:disc_results}. We note that the
inclusion of momenta at the sink leads to 224 $Q^2$ values for the
\texttt{cB211.072.64} ensemble and 262 for the \texttt{cC211.060.80}
and \texttt{cD211.054.96} ensembles when restricting to $Q^2$ values
smaller than $\sim$1 GeV$^2$. These results shown in the table have
been averaged into 50 equal bins within the interval $Q^2 = [0,1)
  \;\rm GeV^2$. More precisely, for bin $i$ starting from $Q^2_i$ and
  ending at $Q^2_{i+1}=Q^2_{i}+\delta Q^2$ we average all form factor
  values with $Q^2_i\le Q^2 < Q^2_{i+1}$ weighted by their errors. We
  use $\delta Q^2$=0.02~GeV$^2$, $Q^2_0$=0.0~GeV$^2$ and
  $Q^2_{50}$=1.0~GeV$^2$.

In addition, in Table~\ref{tab:fitp} we provide the fit parameters and
their associated covariance matrix for the most probable fit to the
$Q^2$-dependence of the form factors, i.e., those used to plot the
bands in Fig.~\ref{fig:GEMpn}.
\begin{table*}[h!]
\centering
\caption{Results for the electromagnetic form factors using the
  \texttt{cB211.072.64} ensemble for the proton, neutron, isoscalar
  connected and isovector connected combinations.}
\label{tab:results_B}
\begin{tabular}{ccccccccc}
\hline
\hline
$Q^2$ & $G_E^p$ & $G_E^n$ & $G_E^{u+d} (\rm conn)$ & $G_E^{u-d}$ & $G_M^p$ & $G_M^n$ & $G_M^{u+d} (\rm conn)$ & $G_M^{u-d}$ \\
\hline
0.0575 & 0.865(10) & 0.0181(76) & 2.634(23) & 0.847(17) & 2.231(52) & -1.425(45) & 2.553(53) & 3.656(96) \\
0.1133 & 0.744(15) & 0.040(14) & 2.328(21) & 0.704(29) & 1.991(37) & -1.276(33) & 2.257(34) & 3.267(69) \\
0.1675 & 0.654(21) & 0.042(18) & 2.060(26) & 0.612(38) & 1.765(26) & -1.120(21) & 2.027(34) & 2.886(46) \\
0.2202 & 0.606(15) & 0.029(12) & 1.873(28) & 0.577(25) & 1.559(20) & -0.987(16) & 1.797(32) & 2.546(35) \\
0.2716 & 0.530(18) & 0.055(17) & 1.718(30) & 0.475(34) & 1.439(21) & -0.908(19) & 1.661(36) & 2.348(38) \\
0.3217 & 0.477(19) & 0.052(16) & 1.551(30) & 0.425(33) & 1.326(23) & -0.835(19) & 1.535(28) & 2.161(41) \\
0.4185 & 0.411(21) & 0.048(17) & 1.341(36) & 0.363(35) & 1.155(25) & -0.720(21) & 1.353(30) & 1.875(45) \\
0.4654 & 0.384(19) & 0.052(15) & 1.273(31) & 0.332(33) & 1.083(25) & -0.675(21) & 1.266(28) & 1.757(45) \\
0.5113 & 0.378(16) & 0.041(13) & 1.221(30) & 0.337(27) & 1.022(23) & -0.639(19) & 1.189(28) & 1.661(41) \\
0.5561 & 0.350(17) & 0.037(13) & 1.128(34) & 0.312(29) & 0.968(23) & -0.598(21) & 1.147(28) & 1.566(43) \\
0.6004 & 0.30(10) & 0.054(91) & 1.020(75) & 0.24(19) & 0.89(15) & -0.54(12) & 1.060(97) & 1.43(27) \\
0.6437 & 0.307(19) & 0.029(14) & 0.976(47) & 0.279(29) & 0.899(20) & -0.560(16) & 1.048(30) & 1.459(35) \\
0.6862 & 0.278(23) & 0.035(18) & 0.908(53) & 0.242(37) & 0.837(25) & -0.516(22) & 0.991(33) & 1.353(46) \\
0.7692 & 0.261(34) & 0.070(23) & 0.964(79) & 0.191(52) & 0.709(48) & -0.431(42) & 0.856(49) & 1.140(88) \\
0.8099 & 0.242(23) & 0.052(19) & 0.851(50) & 0.190(40) & 0.685(37) & -0.409(33) & 0.850(32) & 1.095(69) \\
0.8497 & 0.230(22) & 0.046(19) & 0.802(53) & 0.184(38) & 0.685(34) & -0.415(29) & 0.831(39) & 1.100(62) \\
0.8889 & 0.184(30) & 0.032(25) & 0.62(11) & 0.152(41) & 0.695(33) & -0.419(27) & 0.848(54) & 1.114(58) \\
0.9278 & 0.196(25) & 0.048(23) & 0.707(84) & 0.148(38) & 0.691(33) & -0.422(28) & 0.826(41) & 1.114(60) \\
0.9659 & 0.202(24) & 0.049(21) & 0.727(59) & 0.154(41) & 0.643(32) & -0.387(28) & 0.787(36) & 1.031(59) \\
1.0007 & 0.180(22) & 0.079(20) & 0.750(49) & 0.101(39) & 0.604(37) & -0.353(33) & 0.771(32) & 0.957(69) \\
1.0748 & 0.186(38) & 0.073(19) & 0.75(13) & 0.114(44) & 0.573(81) & -0.336(56) & 0.728(94) & 0.91(14) \\
\hline
\end{tabular}
\end{table*}

\begin{table*}
\centering
\caption{Results for the electromagnetic form factors using the
  \texttt{cC211.060.80} ensemble for the proton, neutron, isoscalar
  connected and isovector connected combinations.}
\label{tab:results_C}
\begin{tabular}{ccccccccc}
\hline
\hline
$Q^2$ & $G_E^p$ & $G_E^n$ & $G_E^{u+d} (\rm conn)$ & $G_E^{u-d}$ & $G_M^p$ & $G_M^n$ & $G_M^{u+d} (\rm conn)$ & $G_M^{u-d}$ \\
\hline
0.0511 & 0.867(13) & 0.030(12) & 2.678(17) & 0.837(24) & 2.398(58) & -1.576(59) & 2.617(32) & 3.97(12) \\
0.1008 & 0.7669(94) & 0.0344(77) & 2.383(14) & 0.733(17) & 2.078(21) & -1.334(18) & 2.357(21) & 3.412(38) \\
0.1492 & 0.665(29) & 0.059(24) & 2.146(25) & 0.607(52) & 1.837(37) & -1.165(22) & 2.124(51) & 3.001(59) \\
0.1964 & 0.614(27) & 0.049(25) & 1.961(22) & 0.565(52) & 1.681(40) & -1.069(24) & 1.929(53) & 2.750(63) \\
0.2425 & 0.560(14) & 0.045(12) & 1.786(21) & 0.515(25) & 1.537(17) & -0.967(13) & 1.792(21) & 2.505(30) \\
0.2876 & 0.512(16) & 0.043(13) & 1.634(24) & 0.469(28) & 1.402(23) & -0.876(20) & 1.648(20) & 2.278(43) \\
0.3752 & 0.4541(89) & 0.0189(64) & 1.388(30) & 0.435(12) & 1.256(13) & -0.7876(92) & 1.464(20) & 2.044(21) \\
0.4174 & 0.4285(90) & 0.0215(60) & 1.319(29) & 0.407(12) & 1.182(13) & -0.742(11) & 1.372(18) & 1.924(23) \\
0.4590 & 0.397(13) & 0.0355(96) & 1.267(32) & 0.362(20) & 1.105(20) & -0.691(15) & 1.288(24) & 1.796(34) \\
0.4999 & 0.383(10) & 0.0319(70) & 1.214(28) & 0.351(15) & 1.051(16) & -0.656(13) & 1.231(19) & 1.707(28) \\
0.5401 & 0.353(18) & 0.006(12) & 1.049(79) & 0.347(15) & 1.008(14) & -0.6403(98) & 1.145(27) & 1.649(22) \\
0.5794 & 0.326(18) & 0.0246(92) & 1.022(60) & 0.301(19) & 0.938(17) & -0.590(13) & 1.084(26) & 1.528(29) \\
0.6181 & 0.311(24) & 0.008(13) & 0.93(11) & 0.304(17) & 0.915(39) & -0.574(25) & 1.058(48) & 1.490(63) \\
0.6933 & 0.172(32) & 0.109(33) & 0.82(13) & 0.063(48) & 0.702(58) & -0.397(43) & 0.948(86) & 1.099(99) \\
0.7315 & 0.246(24) & 0.011(17) & 0.746(97) & 0.235(27) & 0.811(17) & -0.508(13) & 0.938(32) & 1.318(29) \\
0.7676 & 0.245(26) & 0.025(17) & 0.783(82) & 0.220(34) & 0.771(25) & -0.472(22) & 0.924(29) & 1.244(46) \\
0.8032 & 0.198(47) & -0.016(30) & 0.52(20) & 0.214(38) & 0.737(33) & -0.462(22) & 0.851(60) & 1.199(53) \\
0.8394 & 0.208(52) & 0.002(29) & 0.61(20) & 0.206(50) & 0.710(34) & -0.442(27) & 0.830(47) & 1.152(59) \\
0.8739 & 0.25(14) & 0.017(89) & 0.77(69) & 0.231(53) & 0.716(22) & -0.444(15) & 0.838(30) & 1.160(37) \\
0.9066 & 0.225(44) & 0.008(33) & 0.68(23) & 0.218(16) & 0.687(15) & -0.428(11) & 0.800(33) & 1.115(24) \\
0.9739 & 0.2376(78) & 0.0250(50) & 0.767(34) & 0.2126(69) & 0.6602(96) & -0.4103(66) & 0.769(17) & 1.070(15) \\
\hline
\end{tabular}
\end{table*}

\begin{table*}
\centering
\caption{Results for the electromagnetic form factors using the
  \texttt{cD211.054.96} ensemble for the proton, neutron, isoscalar
  connected and isovector connected combinations.}
\label{tab:results_D}
\begin{tabular}{ccccccccc}
\hline
\hline
$Q^2$ & $G_E^p$ & $G_E^n$ & $G_E^{u+d} (\rm conn)$ & $G_E^{u-d}$ & $G_M^p$ & $G_M^n$ & $G_M^{u+d} (\rm conn)$ & $G_M^{u-d}$ \\
\hline
0.0505 & 0.854(11) & 0.0330(95) & 2.648(14) & 0.821(20) & 2.431(61) & -1.584(58) & 2.692(32) & 4.02(12) \\
0.0996 & 0.754(10) & 0.0382(87) & 2.352(14) & 0.715(18) & 2.086(25) & -1.329(22) & 2.391(23) & 3.415(47) \\
0.1475 & 0.671(17) & 0.046(14) & 2.123(19) & 0.625(31) & 1.858(20) & -1.170(20) & 2.163(23) & 3.028(39) \\
0.1942 & 0.610(11) & 0.0338(83) & 1.901(22) & 0.576(19) & 1.688(13) & -1.0552(91) & 1.982(20) & 2.743(21) \\
0.2398 & 0.527(21) & 0.048(17) & 1.692(29) & 0.478(36) & 1.491(23) & -0.914(20) & 1.805(21) & 2.405(42) \\
0.2846 & 0.514(11) & 0.0344(77) & 1.611(23) & 0.479(17) & 1.415(13) & -0.872(11) & 1.691(15) & 2.287(24) \\
0.3710 & 0.386(27) & 0.056(22) & 1.293(46) & 0.331(47) & 1.148(40) & -0.683(37) & 1.443(25) & 1.831(77) \\
0.4130 & 0.359(65) & 0.058(33) & 1.22(11) & 0.301(96) & 1.104(79) & -0.655(64) & 1.389(55) & 1.76(14) \\
0.4542 & 0.376(74) & 0.023(11) & 1.16(21) & 0.352(80) & 1.09(13) & -0.673(84) & 1.28(14) & 1.76(22) \\
0.4948 & 0.363(13) & 0.0246(77) & 1.130(43) & 0.338(16) & 1.052(22) & -0.648(17) & 1.248(20) & 1.701(39) \\
0.5346 & 0.304(30) & 0.038(21) & 0.996(83) & 0.266(43) & 0.953(37) & -0.572(33) & 1.177(30) & 1.525(69) \\
0.5734 & 0.266(37) & 0.032(20) & 0.87(11) & 0.234(47) & 0.869(52) & -0.521(38) & 1.073(65) & 1.390(89) \\
0.6122 & 0.270(24) & 0.019(16) & 0.838(91) & 0.250(28) & 0.876(25) & -0.528(21) & 1.071(29) & 1.404(45) \\
0.6872 & 0.243(28) & 0.022(21) & 0.77(10) & 0.221(36) & 0.785(39) & -0.488(29) & 0.917(57) & 1.273(66) \\
0.7243 & 0.221(35) & 0.0002(256) & 0.64(16) & 0.220(30) & 0.768(28) & -0.470(21) & 0.916(47) & 1.239(47) \\
0.7603 & 0.246(24) & 0.005(17) & 0.73(11) & 0.241(18) & 0.805(15) & -0.486(10) & 0.978(30) & 1.291(24) \\
0.7962 & 0.193(16) & 0.023(18) & 0.621(71) & 0.170(24) & 0.725(23) & -0.441(17) & 0.871(47) & 1.166(37) \\
0.8315 & 0.207(14) & -0.004(13) & 0.582(56) & 0.211(19) & 0.698(23) & -0.435(17) & 0.809(46) & 1.133(37) \\
0.8661 & 0.218(32) & 0.008(13) & 0.65(12) & 0.210(29) & 0.690(24) & -0.421(17) & 0.824(46) & 1.111(39) \\
0.9004 & 0.165(15) & 0.054(18) & 0.634(74) & 0.111(22) & 0.600(32) & -0.344(28) & 0.785(48) & 0.943(59) \\
0.9652 & 0.224(74) & 0.040(26) & 0.77(30) & 0.184(50) & 0.614(39) & -0.369(27) & 0.748(42) & 0.983(66) \\
 \hline
\end{tabular}
\end{table*}

\begin{table*}
\centering
\caption{Results for isoscalar disconnected electromagnetic form
  factors. Results have been binned as explained in the text. Bins
  that do not contain any values are omitted.}
\label{tab:disc_results}.
\begin{tabular}{|c|cc|cc|cc|}
\hline
\hline
\multicolumn{1}{|c|}{} & \multicolumn{2}{|c|}{\texttt{cB211.072.64}} & \multicolumn{2}{c|}{\texttt{cC211.060.80}} & \multicolumn{2}{c|}{\texttt{cD211.054.96}} \\
\hline
$Q^2$ & $G_E^{u+d} (disc)$ & $G_M^{u+d} (disc)$  & $G_E^{u+d} (disc)$ & $G_M^{u+d} (disc)$  & $G_E^{u+d} (disc)$  & $G_M^{u+d} (disc)$ \\
\hline
0.05 & 0.0260(42) & -0.158(21) & 0.0202(39) & -0.196(19) & 0.0188(38) & -0.154(19) \\
0.09 & 0.0219(83) & -0.163(35) & 0.0232(46) & -0.160(17) & 0.0242(30) & -0.121(10) \\
0.11 & 0.0210(36) & -0.126(12) & 0.0294(29) & -0.1476(98) & 0.0264(28) & -0.1190(92) \\
0.13 & - & - & 0.0014(66) & -0.119(22) & 0.0323(58) & -0.097(19) \\
0.15 & 0.0313(81) & -0.086(24) & 0.0175(39) & -0.128(12) & 0.0349(41) & -0.086(12) \\
0.17 & 0.0313(47) & -0.088(12) & 0.0208(97) & -0.104(30) & 0.038(10) & -0.022(27) \\
0.19 & 0.045(14) & -0.072(37) & 0.0311(51) & -0.118(14) & 0.0347(50) & -0.083(12) \\
0.21 & 0.030(11) & -0.089(28) & 0.0328(40) & -0.1228(97) & 0.0372(37) & -0.0967(88) \\
0.23 & 0.0371(50) & -0.077(11) & 0.0227(41) & -0.113(10) & 0.0360(35) & -0.1091(83) \\
0.25 & 0.0374(53) & -0.086(11) & 0.0333(25) & -0.0954(56) & 0.0357(24) & -0.0931(53) \\
0.27 & 0.0365(45) & -0.0788(86) & 0.0260(45) & -0.070(10) & 0.0402(47) & -0.0969(87) \\
0.29 & 0.0390(28) & -0.0796(56) & 0.0258(34) & -0.0719(78) & 0.0394(38) & -0.1008(71) \\
0.31 & 0.0394(58) & -0.087(12) & 0.0321(25) & -0.0774(53) & 0.0434(27) & -0.0840(53) \\
0.33 & 0.0396(42) & -0.0742(81) & 0.0349(86) & -0.064(17) & 0.0085(78) & -0.059(16) \\
0.35 & 0.0341(30) & -0.0645(55) & 0.0156(98) & -0.022(21) & 0.0343(98) & -0.052(17) \\
0.37 & 0.038(14) & -0.012(25) & 0.0269(56) & -0.056(10) & 0.0220(49) & -0.0407(87) \\
0.39 & 0.032(12) & -0.068(24) & 0.0246(61) & -0.061(12) & 0.0289(33) & -0.0359(56) \\
0.41 & 0.0393(71) & -0.059(13) & 0.0334(31) & -0.0550(53) & 0.0283(29) & -0.0446(48) \\
0.43 & 0.0364(74) & -0.023(12) & 0.0307(36) & -0.0478(63) & 0.0280(36) & -0.0228(55) \\
0.45 & 0.0382(48) & -0.0536(84) & 0.0358(35) & -0.0543(58) & 0.0321(26) & -0.0362(37) \\
0.47 & 0.0371(38) & -0.0557(63) & 0.0393(24) & -0.0490(40) & 0.0321(42) & -0.0320(63) \\
0.49 & 0.0257(52) & -0.0570(78) & 0.0237(41) & -0.0507(60) & 0.0312(39) & -0.0251(55) \\
0.51 & 0.0318(35) & -0.0435(51) & 0.0329(29) & -0.0495(45) & 0.0314(29) & -0.0265(39) \\
0.53 & 0.0356(34) & -0.0408(49) & 0.0297(50) & -0.0271(84) & 0.0130(65) & -0.0292(85) \\
0.55 & 0.0291(49) & -0.0303(68) & 0.0271(34) & -0.0447(54) & 0.0326(36) & -0.0340(45) \\
0.57 & 0.0345(45) & -0.0274(57) & 0.0276(32) & -0.0368(46) & 0.0221(51) & -0.0225(64) \\
0.59 & 0.0309(38) & -0.0408(52) & 0.0281(45) & -0.0484(68) & 0.0253(77) & -0.020(11) \\
0.61 & 0.0335(64) & -0.0294(82) & 0.0328(42) & -0.0434(59) & 0.0278(36) & -0.0288(46) \\
0.63 & 0.0298(44) & -0.0387(59) & 0.0232(59) & -0.0387(81) & 0.0210(50) & -0.0255(56) \\
0.65 & 0.0234(67) & -0.0441(89) & 0.0238(40) & -0.0339(52) & 0.0285(42) & -0.0329(48) \\
0.67 & 0.017(11) & -0.045(14) & 0.0262(30) & -0.0364(39) & 0.0349(37) & -0.0269(47) \\
0.69 & 0.0310(54) & -0.0435(69) & 0.0292(36) & -0.0343(45) & 0.0374(35) & -0.0278(46) \\
0.71 & 0.0336(81) & -0.039(10) & 0.0295(29) & -0.0317(35) & 0.0298(27) & -0.0325(36) \\
0.73 & 0.0266(54) & -0.0411(68) & 0.0135(86) & -0.060(12) & 0.0407(72) & -0.0432(83) \\
0.75 & 0.0328(40) & -0.0338(48) & 0.0188(94) & -0.027(11) & 0.0122(77) & -0.056(11) \\
0.77 & 0.0284(49) & -0.0409(60) & 0.0196(73) & -0.0236(89) & 0.0464(82) & -0.0532(94) \\
0.79 & 0.0226(48) & -0.0336(58) & 0.0415(67) & -0.0274(73) & 0.0418(66) & -0.0279(66) \\
0.81 & 0.0282(38) & -0.0297(46) & 0.0103(51) & -0.0283(61) & 0.0357(53) & -0.0338(58) \\
0.83 & 0.043(12) & 0.009(15) & 0.0264(47) & -0.0373(57) & 0.0294(37) & -0.0358(46) \\
0.85 & 0.0174(99) & -0.006(12) & 0.0277(36) & -0.0254(42) & 0.0294(82) & -0.0307(92) \\
0.87 & 0.028(14) & 0.019(17) & 0.0257(58) & -0.0268(62) & 0.0290(50) & -0.0386(57) \\
0.89 & 0.0341(77) & 0.0013(87) & 0.0175(52) & -0.0279(62) & 0.0160(84) & -0.0308(94) \\
0.91 & 0.0230(73) & -0.0155(83) & 0.0239(47) & -0.0194(51) & 0.0274(42) & -0.0203(47) \\
0.93 & 0.0140(66) & -0.0071(75) & -0.0032(92) & -0.031(12) & 0.0383(86) & -0.0352(93) \\
0.95 & 0.0303(55) & -0.0158(56) & 0.0227(43) & -0.0148(47) & 0.0204(45) & -0.0240(49) \\
0.97 & 0.0268(74) & -0.0092(83) & 0.0124(95) & -0.0310(92) & 0.0248(93) & -0.010(11) \\
0.99 & 0.0143(75) & -0.0099(74) & 0.0278(46) & -0.0216(50) & 0.0306(41) & -0.0202(42) \\
\hline
\end{tabular}
\end{table*}

\begin{table*}
\centering
\caption{Fit parameters for the most probable fits shown in
  Fig.~\ref{fig:GEMpn} at $a=0$. For each form factor indicated in the
  first column, the $Q^2_\mathrm{cut}$ and fit form of the most
  probable fit is provided in the second column and the symbols used
  in the text for the fit parameters are provided in the third
  column. The values and errors of the fit parameters are given in the
  fourth column and their corresponding covariance matrix in the
  fifth. }
\label{tab:fitp}
\begin{tabular}{|c|c|c|c|c|}
    \hline
    \hline
    & Fit & Fit parameters, $\vec{p}$ & Results for $\vec{p}$ & Covariance of fit parameters, cov$(\vec{p})$ \\
    \hline
    $G_E^p$ &
    
    z-exp, $Q^2_{\rm cut} = 0.85\;\rm GeV^2$ &

    $\begin{pmatrix}
    c_{1,0}\\ c_{2,0}\\ c_{3,0}
    \end{pmatrix}$ &
    
    $\begin{pmatrix}
    -0.936(84)\\ -1.16(26)\\ -0.54(45)
    \end{pmatrix}$ &
    
    $\begin{pmatrix}
    0.007 & -0.015 & 0.013 \\
    -0.015 & 0.07 & -0.098 \\
    0.013 & -0.098 & 0.203
    \end{pmatrix}$ \\

    \hline

    $G_M^p$ &
    
    z-exp, $Q^2_{\rm cut} = 1\;\rm GeV^2$ &

    $\begin{pmatrix}
    c_{0,0}\\ c_{1,0}\\ c_{2,0}
    \end{pmatrix}$ &
    
    $\begin{pmatrix}
    2.88(10)\\ -2.77(42)\\ -4.17(66)
    \end{pmatrix}$ &
    
    $\begin{pmatrix}
    0.01 & -0.038 & 0.039 \\
    -0.038 & 0.176 & -0.238 \\
    0.039 & -0.238 & 0.441
    \end{pmatrix}$ \\

    \hline

    $G_M^n$ &
    
    z-exp, $Q^2_{\rm cut} = 1\;\rm GeV^2$ &

    $\begin{pmatrix}
    c_{0,0}\\ c_{1,0}\\ c_{2,0}
    \end{pmatrix}$ &
    
    $\begin{pmatrix}
    -1.837(85)\\ 1.97(37)\\ 2.39(58)
    \end{pmatrix}$ &
    
    $\begin{pmatrix}
    0.007 & -0.028 & 0.032 \\
    -0.028 & 0.136 & -0.188 \\
    0.032 & -0.188 & 0.337
    \end{pmatrix}$ \\

    \hline

    $G_E^n$ &
    
    Galster-like, $Q^2_{\rm cut} = 0.3\;\rm GeV^2$ &

    $\begin{pmatrix}
    A\\ B
    \end{pmatrix}$ &
    
    $\begin{pmatrix}
    2.23(73)\\ 23(14)
    \end{pmatrix}$ &
    
    $\begin{pmatrix}
    0.537 & 9.144 \\
    9.144 & 187.346
    \end{pmatrix}$ \\

    \hline
    
\end{tabular}
\end{table*}

\end{document}